\documentclass{aa} 

\newcommand{\todo}{\ifmmode \text{\color{red}\Huge{\(\bullet\)}} \else {\color{red}{\Huge$\bullet$}}\fi}
\newcommand{\tido}{\ifmmode {{\color{red}\bullet}} \else {\color{red}$\bullet$}\fi}

\newcommand{\E        }[1]{\ifmmode 10^{#1} \else $10^{#1}$\fi}
\newcommand{\tE        }[1]{\ifmmode \times10^{#1} \else $\times10^{#1}$\fi}

\newcommand{\pc}	{\ifmmode {\rm pc} \else pc\fi}
\newcommand{\kpc}	{\ifmmode {\rm kpc} \else kpc\fi}
\newcommand{\ld}	{\ifmmode {\rm l.d.} \else l.d.\fi}
\newcommand{\kms}	{\ifmmode {\rm km\,s}^{-1} \else km\,s$^{-1}$\fi}
\newcommand{\cc}	{\ifmmode {\rm cm}^{-3}    \else cm$^{-3}$\fi}
\newcommand{\cmii}	{\ifmmode {\rm cm}^{-2}    \else cm$^{-2}$\fi}
\newcommand{\ergss}	{\ifmmode {\rm erg\,s}^{-1} \else erg s$^{-1}$\fi}
\newcommand{\ergcms}	{\ifmmode {\rm erg\,cm}^{-2}\,{\rm s}^{-1} \else erg\,cm$^{-2}$\,s$^{-1}$\fi}
\newcommand{\ergcmsA}	{\ifmmode {\rm erg\,cm}^{-2}\,{\rm s}^{-1}\,{\rm \AA}^{-1}
\else erg\,cm$^{-2}$\,s$^{-1}$\,\AA$^{-1}$\fi}
\newcommand{  \ergcmsHz  }{\ifmmode{\rm erg\,cm}^{-2}\,{\rm s}^{-1}\,{\rm Hz}^{-1}
                       \else ergs\,cm$^{-2}$\,s$^{-1}$\,Hz$^{-1}$\fi}
\newcommand{\kev}	{\ifmmode {\rm keV} \else keV\fi}

\newcommand{\mic}	{\ifmmode {\rm \mu m} \else $\mu$m\fi}
\newcommand{\vFWHM}	{\ifmmode v_{\mbox{\tiny FWHM}} \else $v_{\mbox{\tiny FWHM}}$\fi}
\newcommand{\vBLR}	{\ifmmode v_{\mbox{\tiny BLR}} \else $v_{\mbox{\tiny BLR}}$\fi}
\newcommand{\sigBLR}	{\ifmmode \sigma_{\mbox{\tiny BLR}} \else $\sigma_{\mbox{\tiny BLR}}$\fi}
\newcommand{\vNLR}	{\ifmmode v_{\mbox{\tiny NLR}} \else $v_{\mbox{\tiny NLR}}$\fi}
\newcommand{\tauBLR}	{\ifmmode \tau_{\mbox{\tiny BLR}} \else $\tau_{\mbox{\tiny BLR}}$\fi}

\newcommand{\Hubble}	{\ifmmode {\rm km\,s}^{-1}\,{\rm Mpc}^{-1} \else km\,s$^{-1}$\,Mpc$^{-1}$\fi}
\newcommand{\NDunit}	{\ifmmode {\rm Mpc}^{-3} \else Mpc$^{-3}$\fi}
\newcommand{\LFunit}	{\ifmmode {\rm Mpc}^{-3}\,{\rm mag}^{-1} \else Mpc$^{-3}$\,mag$^{-1}$\fi}
\newcommand{\MFunit}	{\ifmmode {\rm Mpc}^{-3}\,{\rm dex}^{-1} \else Mpc$^{-3}$\,dex$^{-1}$\fi}

\newcommand{\Msun}{\ifmmode M_{\odot} \else $M_{\odot}$\fi}
\newcommand{\Lsun}{\ifmmode L_{\odot} \else $L_{\odot}$\fi}
\newcommand{\Zsun}{\ifmmode Z_{\odot} \else $Z_{\odot}$\fi}
\newcommand{\mpyr}{\ifmmode \Msun\,{\rm yr}^{-1} \else $\Msun\,{\rm yr}^{-1}$\fi}

\newcommand{\qnote}{\ifmmode q_{0} \else $q_{0}$\fi}
\newcommand{\Hnote}{\ifmmode H_{0} \else $H_{0}$\fi}
\newcommand{\hnote}{\ifmmode h_{0} \else $h_{0}$\fi}
\newcommand{\anote}{\ifmmode a_{0} \else $a_{0}$\fi}
\newcommand{\tnote}{\ifmmode t_{0} \else $t_{0}$\fi}

\def\gsim{\;\rlap{\lower 2.5pt \hbox{$\sim$}}\raise 1.5pt\hbox{$>$}\;}
\def\lsim{\;\rlap{\lower 2.5pt \hbox{$\sim$}}\raise 1.5pt\hbox{$<$}\;}

\newcommand{  \Halpha   }{\ifmmode {\rm H}\alpha \else H$\alpha$\fi}

\newcommand{  \ha       }{\Halpha}
\newcommand{  \Hbeta    }{\ifmmode {\rm H}\beta \else H$\beta$\fi}

\newcommand{  \hb       }{\Hbeta}
\newcommand{  \Hgamma   }{\ifmmode {\rm H}\gamma \else H$\gamma$\fi}
\newcommand{  \Hdelta   }{\ifmmode {\rm H}\delta \else H$\delta$\fi}
\newcommand{  \Lya      }{\ifmmode {\rm Ly}\alpha \else Ly$\alpha$\fi}
\newcommand{  \Lyb      }{\ifmmode {\rm Ly}\beta \else Ly$\beta$\fi}
\newcommand{  \Pa       }{\ifmmode {\rm P}\alpha \else P$\alpha$\fi}
\newcommand{  \Pb       }{\ifmmode {\rm P}\beta \else P$\beta$\fi}
\newcommand{  \Bra      }{\ifmmode {\rm Br}\alpha \else Br$\alpha$\fi}
\newcommand{  \Brg      }{\ifmmode {\rm Br}\gamma \else Br$\gamma$\fi}
\newcommand{  \hii      }{\ifmmode {\rm H}\,\textsc{ii} \else H\,\textsc{ii}\fi}
\newcommand{  \hei      }{\ifmmode {\rm He}\,\textsc{i} \else He\,\textsc{i}\fi}
\newcommand{  \heii     }{\ifmmode {\rm He}\,\textsc{ii} \else He\,\textsc{ii}\fi}
\newcommand{  \HeIIuv   }{\ifmmode {\rm He}\,\textsc{ii}\,\lambda1640 \else He\,\textsc{ii}\,$\lambda1640$\fi}
\newcommand{  \HeIIop   }{\ifmmode {\rm He}\,\textsc{ii}\,\lambda4686 \else He\,\textsc{ii}\,$\lambda4686$\fi}
\newcommand{  \CII	}{\ifmmode \left[{\rm C}\,\textsc{ii}\right]\,\lambda157.74\,\mu{\rm m} \else [C\,{\sc ii}]\ $\lambda157.74\,\mu{\rm m}$\fi}
\newcommand{  \cii	}{\ifmmode \left[{\rm C}\,\textsc{ii}\right] \else [C\,{\sc ii}]\fi}

\newcommand{  \ciii     }{\ifmmode {\rm C}\,\textsc{iii}\right] \else C\,\textsc{iii}]\fi}
\newcommand{  \CIII     }{\ifmmode {\rm C}\,\textsc{iii}\right]\,\lambda1909 \else C\,\textsc{iii}]\,$\lambda1909$\fi}
\newcommand{  \civ      }{\ifmmode {\rm C}\,\textsc{iv}  \else C\,\textsc{iv}\fi}
\newcommand{  \CIV      }{\ifmmode {\rm C}\,\textsc{iv}\,\lambda1549 \else C\,\textsc{iv}\,$\lambda1549$\fi}

\newcommand{\COIItoI}{\ifmmode ^{12}{\rm CO}\,\mbox{(2--1)} \else  $^{12}{\rm CO}$\,\mbox{(2--1)}}
\newcommand{  \NIIopt   }{\ifmmode \left[{\rm N}\,\textsc{ii}\right]\,\lambda6584 \else [N\,\textsc{ii}]\,$\lambda6584$\fi}
\newcommand{  \nii      }{\ifmmode \left[{\rm N}\,\textsc{ii}\right]  \else [N\,\textsc{ii}]\fi}
\newcommand{  \niii     }{\ifmmode {\rm N}\,\textsc{iii} \else N\,\textsc{iii}\fi}
\newcommand{  \NIII     }{\ifmmode {\rm N}\,\textsc{iii}\,\lambda4640 \else N\,\textsc{iii}\,$\lambda4640$\fi}
\newcommand{  \niv      }{\ifmmode {\rm N}\,\textsc{iv}  \else N\,\textsc{iv}\fi}
\newcommand{  \NIVuv    }{\ifmmode {\rm N}\,\textsc{iv}\,\lambda1486 \else N\,\textsc{iv}\,$\lambda1486$\fi}
\newcommand{  \nv       }{\ifmmode {\rm N}\,\textsc{v}   \else N\,\textsc{v}\fi}
\newcommand{\oi}{\ifmmode \left[{\rm O}\,\textsc{i}\right] \else [O\,{\sc i}]\fi}
\newcommand{\OI}{\ifmmode \left[{\rm O}\,\textsc{i}\right]\,\lambda6300 \else [O\,{\sc i}]$\,\lambda6300$\fi}
\newcommand{\oii}{\ifmmode \left[{\rm O}\,\textsc{ii}\right] \else [O\,{\sc ii}]\fi}
\newcommand{\OII}{\ifmmode \left[{\rm O}\,\textsc{ii}\right]\,\lambda3727 \else [O\,{\sc ii}]\,$\lambda3727$\fi}
\newcommand{\oiii}{\ifmmode \left[{\rm O}\,\textsc{iii}\right] \else [O\,{\sc iii}]\fi}
\newcommand{\OIII}{\ifmmode \left[{\rm O}\,\textsc{iii}\right]\,\lambda5007 \else [O\,{\sc iii}]\,$\lambda5007$\fi}
\newcommand{  \OIIIbf   }{\ifmmode {\rm O}\,\textsc{iii}\,\lambda3133 \else O\,\textsc{iii}\,$\lambda3133$\fi}
\newcommand{  \OIIIuv   }{\ifmmode {\rm O}\,\textsc{iii}\,\lambda1663 \else O\,\textsc{iii}\,$\lambda1663$\fi}
\newcommand{  \oiv      }{\ifmmode {\rm O}\,\textsc{iv}  \else O\,\textsc{iv}\fi}
\newcommand{  \OIVuv    }{\ifmmode {\rm O}\,\textsc{iv}\,\lambda1402  \else O\,\textsc{iv}\,$\lambda1402$\fi}
\newcommand{  \OIVIR    }{\ifmmode {\rm O}\,\textsc{iv}\,25.9\,\mu {\rm m} \else O\,\textsc{iv}\,$25.9\,\mu$m\fi}
\newcommand{  \ovi      }{\ifmmode {\rm O}\,\textsc{vi}   \else O\,\textsc{vi}\fi}
\newcommand{  \Ovi      }{\ifmmode {\rm O}\,\textsc{vi}\,\lambda1035 \else O\,\textsc{vi}\,$\lambda1035$\fi}
\newcommand{  \nei      }{\ifmmode {\rm Ne}\,\textsc{i}   \else Ne\,\textsc{i}\fi}
\newcommand{  \neii     }{\ifmmode {\rm Ne}\,\textsc{ii}  \else Ne\,\textsc{ii}\fi}
\newcommand{  \NeiiIR   }{\ifmmode {\rm Ne}\,\textsc{ii}\,12.8\,\mu {\rm m} \else Ne\,\textsc{ii}\,$12.8\,\mu$m\fi}
\newcommand{  \neiii    }{\ifmmode {\rm Ne}\,\textsc{iii} \else Ne\,\textsc{iii}\fi}
\newcommand{  \neiv     }{\ifmmode {\rm Ne}\,\textsc{iv}  \else Ne\,\textsc{iv}\fi}
\newcommand{  \nev      }{\ifmmode {\rm Ne}\,\textsc{v}   \else Ne\,\textsc{v}\fi}
\newcommand{  \NevIR    }{\ifmmode {\rm Ne}\,\textsc{v}\,24.3\,\mu {\rm m} \else Ne\,\textsc{v}\,$24.3\,\mu$m\fi}
\newcommand{  \nevi     }{\ifmmode {\rm Ne}\,\textsc{vi}  \else Ne\,\textsc{vi}\fi}
\newcommand{  \mgi      }{\ifmmode {\rm Mg}\,\textsc{i} \else Mg\,\textsc{i}\fi}
\newcommand{  \mgii     }{\ifmmode {\rm Mg}\,\textsc{ii} \else Mg\,\textsc{ii}\fi}
\newcommand{  \MgII     }{\ifmmode {\rm Mg}\,\textsc{ii}\,\lambda2798 \else Mg\,\textsc{ii}\,$\lambda2798$\fi}
\newcommand{  \sii      }{\ifmmode {\rm S}\,\textsc{ii} \else S\,\textsc{ii}\fi}
\newcommand{  \siii     }{\ifmmode {\rm S}\,\textsc{iii} \else S\,\textsc{iii}\fi}
\newcommand{  \siv      }{\ifmmode {\rm S}\,\textsc{iv} \else S\,\textsc{iv}\fi}
\newcommand{  \sili     }{\ifmmode {\rm Si}\,\textsc{i}   \else Si\,\textsc{i}\fi}
\newcommand{  \silii    }{\ifmmode {\rm Si}\,\textsc{ii}  \else Si\,\textsc{ii}\fi}
\newcommand{  \Siliv    }{\ifmmode {\rm Si}\,\textsc{iv}  \else Si\,\textsc{iv}\fi}
\newcommand{  \SilIVuv  }{\ifmmode {\rm Si}\,\textsc{iv}\,\lambda1400  \else Si\,\textsc{iv}\,$\lambda1400$\fi}
\newcommand{  \AlIII   }{\ifmmode {\rm Al}\,\textsc{iii}\,\lambda1857 \else Al\,\textsc{iii}\,$\lambda1857$\fi}
\newcommand{  \Aliii   }{\ifmmode {\rm Al}\,\textsc{iii} \else Al\,\textsc{iii}\fi}
\newcommand{  \caii     }{\ifmmode {\rm Ca}\,\textsc{ii} \else Ca\,\textsc{ii}\fi}
\newcommand{  \feii     }{\ifmmode {\rm Fe}\,\textsc{ii} \else Fe\,\textsc{ii}\fi}
\newcommand{  \feiii    }{\ifmmode {\rm Fe}\,\textsc{iii} \else Fe\,\textsc{iii}\fi}
\newcommand{  \Kalpha   }{\ifmmode {\rm K}\alpha \else K$\alpha$\fi}

\newcommand{ \Lhb   }{\ifmmode L_{\hb} \else $L_{\hb}$\fi}
\newcommand{ \Lha   }{\ifmmode L_{\ha} \else $L_{\ha}$\fi}
\newcommand{ \fwhb  }{\ifmmode {\rm FWHM}\left(\hb\right) \else FWHM(\hb)\fi}
\newcommand{\sighb  }{\ifmmode \sigma\left(\hb\right) \else $\sigma\left(\hb\right)$\fi}
\newcommand{ \ewhb  }{\ifmmode {\rm EW}\left(\hb\right) \else EW(\hb)\fi}
\newcommand{ \fwha  }{\ifmmode {\rm FWHM}\left(\ha\right) \else FWHM(\ha)\fi}
\newcommand{ \ewha  }{\ifmmode {\rm EW}\left(\ha\right) \else EW(\ha)\fi}
\newcommand{ \Lmg   }{\ifmmode L\left(\mgii\right) \else $L\left(\mgii\right)$\fi}
\newcommand{ \fwmg  }{\ifmmode {\rm FWHM}\left(\mgii\right) \else FWHM(\mgii)\fi}
\newcommand{ \Lciv  }{\ifmmode L\left(\civ\right) \else $L\left(\civ\right)$\fi}
\newcommand{ \fwciv }{\ifmmode {\rm FWHM}\left(\civ\right) \else FWHM(\civ)\fi}
\newcommand{ \fwhm  }{\ifmmode {\rm FWHM} \else FWHM\fi} 
\newcommand{ \voff  }{\ifmmode v_{\rm off} \else $v_{\rm off}$\fi} 
\newcommand{ \vmax  }{\ifmmode v_{\rm max} \else $v_{\rm max}$\fi} 

\newcommand{ \mumg  }{\ifmmode \mu\left(\mgii\right) \else $\mu\left(\mgii\right)$\fi}
\newcommand{ \fmg   }{\ifmmode f\left(\mgii\right) \else $f\left(\mgii\right)$\fi}
\newcommand{ \muciv }{\ifmmode \mu\left(\civ\right) \else $\mu\left(\civ\right)$\fi}
\newcommand{ \fciv  }{\ifmmode f\left(\civ\right) \else $f\left(\civ\right)$\fi}

\newcommand{  \auvo     }{\ifmmode \alpha_{\nu,{\rm UVO}} \else $\alpha_{\nu,{\rm UVO}}$\fi}
\newcommand{  \Ledd     }{\ifmmode L_{\rm Edd} \else $L_{\rm Edd}$\fi}
\newcommand{  \lamLlam  }{\ifmmode \lambda L_{\lambda} \else $\lambda L_{\lambda}$\fi}
\newcommand{  \lLl      }{\ifmmode \lambda L_{\lambda} \else $\lambda L_{\lambda}$\fi}
\newcommand{  \nuLnu    }{\ifmmode \nu L_{\nu} \else $\nu L_{\nu}$\fi}
\newcommand{  \nLn      }{\ifmmode \nu L_{\nu} \else $\nu L_{\nu}$\fi}
\newcommand{  \Luv      }{\ifmmode L_{1450} \else $L_{1450}$\fi}
\newcommand{  \Lop      }{\ifmmode L_{5100} \else $L_{5100}$\fi}
\newcommand{  \lLop     }{\ifmmode \log\left(\Lop/\ergs\right) \else $\log\left(\Lop/\ergs\right)$\fi}
\newcommand{  \Lthree   }{\ifmmode L_{3000} \else $L_{3000}$\fi}
\newcommand{  \lLthree  }{\ifmmode \log\left(\Lthree/\ergs\right) \else $\log\left(\Lthree/\ergs\right)$\fi}
\newcommand{  \Lsix      }{\ifmmode L_{6200} \else $L_{6200}$\fi}
\newcommand{  \lLisx     }{\ifmmode \log\left(\Lop/\ergs\right) \else $\log\left(\Lop/\ergs\right)$\fi}
\newcommand{  \Lxray    }{\ifmmode L_{\rm X} \else $L_{\rm X}$\fi}

\newcommand{  \Lhard    }{\ifmmode L_{\rm 2-10} \else $L_{\rm 2-10}$\fi}
\newcommand{  \Lsoft    }{\ifmmode L_{\rm 0.5-2} \else $L_{\rm 0.5-2}$\fi}

\newcommand{\Fthree}{\ifmmode F_{3000} \else $F_{3000}$\fi}
\newcommand{\fuv}{\ifmmode f_{\lambda}\left(1450{\rm \AA}\right) \else $f_{\lambda}\left(1450 {\rm \AA}\right)$\fi}
\newcommand{\fthree}{\ifmmode f_{\lambda}\left(3000{\rm \AA}\right) \else $f_{\lambda}\left(3000{\rm \AA}\right)$\fi}
\newcommand{\fH}{\ifmmode f_{\lambda}\left(1.65\micron\right) \else
$f_{\lambda}\left(1.65\micron\right)$\fi}

\newcommand{\fbol}{\ifmmode f_{\rm bol} \else $f_{\rm bol}$\fi}
\newcommand{\fbolwv}{\ifmmode f_{\rm bol}\left(\lambda\right) \else $f_{\rm bol}\left(\lambda\right)$\fi}
\newcommand{\fbolopt}{\ifmmode f_{\rm bol}\left(5100{\rm \AA}\right) \else $f_{\rm bol}\left(5100{\rm \AA}\right)$\fi}
\newcommand{\fbolthree}{\ifmmode f_{\rm bol}\left(3000{\rm \AA}\right) \else $f_{\rm bol}\left(3000{\rm \AA}\right)$\fi}
\newcommand{\fboluv}{\ifmmode f_{\rm bol}\left(1450{\rm \AA}\right) \else $f_{\rm bol}\left(1450{\rm \AA}\right)$\fi}

\newcommand{\fbolbat}{\ifmmode f_{\rm bol}\left(14-150\,\kev\right) \else $f_{\rm bol}\left(14-150\,\kev\right)$\fi}
\newcommand{\fbolhard}{\ifmmode f_{\rm bol}\left(2-10\,\kev\right) \else $f_{\rm bol}\left(2-10\,\kev\right)$\fi}

\newcommand{\fobs}{\ifmmode f_{\rm obs} \else $f_{\rm obs}$\fi}

\newcommand{  \mbh      }{\ifmmode M_{\rm BH} \else $M_{\rm BH}$\fi}
\newcommand{  \lmbh     }{\ifmmode \log\left(\mbh/\Msun\right) \else $\log\left(\mbh/\Msun\right)$\fi} 
\newcommand{  \lledd    }{\ifmmode L/L_{\rm Edd} \else $L/L_{\rm Edd}$\fi}
\newcommand{  \mmedd    }{\ifmmode \dot{m}/\dot{m}_{\rm \,Edd} \else $\dot{m}/\dot{m}_{\rm \,Edd}$\fi}
\newcommand{  \Lbol     }{\ifmmode L_{\rm bol} \else $L_{\rm bol}$\fi}
\newcommand{  \lbol     }{\ifmmode L_{\rm bol} \else $L_{\rm bol}$\fi}
\newcommand{  \lLbol    }{\ifmmode \log\left(\Lbol/\ergs\right) \else $\log\left(\Lbol/\ergs\right)$\fi} 
\newcommand{  \Lagn     }{\ifmmode L_{\rm AGN} \else $L_{\rm AGN}$\fi}
\newcommand{  \lagn     }{\ifmmode L_{\rm AGN} \else $L_{\rm AGN}$\fi}

\newcommand{  \tgrow     }{\ifmmode t_{\rm growth} \else $t_{\rm growth}$\fi}
\newcommand{  \tAD     }{\ifmmode t_{\rm acc} \else $t_{\rm acc}$\fi}
\newcommand{  \tacc    }{\ifmmode t_{\rm acc} \else $t_{\rm acc}$\fi}
\newcommand{  \tUni      }{\ifmmode t_{\rm Universe} \else $t_{\rm Universe}$\fi}

\newcommand{  \Mdotin	}{\ifmmode \dot{M}_{\rm infall} \else $\dot{M}_{\rm infall}$\fi}
\newcommand{  \Mdotbh	}{\ifmmode \dot{M}_{\rm BH} \else $\dot{M}_{\rm BH}$\fi}
\newcommand{  \Mdotad	}{\ifmmode \dot{M}_{\rm AD} \else $\dot{M}_{\rm AD}$\fi}
\newcommand{  \Mdotacc	}{\ifmmode \dot{M}_{\rm acc} \else $\dot{M}_{\rm acc}$\fi}
\newcommand{  \Mdotthin	}{\ifmmode \dot{M}_{\rm thin} \else $\dot{M}_{\rm thin}$\fi}
\newcommand{  \Mdotdisk	}{\ifmmode \dot{M}_{\rm disk} \else $\dot{M}_{\rm disk}$\fi}

\newcommand{  \Mindot	}{\ifmmode \dot{M}_{\rm infall} \else $\dot{M}_{\rm infall}$\fi}
\newcommand{  \Mbhdot	}{\ifmmode \dot{M}_{\rm BH} \else $\dot{M}_{\rm BH}$\fi}
\newcommand{  \Maddot	}{\ifmmode \dot{M}_{\rm AD} \else $\dot{M}_{\rm AD}$\fi}
\newcommand{  \Maccdot	}{\ifmmode \dot{M}_{\rm acc} \else $\dot{M}_{\rm acc}$\fi}
\newcommand{  \Mthdot	}{\ifmmode \dot{M}_{\rm thin} \else $\dot{M}_{\rm thin}$\fi}
\newcommand{  \Mdsdot	}{\ifmmode \dot{M}_{\rm disk} \else $\dot{M}_{\rm disk}$\fi}

\newcommand{  \as	}{\ifmmode a_{\rm *} \else $a_{\rm *}$\fi}
\newcommand{  \avec	}{\ifmmode \vec{a}_{\rm *} \else $\vec{a}_{\rm *}$\fi}
\newcommand{  \re	}{\ifmmode \eta      	 \else $\eta$\fi}
\newcommand{  \RISCO	}{\ifmmode R_{\rm ISCO}  \else $R_{\rm ISCO}$\fi}

\newcommand{  \mseed    }{\ifmmode M_{\rm seed} \else $M_{\rm seed}$\fi}
\newcommand{  \mbul     }{\ifmmode M_{\rm bulge} \else $M_{\rm bulge}$\fi} 
\newcommand{  \mstar    }{\ifmmode M_{*} \else $M_{*}$\fi} 
\newcommand{  \mgal     }{\ifmmode M_{*} \else $M_{*}$\fi} 
\newcommand{  \mhost    }{\ifmmode M_{\rm host} \else $M_{\rm host}$\fi}
\newcommand{  \mmsmall  }{\ifmmode M_{\rm BH}/M_{*} \else $M_{\rm BH}/M_{*}$\fi}
\newcommand{  \mmlarge  }{\ifmmode M_{*}/M_{\rm BH} \else $M_{*}/M_{\rm BH}$\fi}

\newcommand{  \mmdotlarge}{\ifmmode \dot{M}_*/\Mbhdot \else $\dot{M}_*/\Mbhdot$\fi}
\newcommand{  \mmdotsmall}{\ifmmode \Mbhdot/\dot{M}_* \else $\Mbhdot/\dot{M}_*$\fi}

\newcommand{  \mmwp     }{\ifmmode \left(M_{*}/M_{\rm BH}\right) \else $\left(M_{*}/M_{\rm BH}\right)$\fi}
\newcommand{  \ml       }{\ifmmode M_{*}/L_{*} \else $M_{*}/L_{*}$\fi}
\newcommand{  \mlwp     }{\ifmmode \left(M_{*}/L\right) \else $\left(M_{*}/L\right)$\fi}
\newcommand{  \mlk      }{\ifmmode \left(M_{*}/L_{K}\right) \else $\left(M_{*}/L_{K}\right)$\fi}
\newcommand{  \sigs     }{\ifmmode \sigma_{*} \else $\sigma_{*}$\fi}
\newcommand{  \Reff     }{\ifmmode R_{\rm e} \else $R_{\rm e}$\fi}
\newcommand{  \Rvir     }{\ifmmode R_{\rm vir} \else $R_{\rm vir}$\fi}
\newcommand{  \Rtwo     }{\ifmmode R_{200} \else $R_{200}$\fi}
\newcommand{  \Rfive    }{\ifmmode R_{500} \else $R_{500}$\fi}
\newcommand{  \Rgrp     }{\ifmmode R_{\rm grp} \else $R_{\rm grp}$\fi}
\newcommand{  \nser     }{\ifmmode n_{\rm s} \else $n_{\rm s}$\fi}
\newcommand{  \LSF      }{\ifmmode L_{\rm SF}  \else $L_{\rm SF}$\fi}
\newcommand{  \LFIR     }{\ifmmode L_{\rm FIR} \else $L_{\rm FIR}$\fi}
\newcommand{  \Lfir     }{\ifmmode L_{\rm FIR} \else $L_{\rm FIR}$\fi}
\newcommand{  \LTIR     }{\ifmmode L_{\rm TIR} \else $L_{\rm TIR}$\fi}
\newcommand{  \Ltir     }{\ifmmode L_{\rm TIR} \else $L_{\rm TIR}$\fi}

\newcommand{  \mdyn     }{\ifmmode M_{\rm dyn} \else $M_{\rm dyn}$\fi} 
\newcommand{  \mgas     }{\ifmmode M_{\rm gas} \else $M_{\rm gas}$\fi} 
\newcommand{  \mh       }{\ifmmode M_{\rm h} \else $M_{\rm h}$\fi}
\newcommand{  \mhalo    }{\ifmmode M_{\rm halo} \else $M_{\rm halo}$\fi}
\newcommand{  \sfr      }{\ifmmode {\rm SFR} \else SFR\fi}

\newcommand{ \Lcii     }{\ifmmode L_{\cii} \else $L_{\cii}$\fi}
\newcommand{ \fwcii  }{\ifmmode {\rm FWHM}\cii \else FWHM\cii\fi}

\newcommand{\bj}{\ifmmode b_{\rm J} \else $b_{\rm J}$\fi}

\newcommand{\iab}{\ifmmode i_{\rm AB} \else $i_{\rm AB}$\fi}

\newcommand{\jab}{\ifmmode J_{\rm AB} \else $J_{\rm AB}$\fi}
\newcommand{\hab}{\ifmmode H_{\rm AB} \else $H_{\rm AB}$\fi}
\newcommand{\kab}{\ifmmode K_{\rm AB} \else $K_{\rm AB}$\fi}

\newcommand{\jveg}{\ifmmode J_{\rm Vega} \else $J_{\rm Vega}$\fi}
\newcommand{\hveg}{\ifmmode H_{\rm Vega} \else $H_{\rm Vega}$\fi}
\newcommand{\kveg}{\ifmmode K_{\rm Vega} \else $K_{\rm Vega}$\fi}

\def\arcmin{\hbox{$^\prime$}}
\def\arcsec{\hbox{$^{\prime\prime}$}}

\newcommand{  \Chisq    }{\ifmmode \chi^{2} \else $\chi^{2}$}
\newcommand{  \nelec    }{\ifmmode n_{e} \else $n_{e}$\fi}     
\newcommand{  \nh       }{\ifmmode n_{\rm H} \else $n_{\rm H}$\fi}     
\newcommand{  \Ncol     }{\ifmmode N_{\rm col} \else $N_{\rm col}$\fi} 
\newcommand{  \NH       }{\ifmmode N_{\rm H} \else $N_{\rm H}$\fi}     

\def\deg{\hbox{$^\circ$}}

\def\arcmin{\hbox{$^\prime$}}
\def\arcsec{\hbox{$^{\prime\prime}$}}
\def\ion#1#2{#1$\;${\small\rm\@Roman{#2}}\relax}

\newcommand{\SiX}{\ifmmode \left[{\rm Si}\,\textsc{x}\right]\,\lambda14300 \else [Si\,{\sc x}]\,$\lambda14300$\fi}
\newcommand{\SiVI}{\ifmmode \left[{\rm Si}\,\textsc{vi}\right]\,\lambda19640 \else [Si\,{\sc vi}]\,$\lambda19640$\fi}
\newcommand{\SXI}{\ifmmode \left[{\rm S}\,\textsc{xi}\right]\,\lambda19196 \else [S\,{\sc xi}]\,$\lambda19196$\fi}
\newcommand{\SVIII}{\ifmmode \left[{\rm S}\,\textsc{viii}\right]\,\lambda9915 \else [S\,{\sc viii}]\,$\lambda9915$\fi}
\newcommand{\SIX}{\ifmmode \left[{\rm S}\,\textsc{ix}\right]\,\lambda12520 \else [S\,{\sc ix}]\,$\lambda12520$\fi}
\newcommand{\FeXIII}{\ifmmode \left[{\rm Fe}\,\textsc{xiii}\right]\,\lambda10747 \else [Fe\,{\sc xiii}]\,$\lambda10747$\fi}

\newcommand{  \hi       }{\ifmmode {\rm H}\,\textsc{i} \else H\,\textsc{i}\fi}

\usepackage{graphicx}	
\usepackage{graphics}
\usepackage{amsmath}	
\usepackage{amssymb}	
\usepackage{bm}	        
\usepackage{csvsimple}
\usepackage{multirow}
\usepackage{siunitx} 
\usepackage[utf8]{inputenc}
\usepackage{dirtytalk}
\usepackage{xparse}
\usepackage{txfonts}
\newcommand{\orcid}[1]{\href{https://orcid.org/#1}{\includegraphics[width=10pt]{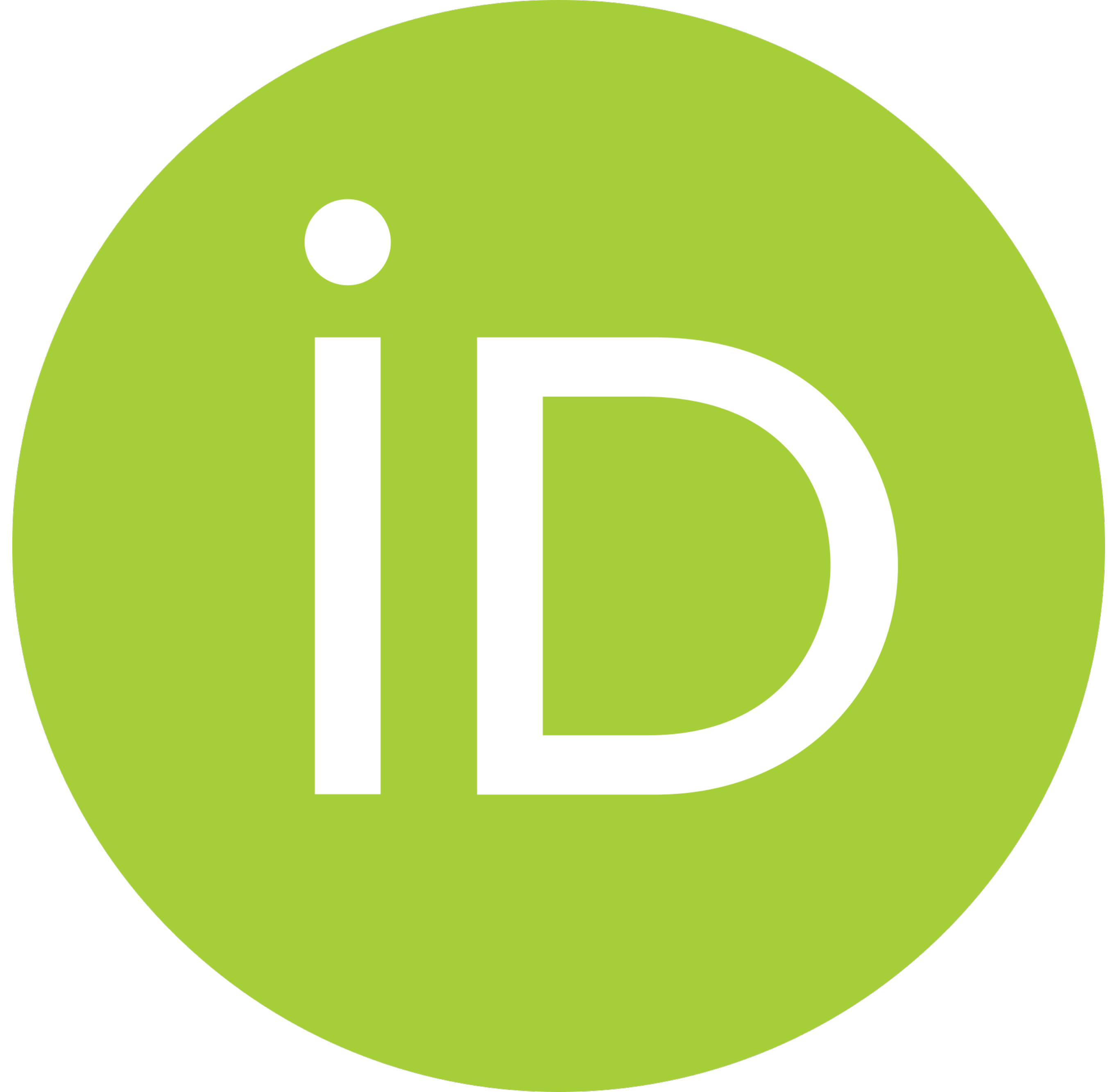}}}
\usepackage[pdfpagelabels=false]{hyperref}	
\hypersetup{colorlinks=true,linkcolor=blue,citecolor=blue,filecolor=blue,urlcolor=blue,}
\begin{document} 
\renewcommand{\figureautorefname}{Fig.} 
\renewcommand{\equationautorefname}{Eq.} 
\renewcommand{\sectionautorefname}{Section} 
\renewcommand{\subsectionautorefname}{Section}
\renewcommand{\subsubsectionautorefname}{Section}
\renewcommand{\appendixautorefname}{Appendix} 



\newcommand*{\species}[1]{%
    \ensuremath{\IfBold{\bm{\mathrm{#1}}}{\mathrm{#1}}}%
}

\newcommand*{\trans}[1]{%
    \ensuremath{\IfBold{\bm{\mytrans{#1}}}{\mytrans{#1}}}%
}

\newcommand{\delim}{{-}}
\newcommand*{\mytrans}[1]{%
    \ifnum#1=10 (1\delim0)\else%
    \ifnum#1=21 (2\delim1)\else%
    \ifnum#1=32 (3\delim2)\else%
    \ifnum#1=43 (4\delim3)\else%
    \ifnum#1=54 (5\delim4)\else%
    \ifnum#1=65 (6\delim5)\else%
    \ifnum#1=76 (7\delim6)\else%
    \ifnum#1=87 (8\delim7)\else%
    \ifnum#1=98 (9\delim8)\else%
    \ifnum#1=109 (10\delim9)\else#1%
    \fi\fi\fi\fi\fi\fi\fi\fi\fi\fi%
}

\makeatletter
\newcommand*{\IfBold}{%
  \ifx\f@series\my@test@bx
    \expandafter\@firstoftwo
  \else
    \expandafter\@secondoftwo
  \fi
}
\newcommand*{\my@test@bx}{bx}
\makeatother

\DeclareDocumentCommand{\chem}{ m g }{%
    {\species{#1}%
        \IfNoValueF {#2} {\,\trans{#2}}%
    }%
}

	\vspace{-100pt}
   \title{A \textit{CO isotopologue Line Atlas within the Whirlpool galaxy Survey} (CLAWS)}


   \author{Jakob S. den Brok \inst{1}\orcid{0000-0002-8760-6157} \thanks{\email{jdenbrok@astro.uni-bonn.de}} 
          \and
          Frank Bigiel \inst{1} \orcid{0000-0003-0166-9745}
          \and
          Kazimierz~Sliwa
          \and
          Toshiki Saito \inst{2, 3, 4} \orcid{0000-0002-2501-9328}
          \and 
          Antonio Usero \inst{5} \orcid{0000-0003-1242-505X}
          \and 
          Eva~Schinnerer \inst{2} \orcid{0000-0002-3933-7677}
          \and 
          Adam~K.~Leroy \inst{6} \orcid{0000-0002-2545-1700}
          \and 
          Mar\'ia~J.~Jim\'enez-Donaire \inst{5, 7} \orcid{0000-0002-9165-8080}
          \and
          Erik~Rosolowsky \inst{8} \orcid{0000-0002-5204-2259}
          \and
          Ashley T. Barnes \inst{1} \orcid{0000-0003-0410-4504}
          \and 
          Johannes Puschnig \inst{1} \orcid{0000-0003-1111-3951}
          \and 
          Jérôme Pety \inst{9, 10} \orcid{0000-0003-3061-6546}
          \and
          Andreas Schruba \inst{11}
          \and 
          Ivana~Be\v{s}li\'c \inst{1} \orcid{0000-0003-0783-0157 }
          \and 
          Yixian Cao \inst{11} \orcid{0000-0001-5301-1326}
          \and 
          Cosima Eibensteiner \inst{1} \orcid{0000-0002-1185-2810}
          \and
          Simon~C.~O.~Glover \inst{12} \orcid{0000-0001-6708-1317}
          \and 
          Ralf~S.~Klessen \inst{12, 13} \orcid{0000-0002-0560-3172}
          \and 
          J.~M.~Diederik Kruijssen \inst{14} \orcid{0000-0002-8804-0212}
          \and
          Sharon~E.~Meidt \inst{15} \orcid{0000-0002-6118-4048}
          \and 
          Lukas Neumann \inst{1} \orcid{0000-0001-9793-6400}
          \and 
          Neven Tomi\v{c}i\'{c} \inst{16} \orcid{0000-0002-8238-9210}
          \and 
          Hsi-An Pan \inst{2, 17} \orcid{0000-0002-1370-6964}
          \and
          Miguel Querejeta \inst{5} \orcid{0000-0002-0472-1011}
          \and 
          Elizabeth Watkins \inst{14} \orcid{0000-0002-7365-5791}
          \and 
          Thomas G. Williams \inst{2} \orcid{0000-0002-0012-2142}
          \and 
          David~Wilner \inst{18} \orcid{0000-0003-1526-7587}
          }

   \institute{
        Argelander-Institut für Astronomie, Universität Bonn, Auf dem Hügel 71, 53121 Bonn, Germany
         \and
             Max-Planck-Institut f{\"u}r Astronomie, K{\"o}nigstuhl 17, D-69117 Heidelberg, Germany
        \and 
            Department of Physics, General Studies, College of Engineering, Nihon University, 1 Nakagawara, Tokusada, Tamuramachi, Koriyama, Fukushima, 963-8642, Japan
        \and 
            National Astronomical Observatory of Japan, 2-21-1 Osawa, Mitaka, Tokyo, 181-8588, Japan
        \and
            Observatorio Astron{\'o}mico Nacional (IGN), C/ Alfonso XII 3, E-28014 Madrid, Spain
        \and
            Department of Astronomy, The Ohio State University, 4055 McPherson Laboratory, 140 West 18th Avenue, Columbus, OH 43210, USA
        \and
            Centro de Desarrollos Tecnológicos, Observatorio de Yebes (IGN), 19141 Yebes, Guadalajara, Spain
        \and
            4-183 CCIS, University of Alberta, Edmonton, Alberta, T6G 2E1, Canada
        \and 
           Institut de Radioastronomie Millim\'{e}trique (IRAM), 300 Rue de la Piscine, F-38406 Saint Martin d'H\`{e}res, France
        \and
          LERMA, Observatoire de Paris, PSL Research University, CNRS, Sorbonne Universit\'es, 75014 Paris
        \and 
            Max-Planck-Institut f{\"u}r extraterrestrische Physik, Giessenbachstra{\ss}e~1, D-85748 Garching, Germany
        \and
            Universit\"{a}t Heidelberg, Zentrum f\"{u}r Astronomie, Institut f\"{u}r Theoretische Astrophysik, Albert-Ueberle-Str 2, D-69120 Heidelberg, Germany
        \and 
            Universit\"{a}t Heidelberg, Interdisziplin\"{a}res Zentrum f\"{u}r Wissenschaftliches Rechnen, Im Neuenheimer Feld 205, D-69120 Heidelberg, Germany
        \and 
            Astronomisches Rechen-Institut, Zentrum f\"{u}r Astronomie der Universit\"{a}t Heidelberg, M\"{o}nchhofstra\ss e 12-14, D-69120 Heidelberg, Germany
        \and 
            Sterrenkundig Observatorium, Universiteit Gent, Krijgslaan 281 S9, B-9000 Gent, Belgium 
        \and 
            INAF-Osservatorio Astronomico di Padova, Vicolo Osservatorio 5, 35122 Padova, Italy
        \and 
            Department of Physics, Tamkang University, No.151, Yingzhuan Rd., Tamsui Dist., New Taipei City 251301, Taiwan
        \and
            Center for Astrophysics \textbar\, Harvard \& Smithsonian, 60 Garden St., Cambridge, MA 02138, USA \\
             }

   \date{Received 17 September 2021; accepted 23 December 2021 }

 
  \abstract{We present the \textit{CO isotopologue Line Atlas within the Whirpool galaxy Survey} (CLAWS) based on an IRAM \mbox{30-m} large programme which provides a benchmark study of numerous, faint CO isotopologues in the mm-wavelength regime across the full disc of the nearby grand-design spiral galaxy M51 (NGC~5194). The survey's core goal is to use the low-\textit{J} CO isotopologue lines to constrain CO excitation and chemistry, and therefrom the local physical conditions of the gas. In this survey paper, we describe the CLAWS observing and data reduction strategies. We map the $J=1\!\rightarrow\!0$ and $2\!\rightarrow\!1$ transitions of the CO isotopologues \chem{^{12}CO}, \chem{^{13}CO}, \chem{C^{18}O} and \chem{C^{17}O}, as well as several supplementary lines within the 1\,mm and 3\,mm window (\chem{CN}{10}, \chem{CS}{21}, \chem{CH_3OH}{21}, \chem{N_2H^+}{10}, \chem{HC_3N}{109}) at ${\sim}1$\,kpc resolution. A total observation time of 149\,h offers unprecedented sensitivity.  
We use these data to explore several CO isotopologue line ratios in detail, study their radial (and azimuthal) trends and investigate whether changes in line ratios stem from changes in ISM properties such as gas temperatures, densities or chemical abundances.  
For example, we find negative radial trends for the $\chem{^{13}CO}/\chem{^{12}CO}$, $\chem{C^{18}O}/\chem{^{12}CO}$ and $\chem{C^{18}O}/\chem{^{13}CO}$ line ratios in their $J=1\!\rightarrow\!0$ transitions. We also find variations with local environment, such as higher $\chem{^{12}CO}{21}/\trans{10}$ or $\chem{^{13}CO}/\chem{^{12}CO}{10}$ line ratios in interarm regions compared to spiral arm regions.
We propose that these aforementioned variations of CO line ratios are most likely due to a variation of the optical depth, while abundance variations due to selective nucleosynthesis on a galaxy-wide scale could also play a role. We also study the CO spectral line energy distribution (SLED) using archival JCMT \chem{^{12}CO}{32} data and find a variation of the SLED shape with local environmental parameters further underlying changes in optical depth, gas temperatures or densities.}


   \maketitle
%


\section{Introduction}

A key to our understanding of the interstellar medium (ISM) and its chemical evolution is the study of emission from the second most abundant molecule after H$_2$, carbon monoxide (CO) and its isotopologues. Such isotopologue studies allow us to examine the physical conditions within the gas, study the enrichment of the ISM and open up the potential of deciphering the star formation history of a galaxy.
Due to CO's permanent dipole moment and low mass, it has low-energy rotational transitions. Consequently, the emission from these rotational transitions is excited and can be observed at low temperatures (${<}10$\,K) -- unlike for H$_2$, which is hardly excited and thus not observable under typical ISM conditions. While the low-\textit{J} CO transitions of the main isotopologue, \chem{^{12}CO}, are known to be optically thick, a relation of their emission with the molecular gas mass has been found via the CO-to-H$_2$ conversion factor $\alpha_{\rm CO}$ (e.g. \citealt{Solomon1987,Nakai1995,Leroy2011_aCO, Sandstrom2013}; or see review by \citealt{Bolatto2013})\\

The \chem{^{12}CO} line brightness temperature ratios between different rotational transitions are generally of great interest. High-$z$ observations typically observe higher-\textit{J} \chem{^{12}CO} lines \citep{Carilli2013}. By assuming line ratios, such studies can calculate an equivalent \chem{^{12}CO}{10} brightness temperature \citep[e.g.][]{Tacconi2008, Genzel2012, Canameras2018} and then convert to physical parameters such as the molecular gas mass using $\alpha_{\rm CO}$. Such studies often adopt CO line ratios and a CO-to-H$_2$ conversion factor measured in the local universe \citep[e.g.][]{Tacconi2008, Schruba2012, Sandstrom2013}, see also reviews by \citealt{Solomon2005, Carilli2013}). 
However, recent studies find variations in the line ratio within and among nearby spiral galaxies \citep{denbrok2021, Yajima2021, Leroy2021_ratio}, which have consequences for the down-conversion of high-\textit{J} CO transitions and the conversion to ISM physical parameters. Such variations are also expected from simulations. Modelling individual giant molecular clouds, \citet{Penaloza2018} find variations of the \chem{^{12}CO} line brightness temperature ratios of order $0.3$~dex and attribute these changes to varying environmental conditions including cloud mass and density, the interstellar radiation field, or the cosmic ray ionisation rate. Furthermore, the CO-to-H$_2$ conversion factor itself is subject to environmental variations (e.g. \citealt{Young1982, Sandstrom2013, Accurso2017}; see also simulations, e.g. \citealt{Shetty2011, Shetty2011_b, Gong2018, Gong2020}).  
The value for $\alpha_{\rm CO}$ is empirically calibrated either using many Milky Way clouds \citep{Solomon1987}, from CO, \hi\ and dust mass observations in external galaxies \citep{Sandstrom2013}, or using \textsc{[C\,ii]} emission \citep{Madden2020}. When comparing $\alpha_{\rm CO}$ to other galaxies, metallicity, presence of CO-dark gas and temperature variations relative to the Milky Way should be taken into account. Thus, it is important to constrain variations of the CO line ratio and conversion factor and understand their dependencies on the galactic environment and ISM conditions.

CO isotopologue transitions help to study the conditions of the ISM. Whereas the low \chem{^{12}CO} transitions usually remain optically thick, \chem{C^{18}O} and \chem{C^{17}O} lines stay optically thin over large parts of the galaxy. By contrast, the \chem{^{13}CO} emission can be optically thin or has a moderate optical depth, depending on its relative abundance \citep[see review by][]{Heyer2015}. Comparing two optically thick lines gives insight into the physical conditions of the emitting gas, such as its temperature or density \citep{Leroy2017_density, Donaire2019,denbrok2021}. Contrasting optically thin to optically thick lines allows us to analyse the optical depth of the gas and investigate the gas column and volume densities of the molecular gas \citep{Young1982,Pineda2008,Wilson2009}. Finally, studying the ratio of two optically thin lines can be used to study abundance variations within the Milky Way \citep{Langer1990,Milam2005} or across galaxy discs \citep{Donaire2017}.

The study of CO isotopologues can also be used to investigate the chemical enrichment of the molecular gas. C~and O~isotopes -- and consequently CO isotopologues -- are a direct byproduct of stellar evolution via the CNO cycle. By studying their abundances, the physical processes that generate the various CO isotopologue species can be analysed.
For instance, the $^{13}$C isotope is primarily produced in low-mass stars \citep{Wilson1994}, while $^{18}$O is mainly replenished due to massive stars \citep{Henkel1994}. This makes the CO isotopologues a useful diagnostic tool to study stellar populations \citep{Sliwa2017a,Sliwa2017b,Zhang2018,Brown2019}.
Previous studies of CO isotopologues and other C, N and O isotope ratios have already been carried out extensively for the Milky Way \citep{Langer1990,Wilson1992, Wilson1994,Henkel1994,Milam2005}. The past decade has also seen an increase in the study of CO isotopologues in extragalactic sources \citep{Martin2010,Henkel2014,Meier2015,Cao2017,Donaire2017_a,Donaire2017,Donaire2019,denbrok2021,Yajima2021}. 

The low-\textit{J} \chem{^{12}CO} transitions produce the brightest molecular lines (e.g.\ at $30$\,arcsec, we find brightness temperatures of around $60$\,K\,km\,s$^{-1}$ in the centre of M51) and consequently have been covered in numerous previous studies carried out with the IRAM \mbox{30-m} telescope and IRAM Northern Extended Millimeter Array (NOEMA), the Atacama Large Millimeter/\linebreak[0]{}submillimeter Array (ALMA),  and further mm-wavelength observatories \citep{Hasegawa1997,Hasegawa1997_2,Sakamoto1997,Leroy2009,Koda2011,Koda2020}. CO isotopologue ratios are harder to observe, as due to their lower abundance, the emission is of order ${\sim}10$ to ${\sim}50$ times fainter for \chem{^{13}CO} or \chem{C^{18}O}, respectively, compared to the \chem{^{12}CO} emission (see \autoref{tab:lines}). Therefore in the past, these transitions were usually either studied in the Milky Way \citep{Langer1990, Wilson1994, Sawada2001,Yoda2010} or in strongly active star-forming galaxies, such as starburst galaxies or \mbox{(ultra)}\linebreak[0]{}luminous infrared galaxies \citep[ULIRGS;][]{Meier2004,Costagliola2011,Aladro2013,Sliwa2017,Brown2019}. Only in recent years, due to the advancement of state-of-the-art receivers and the carrying out of large programs, have we seen an increase in CO isotopologue line surveys of star-forming, spiral galaxies.

The \textit{EMIR Multiline Probe of the ISM Regulating Galaxy Evolution} (EMPIRE) survey targeted nine galaxies and covered the \chem{^{13}CO} and \chem{C^{18}O} $J = 1\!\rightarrow\!0$ transitions \citep{Donaire2019}. It has been found, for example, that the \chem{^{13}CO} to \chem{C^{18}O} line ratio is much lower in ULIRGs and star-bursting systems than compared to the Milky Way or nearby normal star-forming galaxies \citep{Greve2009,Matsushita2009, Donaire2017, Brown2019}, consistent with recent or ongoing star formation as well as a top-heavy stellar initial mass function \citep{Brown2019}.
Furthermore, \cite{Cormier2018} used the optically thin \chem{^{13}CO}{10} line from EMPIRE to derive a spatially resolved \chem{^{13}CO}-to-H$_2$ conversion factor for nearby galaxies.  

With the \textit{CO Isotopologue Line Atlas within the Whirlpool galaxy Survey} (CLAWS), we use the IRAM \mbox{30-m} telescope to provide an in-depth analysis of low-\textit{J} transitions of CO and several isotopologues (\chem{^{13}CO}, \chem{C^{18}O} and \chem{C^{17}O}) over the entire star-forming and molecular disc ($6.6\,{\rm arcmin} \times 6.6\,{\rm arcmin}$) of the grand-design spiral galaxy M51 (NGC~5194). Due to its relative proximity ($D = 8.6$\,Mpc; \citealp{McQuinn2016}) and high surface density, it is routinely observed from high-energetic $X$-ray to radio wavelengths, thus providing a wealth of ancillary data. Previous wide field-of-view imaging observations have targeted different low-\textit{J} \chem{^{12}CO} and \chem{^{13}CO} emission line transitions (e.g. \citealt{Koda2011}; \citealt{Pety2013} and \citealt{Schinnerer2013} as part of PAWS; \citealt{Donaire2019} as part of EMPIRE) and even \chem{C^{18}O}  was observed towards a few bright regions inside M51 \citep[see][]{Schinnerer2010,Tan2011,Watanabe2014,Watanabe2016}. We will complement these studies and provide the currently most complete extragalactic CO isotopologue line atlas. 

The main goal of this project is to use the large number of CO isotopologues, to study the dependence of excitation on galactic environment and investigate isotopic abundance variations. In addition, we use the CO isotopologues to constrain the spatial variation of the CO-to-H$_2$ conversion factor. 

\begin{table}
    \begin{center}
    \caption{M51 source description}
    \label{tab:target_descr}
    \begin{tabular}{l l}
    \hline \hline
     Property    & Value  \\ \hline
     Name         & NGC~5194 (M51) \\
     R.A. (J2000)$^{\rm (a)}$ & $13^\mathrm{h}29^\mathrm{m}52\fs7$ \\
     Decl (J2000)$^{\rm (a)}$ & $47\degr11\arcmin43\arcsec$ \\
     $i^{\rm (b)}$ & {$22^\circ$} \\
     P.A.$^{\rm (c)}$ & $172^\circ$ \\
     $r_{25}^{\rm (d)}$ & $3.9^\prime$ \\
     $D^{\rm (e)}$ & $8.6$\,Mpc \\
     $V_{\rm hel}^{\rm (f)}$ & $456.2\,$km\,s$^{-1}$\\
     Metallicity$^{\rm (g)}$ [$12+\log(\rm O/H)$] & 8.55 \\
     Morphology$^{\rm (h)}$ & SAbc \\
     $\langle\Sigma_{\rm SFR}\rangle^{\rm (i)}$  & $20\times 10^{-3}$\, M$_\odot$\,yr$^{-1}$\,kpc$^{-2}$ \\
     $\log_{10}(M_\star/\mathrm{M}_\odot)^{\rm (j)}$ & 10.5 \\ \hline
    \end{tabular}
     \end{center}
    \raggedright{ {\bf Notes:} (a) Coordinates of the centre of the galaxy adopted from the NASA Extragalactic Database (NED); From \cite{Shetty2007}: (b) inclination of the galaxy with respect to the plane of the sky, {{from \cite{Colombo2014}}}, (c) position angle of the galaxy; (d) 25th magnitude isophote radius of the $B$-band taken from the Extragalactic Distance Database \citep[EDD;][]{Tully2009}; (e) distance to the galaxy from \citep{McQuinn2016}; (f) heliocentric systemic velocity from \cite{Walter2008}; (g) metallicity averaged across the full galaxy from \citet{Moustakas2010}; (h) morphological type as given in \citet{Leroy2013}; Adopted from \citet{Dale2009}: (i) the average SFR surface density within $0.75 r_{25}$, and (j) integrated stellar mass derived from 3.6\,$\mu$m emission.}
\end{table}
\begin{figure}
    \centering
    \includegraphics[width = \columnwidth]{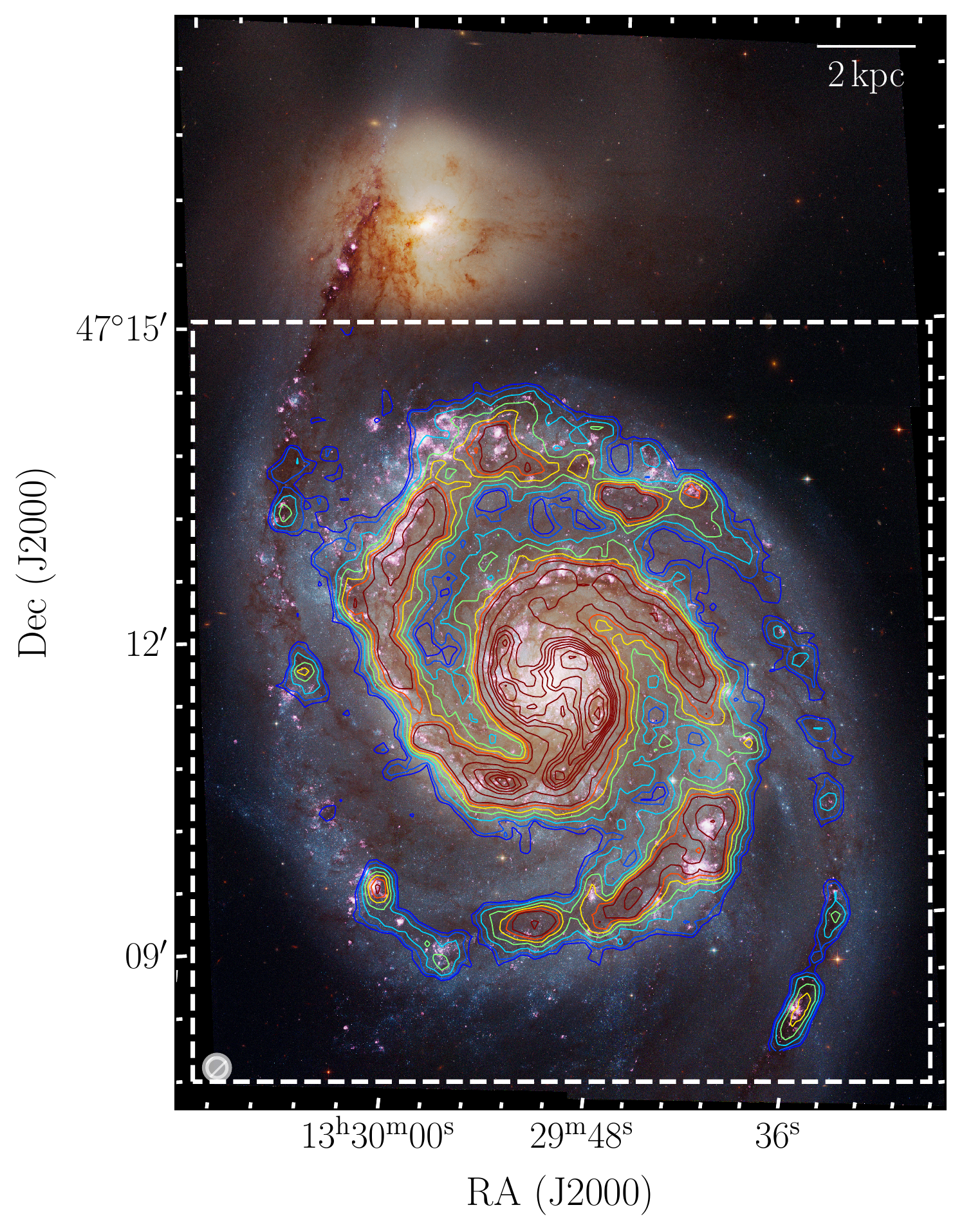}
    \caption{{\bf RGB image of M51 with \chem{^{12}CO} emission.} Colour \textit{Hubble Space Telescope} image composed of $B$, $V$, $I$ filter images taken from \citet{Mutchler2005}. Overlaid as contours is the IRAM \mbox{30-m} \chem{^{12}CO}{21} integrated line brightness temperature (at a resolution of $13$\,arcsec, indicated by circle in the lower left). The contours show signal-to-noise levels from 8 to 70 from blue to brown. The beam size of the \chem{^{12}CO}{21} observations is indicated on the bottom left of the figure. The dashed region shows the field-of-view of the IRAM \mbox{30-m} observations.}
    \label{fig:RGBimage}
\end{figure}
This paper is organised as follows: In \autoref{sec:data}, we present and describe the IRAM \mbox{30-m} observations as well as the ancillary data which are used in this paper. \autoref{lab:analysis} explains how we convert observational measurements to physical quantities. The main results of the paper, which include the different CO line ratios as well as their spatial variations, are presented in \autoref{sec:result}. Finally, \autoref{sec:discussion} discusses our findings and provides an outlook on future projects that can be done with the data from this project.



\section{Observations and Data Reduction}
\label{sec:data}

\subsection{Target}

M51 (NGC~5194) is a prime target for the study of faint CO isotopologues in extragalactic systems. With a distance of $D \approx 8.6$\,Mpc \citep{McQuinn2016} -- so an angular scale of $1$\,arcsec corresponds to physical scale of ${\sim}40$\,pc -- it is one of the brightest nearby grand-design spiral galaxies. It is a tidally interacting, active star-forming galaxy, { with the molecular gas dominating the inner ${\sim}5-6$\,kpc \citep{Schuster2007,Leroy2008}} 
Furthermore, the galaxy hosts an active galactic nucleus \citep{Querejeta2016}. \hyperref[fig:RGBimage]{Figure~\ref*{fig:RGBimage}} shows the galaxy in the optical using a \textit{Hubble Space Telescope} (\textit{HST}) image. The \chem{^{12}CO}{21} emission observed as part of this programme is overlaid. The galaxy is close-to face on ($i=22^\circ$; \citealt{Colombo2014}) and
the target's key parameters are listed in \autoref{tab:target_descr}.
M51 is a template for active star-forming galaxies, where we can resolve discrete environments, know molecular cloud properties (e.g. from PAWS; \citealt{Colombo2014}) and have a wealth of ancillary data and observations from all wavelength regimes. The ancillary data used in this study are provided and described in the following sections. 

\begin{table*}
    \begin{center}
    \caption{Summary of the lines covered in CLAWS, the key observational parameters, as well as key characteristics of the extracted data products.}
    \label{tab:obs_lines}
    \begin{tabular}{c c c c c c c c c c}
         \hline \hline 
         Setup & Band & Line & $\nu_{\rm rest}$ & \multicolumn{2}{c}{Beam size} & On-source time & $\langle T_{\rm sys}\rangle$ & $\langle$pwv$\rangle$\\
         &&& [GHz] &[$\arcsec$]&[kpc]& [hr] & [K]&[mm]\\
         &&&&(1)&(1)&(2)&(3)&(4)\\\hline
         \multirow{8}{*}{1}&\multirow{8}{*}{E0 (3\,mm)}&\chem{CN}{10}&113.250&26.1&1.1&\multirow{8}{*}{{44.5}}&\multirow{8}{*}{111}&\multirow{8}{*}{1.8}\\
         &&\chem{C^{17}O}{10}&112.359&26.3&1.1\\
         &&\chem{^{13}CO}{10}&110.201&26.8&1.1\\
         &&\chem{C^{18}O}{10}&109.782&26.9&1.1\\
         &&\chem{CS}{21}&97.981&30.2&1.3\\
         &&\chem{CH_3OH}{21}&96.700&30.6&1.3\\
         &&\chem{N_2H^+}{10}&93.173&31.7&1.3\\
         &&\chem{HC_3N}{109}&90.897&32.5&1.4\\ \hline
         \multirow{3}{*}{2}&\multirow{3}{*}{E2 (1.3\,mm)}&\chem{^{12}CO}{21}&230.538&12.8&0.53&\multirow{3}{*}{{ 20.9}}&\multirow{3}{*}{217}&\multirow{3}{*}{1.7}\\
         &&\chem{^{13}CO}{21}&220.399&13.4&0.56\\
         &&\chem{C^{18}O}{21}&219.560&13.5&0.56\\ \hline
    \end{tabular}
    \end{center}
    {\raggedright {\bf Notes:} (1) Beam size of the final data cube after reduction. (2) Total on-source time (spectral time) excluding additional telescope overheads. (3) Average system temperature. (4) Average precipitable water vapour (pwv) during observations.}
\end{table*}

\subsection{Observations}
As part of an IRAM \mbox{30-m} large program (\#055-17), the EMIR receiver was used to map emission lines in the 1\,mm (220\,GHz) and 3\,mm (100\,GHz) windows in dual polarisation from the entire disc of M51 for a total of 149\,h (109.2\,h on-source time) between 2017 and 2019. The receiver has an instantaneous bandwidth of 15.6\,GHz per polarization. The observations were split into two parts by implementing E90 LO/LI + UI/UO (Setup~1) as well as E230 LO/LI + UI/UO (Setup~2). The first setup covers the 3\,mm range and was carried out under good atmospheric conditions (1.8\,mm of precipitable water vapour (pwv) and mean $T_{\rm sys} = 111$\,K\,[$T_\mathrm{a}^\star$]). The total on-source time accumulated to 65.9\,h. The second setup observes $J = 2\!\rightarrow\!1$ lines in the 1\,mm regime. For this setup, good winter conditions were required (1.7\,mm pvw and $T_{\rm sys} = 217$\,K [antenna temperature $T_\mathrm{a}^\star$] mean) and the final total on-source time amounted to 43.3\,h.
The Fast Fourier Transform spectrometers with $195$\,kHz spectral resolution (FTS200) were used for both setups, to provide a spectral resolution of ${\sim}0.5$\,km\,s$^{-1}$ for the E090 and ${\sim}0.2$\,km\,s$^{-1}$ for the E230 band. \autoref{tab:obs_lines} lists the lines covered. 

A field of $6.6\,\mathrm{arcmin} \times 6.6\,\mathrm{arcmin}$ (around $44\,\mathrm{arcmin}^2$) was mapped in the on-the-fly/\linebreak[0]{}position switching (OTF-PSW) mode and included two emission-free reference positions nearby. The mapping approach is similar to the one used in the EMPIRE survey \citep[see][]{Donaire2019}. For each spectral setup, a scan of $8$\,arcsec\,s$^{-1}$ is performed using multiple paths that are each offset by $8$\,arcsec from each other. For each execution of the mapping script, the scanned box is shifted by $\sqrt{2} \times (0,2,4,6)$ along the diagonal. So in the end, M51 is covered with a much finer, $2$\,arcsec instead of $8$\,arcsec, grid. The read-out dump time is $0.5$\,s, the final spacing between data points is $4$\,arcsec.
The focus of the telescope was determined using observations of bright quasars or planets at the beginning of each observation session which had typical length of $2{-}3$\,h. In the case of longer sessions, the focus was corrected every 3\,h, and in addition after sunset and sunrise. Every $1{-}1.5$\,h, the pointing of the telescope was adjusted using either a nearby quasar or planet. In order to properly perform the antenna temperature ($T_\mathrm{a}^\star$) calibration, a chopper-wheel calibration was done repeatedly every $10{-}15$ minutes using hot-/\linebreak[0]{}cold-load absorber and sky measurements. Finally, line calibrators (IRC+10216, W3OH and W51D) were routinely observed to monitor systematic error in amplitude and the flux calibration.

\subsection{Data Reduction}
The data reduction is performed automatically using the scripts and pipeline used for the EMPIRE survey \citep[see description in][]{Donaire2019}. Basic calibration is done using \texttt{MRTCAL}\footnote{\url{ https://www.iram-institute.org/medias/uploads/mrtcal-check.pdf}}. The first step consists of converting the spectrum to the antenna temperature scale. For this, each science scan is combined with the last previous calibration scan. Next, we subtract from the calibrated spectrum the OFF measurement. These steps constitute the most basic calibration. The target lines are then extracted using the Continuum and Line Analysis Single-dish Software (\texttt{CLASS}\footnote{\url{https://www.iram.fr/IRAMFR/GILDAS/doc/html/class-html/class.html}}). A zeroth-order baseline is subtracted, omitting the range of 50 to 300\,km\,s$^{-1}$ around the centre of the line in the fit. The individual spectra are then regridded to have a 4\,km\,s$^{-1}$ channel width across the full bandpass. The spectra are saved as FITS files for further processing.

In order to monitor the stability of the flux calibration, spectra of line-calibrator sources (e.g. IRC+10216) were further obtained. From these we find a maximum day-to-day variation in amplitude of ${\sim}7.5$~per~cent over all observations. The 1$\sigma$ variation is ${\sim}2.4$~per~cent only.

Subsequent data reduction is performed using a custom \texttt{IDL} routine based on the HERACLES data reduction pipeline \citep{Leroy2009}. This routine removes pathological data such as bad scans or spectra. Platforming correction at the edges of the FTS units is also corrected for in the EMPIRE pipeline.
The baseline fitting is performed again excluding a generous line window using the \chem{^{12}CO}{10} line emission from PAWS as a prior: Around the mean \chem{^{12}CO}{10} velocity, a window is placed which full width ranges between 50 and 300\,km\,s$^{-1}$, depending on the width of the line for each pixel. Two further windows of the same width are defined adjacent to the central window and a second-order polynomial fit of the baseline is performed in these windows. The resulting baseline is subtracted from the entire spectrum.

After these steps, we check for further pathological spectra. These are rejected by sorting the remaining spectra by their rms, which is calculated from the line-free windows after the baseline subtraction, and the highest 10~per~cent are rejected. Upon careful inspection by eye, additional spectra were discarded if they showed platforming or other potential issues.

The antenna temperature scale ($T_\mathrm{a}^\star$) is converted to main beam temperature ($T_{\rm mb}$) using a cubic interpolation of the forward ($F_{\rm eff}$) and beam ($B_{\rm eff}$) efficiencies from the IRAM documentation\footnote{The online IRAM documentation can be found at \url{http://www.iram.es/IRAMES/mainWiki/Iram30mEfficiencies}} as function of the observing frequency.  In particular, the conversion is performed using the following equation:
\begin{equation}
    T_{\rm mb} = \frac{F_{\rm eff}}{B_{\rm eff}} T_\mathrm{a}^\star~.
\end{equation}
The $F_{\rm eff}/B_{\rm eff}$ ratio adopted for our observing programme was 1.2 for Setup~1 (3\,mm regime) and 1.6 for Setup~2 (1\,mm regime). In this study, we will exclusively use the main beam temperature $T_{\rm mb}$.

The final data cube is generated by gridding the spectra onto a $2$\,arcsec spaced Cartesian grid. Consequently, the final resolution is coarser than the IRAM \mbox{30-m} native resolution due to the gridding kernel by a factor of $1.2$.

We do not correct for the contribution from the IRAM \mbox{30-m} error beam to the observed main beam temperature. We discuss this effect in \hyperref[sec:errcontr]{Appendix~\ref{sec:errcontr}}. In short, emission can enter our detection via the telescope's error beam, thus increasing the observed flux. Regions with faint emission in the galaxy are most likely affected by this. The exact shape of the IRAM \mbox{30-m} error beam is difficult to determine and it fluctuates depending on the telescope's elevation. Consequently, we can only estimate the impact. We see that in the 3\,mm regime, the contribution leads to an additional 10~per~cent increase in flux in faint regions. In the 1\,mm regime, the impact is larger with a contribution of up to $30$ to $40$~per~cent in certain regions. For a more detailed discussion of the estimation of error beam contributions, we refer the reader to \hyperref[sec:errcontr]{Appendix~\ref{sec:errcontr}}.

\begin{figure*}
\centering
\includegraphics[width = \textwidth]{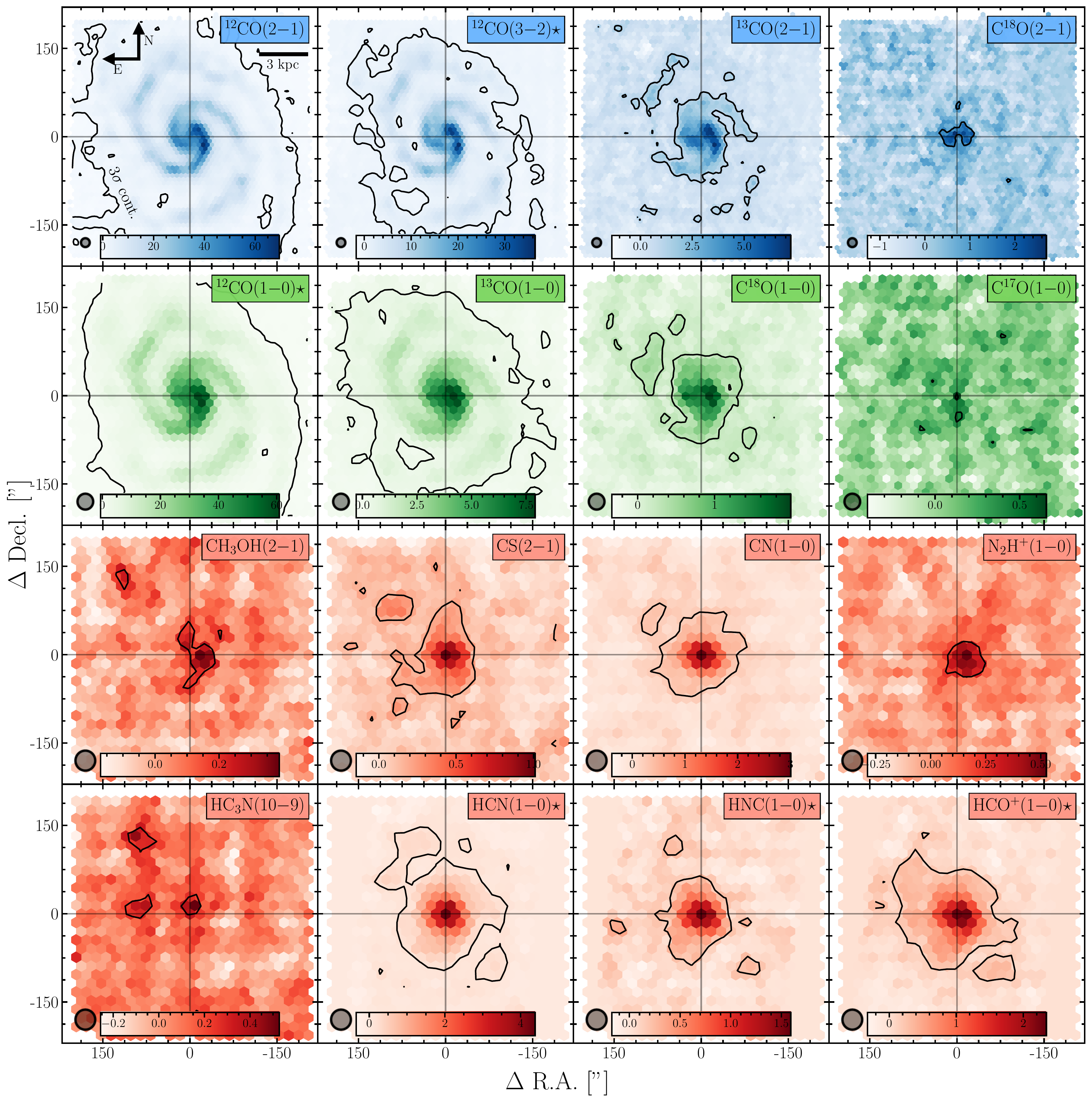}
\caption{{\bf Maps of the velocity-integrated brightness temperature for all lines used in this study.} The maps are convolved to certain common beam sizes and regridded onto a common hexagonal grid, which allows a more uniform sampling. The top row (blue) is at $15$\,arcsec resolution, the second row (green) at $27$\,arcsec, and the remaining two rows (red) are convolved to $34$\,arcsec (corresponding to $0.62, 1.1 \text{ and } 1.4$\,kpc, respectively). The grid spacing is chosen to be half the beam size. The velocity-integrated brightness temperature is in units of K\,km\,s$^{-1}$. The coordinates are relative to the centre coordinates given in \autoref{tab:target_descr}. The black line indicates the $\mathrm{S/N} = 3$ contour. Lines observed by other programmes are indicated by an asterisk after the line name in each panel. \chem{^{12}CO}{10} is part of the PAWS survey \citep{Schinnerer2013,Pety2013}, \chem{^{12}CO}{32} is taken from the NGLS survey \citep{Wilson2012}, and \chem{HCN}{10}, \chem{HNC}{10} and \chem{HCO^+}{10} are emission lines obtained by the EMPIRE survey \citep{Donaire2019}.}
\label{fig:SB_maps}
\end{figure*}

\begin{table*}
    \centering
    \caption{Comparing channel rms sensitivities of lines that were observed in other programmes before. The channel rms is computed at $27$\,arcsec resolution and $4$\,km\,s$^{-1}$ channel width.}
    \label{tab:comp_other_prog}
    \begin{tabular}{l l l c | c}
        \hline
         Programme & Line Transition & Telescope/Instrument & $\langle{\rm rms}\rangle$ & $\langle{\rm rms}\rangle$ CLAWS  \\
         &&& [mK] & [mK] \\\hline \hline
         HERACLES & \chem{^{12}CO}{21} & IRAM \mbox{30-m} / HERA & 7 & 11\\
         PAWS & \chem{^{13}CO}{10} & IRAM \mbox{30-m} / EMIR& 4 & 5\\ \hline
    \end{tabular}
\end{table*}

\subsection{Ancillary Data}
\label{sec:anc_data}


\subsubsection{PAWS \texorpdfstring{\textsuperscript{12}CO\,\mbox{(1--0)}}{Lg} Emission Line Data}

The \textit{PdBI Arcsecond Whirlpool Survey} (PAWS)\footnote{Data can be downloaded from \url{https://www2.mpia-hd.mpg.de/PAWS/PAWS/Data.html}.} covers the \chem{^{12}CO}{10} line emission at $1$\,arcsec ($\approx 40$\,pc) resolution across the full disc of M51 \citep{Schinnerer2013}. The survey combines observations from IRAM's Plateau de~Bure Interferometer (PdBI) and the IRAM \mbox{30-m} single dish telescope. We only use the IRAM \mbox{30-m} observations for this work, as we do not require high spatial resolution. The data reduction is described in \citet{Pety2013}. 
Observations were carried out in 2010 and the data have a native spatial resolution of $23$\,arcsec ($\approx 1.0$\,kpc) with an $1\sigma$ noise level of 16\,mK at 5\,km\,s$^{-1}$ spectral resolution.

\subsubsection{NGLS \texorpdfstring{\textsuperscript{12}CO\,\mbox{(3--2)}}{Lg} Emission Line Data}

As part of the James Clerk Maxwell Telescope (JCMT) \textit{Nearby Galaxy Legacy Survey} \citep[NGLS;][]{Wilson2012}, the \chem{^{12}CO}{32} emission across the full disc of M51 was mapped \citep{Vlahakis2013}. Observations with an angular resolution of $14.5$\,arcsec ($\approx 600$\,pc) were carried out with the 16 pixel array receiver \mbox{HARP-B} at the JCMT between 2007 and 2009. 

\subsubsection{THINGS 21-cm \texorpdfstring{\hi}{Lg} Emission Line Data}

The \hi\ data\footnote{Available at \url{https://www2.mpia-hd.mpg.de/THINGS/Data.html}.} from \textit{The \hi\ Nearby Galaxy Survey} \citep[THINGS;][]{Walter2008} is used to map the atomic gas content across the full disc of M51. The \hi\ emission is of particular interest, as it is very extended, making it a good prior for masking emission regions outside the central region of the galaxy.  The survey employed the Very Large Array (VLA). The natural weighted data used in this study have an angular resolution of ${\sim}10$\,arcsec ($\approx 400$\,pc) and a spectral resolution of ${\sim}5$\,km\,s$^{-1}$.

\subsubsection{EMPIRE High Density Emission Line Data}

The \textit{EMIR Multiline Probe of the ISM Regulating Galaxy Evolution} \citep[EMPIRE;][]{Donaire2019} survey covered the emission of a number of high-density emission lines in the 3\,mm regime, such as \chem{HNC}{10}, \chem{HCN}{10} or \chem{HCO^+}{10}, across the entire star-forming disc of a sample of nine nearby spiral galaxies. The survey uses observations\footnote{Details on the survey and the data can be found at \url{https://empiresurvey.webstarts.com/index.html}} carried out with the EMIR receiver at the IRAM \mbox{30-m} single dish telescope. The science goal of the survey was to take deep and extended intensity maps of high critical density lines tracing the dense gas content in the galaxy. While the $J=1\!\rightarrow\!0$ transition of \chem{^{13}CO} is also covered in the survey, we will only consider the high density lines \chem{HCN}{10}, \chem{HNC}{10} and \chem{HCO^+}{10} in our study and rely on our own \chem{^{13}CO}{10} observation (since our observations are deeper). 

\subsubsection{VNGS Infrared Data}

Infrared broadband data in the range of $3.6$ to $500\,\mu$m are taken from the \textit{Very Nearby Galaxy Survey} \citep[VNGS;][]{Parkin2013}. These observations were carried out using the MIPS instrument on board the \textit{Spitzer Space Telescope}, as well as PACS and SPIRE instruments on board of the \textit{Herschel Space Observatory}. The infrared bands are used to estimate the total infrared emission (TIR) and star formation rate (SFR; see \autoref{sec:sfr}).

\subsection{Final Data Products}

\begin{table*}
    \begin{center}
    \caption{List of lines covered as part of CLAWS with measurements regridded onto the same hexagonal grid and convolved to a common working resolution $15, 27 \text{ or } 34$\, arcsec.}
    \label{tab:lines}
    
    \resizebox{\textwidth}{!}{%
    \begin{tabular}{l l c c c | c c c|c c c}
         \hline \hline 
         &&&&&\multicolumn{3}{c|}{$35''/1.4{\rm \, kpc}$ Aperture$^{\rm a}$}&\multicolumn{3}{c}{$60''/2.4{\rm \, kpc}$ Aperture$^{\rm a}$}\\
         Band & Line &  Resolution &T$_{\rm peak}$ & $\langle$rms$\rangle$ &$W_{\rm line}$ & $\mathrm{S/N}_{\rm line}$& $ W_{\rm line} / W_{\chem{^{12}CO}{21}}$&$W_{\rm line}$ & $\mathrm{S/N}_{\rm line}$& $ W_{\rm line} / W_{\chem{^{12}CO}{21}}$  \\
         &&  [$^{\prime \prime}$]&[mK] & [mK] & [K\,km\,s$^{-1}$]&&&[K\,km\,s$^{-1}$]&& \\
         &&(1)&(2)&(3)&(4)&(5)&(6)&(7)&(8)&(9)\\\hline \hline
         \multirow{11}{*}{3\,mm}&\chem{CN}{10}&34&32&4.4& 1.7&66&0.05&1.1&50&0.04\\ 
         &\multirow{2}{*}{\chem{C^{17}O}{10}}&27&26&5.8&\multirow{2}{*}{0.23}&\multirow{2}{*}{10}&\multirow{2}{*}{0.006}&\multirow{2}{*}{0.1}&\multirow{2}{*}{11}&\multirow{2}{*}{0.007}\\
         &&34&20&4.2&&&&&&\\ 
         &\multirow{2}{*}{\chem{^{13}CO}{10}}&27&142&4.9&\multirow{2}{*}{5.5}&\multirow{2}{*}{250}&\multirow{2}{*}{0.15}&\multirow{2}{*}{4.2}&\multirow{2}{*}{280}&\multirow{2}{*}{0.15}\\
         &&34&116&3.6&&&&&&\\ 
         &\multirow{2}{*}{\chem{C^{18}O}{10}}&27&40&4.7&\multirow{2}{*}{1.2}&\multirow{2}{*}{52}&\multirow{2}{*}{0.033}&\multirow{2}{*}{0.9}&\multirow{2}{*}{47}&\multirow{2}{*}{0.030}\\
         &&34&33&3.4&&&&&&\\
         &\chem{CS}{21}&34&15&3.0&0.6&35&0.016&0.4&36&0.014\\
         &\chem{CH_3OH}{21}&34&13&2.8& 0.2&9&0.004&0.1&7&0.004\\
         &\chem{N_2H^+}{10}&34&14&3.0&0.3&19&0.007&0.2&11&0.006\\
         &\chem{HC_3N}{109}&34&16&3.7&0.2&10&0.006&0.2&15&0.007\\ \hline \hline
         \multirow{9}{*}{1\,mm}&\multirow{3}{*}{\chem{^{12}CO}{21}}&15&1420&20&\multirow{3}{*}{37}&\multirow{3}{*}{710}&\multirow{3}{*}{1.0}&\multirow{3}{*}{29}&\multirow{3}{*}{750}&\multirow{3}{*}{1.0}\\
         &&27&850&10.6&&&&&&\\
         &&34&730&8.8&&&&&&\\
         &\multirow{3}{*}{\chem{^{13}CO}{21}}&15&185&17.6&\multirow{3}{*}{3.3}&\multirow{3}{*}{67}&\multirow{3}{*}{0.09}&\multirow{3}{*}{2.3}&\multirow{3}{*}{73}&\multirow{3}{*}{0.08}\\
         &&27&103&9.2&&&&&&\\
         &&34&88&7.7&&&&&&\\
         &\multirow{3}{*}{\chem{C^{18}O}{21}}&15&88&18.3&\multirow{3}{*}{0.9}&\multirow{3}{*}{21}&\multirow{3}{*}{0.03}&\multirow{3}{*}{0.6}&\multirow{3}{*}{21}&\multirow{3}{*}{0.02}\\
         &&27&51&9.4&&&&&&\\
         &&34&41&7.8&&&&&&\\\hline
    \end{tabular}
    }
    \end{center}
    {\raggedright {\bf Notes:} (a) Emission of the central region convolved to the given aperture size. (1) Working resolutions used for each molecular line; (2) maximum peak temperature observed across the full map of M51 for a given emission line at a given working resolution; (3) average channel rms sensitivity at 4\,km\,s$^{-1}$ channel width; (4,~7) velocity-integrated brightness temperature within the central aperture; (5,~8) $\mathrm{S/N}$ within the central aperture; (6,~9) line ratio of a given line with respect to the velocity-integrated \chem{^{12}CO}{21} brightness temperature.}
\end{table*}

\begin{figure}
    \centering
    \includegraphics[width = 0.9\columnwidth]{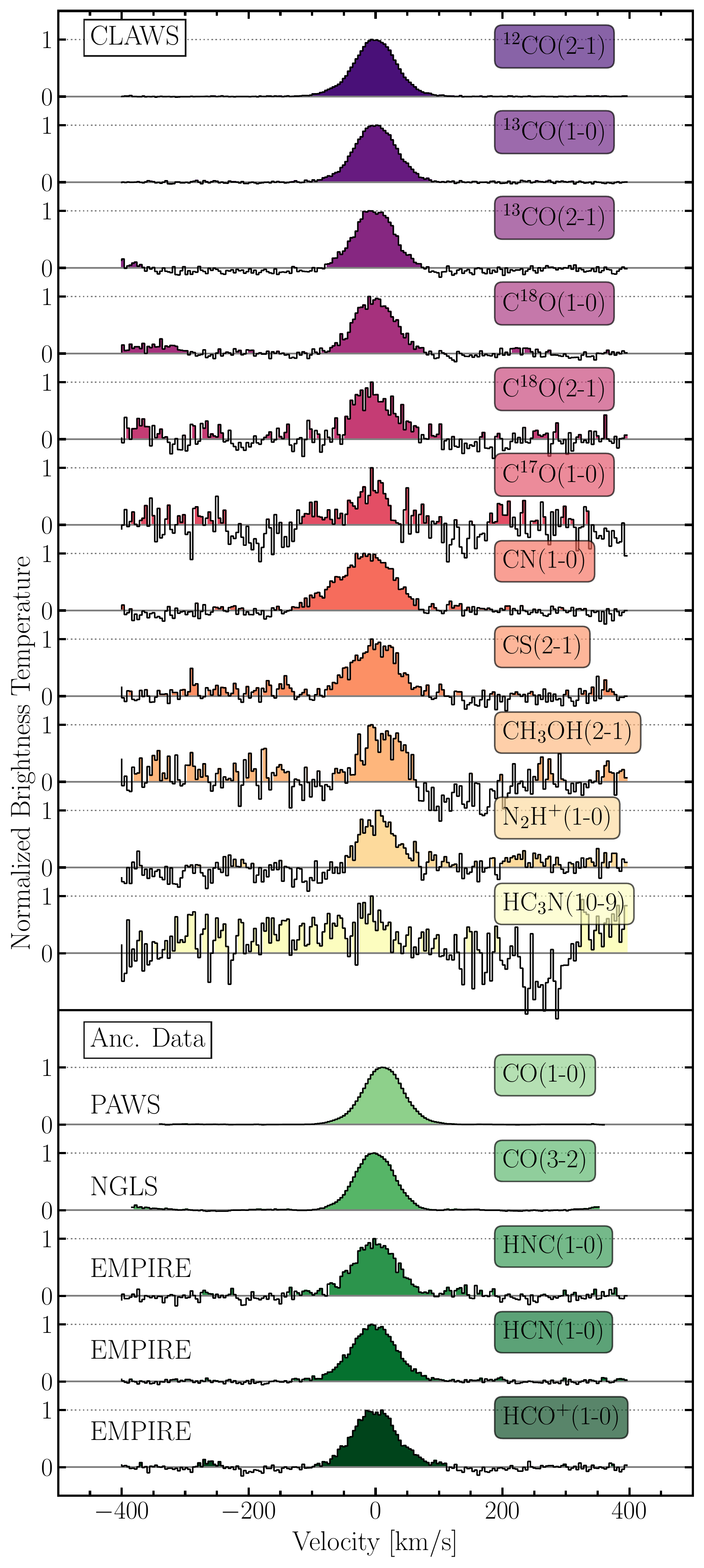}
    \caption{{\bf  Spectra of emission lines covered by this programme stacked over the central 1.5\,kpc region at 34\arcsec.} The emission lines are normalised to the maximum in the spectrum. The five bottom spectra are taken from different observing programmes (see \autoref{sec:anc_data}), while the spectra above come from this project. For the absolute intensities we refer the reader to \autoref{tab:lines}. We note that we employ a hexagonal, half beam sized sampling. The angular resolution of 34\arcsec corresponds to a physical scale of around 1.4\,kpc. So the shown stacked spectra over the central 1.5\,kpc region are thus the combination of the spectra of the seven most central sampling points.  }
    \label{fig:central_sightline}
\end{figure}

Due to the large wavelength range covered, we have a variety of spatial resolutions for our observations, ranging from $12.8\,\mathrm{arcsec}/500$\,pc for \chem{^{12}CO}{21} to $32.5\,\mathrm{arcsec}/1.3$\,kpc for \chem{HC_3N}{109}. To properly match and compare observations of different emission lines, we convolve to the same working resolution and reproject all data to a common grid. In this study, we will use three different working resolutions:
\begin{enumerate}
    \item $15\,\mathrm{arcsec}/600$\,pc: Analysis involving \chem{^{12}CO}{21} and \mbox{(3--2)}. 
    \item $27\,\mathrm{arcsec}/1.1$\,kpc: Analysis involving all CO isotopologues.
    \item $34\,\mathrm{arcsec}/1.4$\,kpc: Analysis involving all molecular lines.
\end{enumerate}

\autoref{tab:lines} lists the working resolutions used for every line in our sample. The data are finally resampled onto a hexagonal grid with a grid size of half the beam size (see \autoref{fig:SB_maps}, note that lines observed as part of other programmes than CLAWS are marked with an asterisk). Regridding onto a hexagonal grid is commonly done in the literature \citep[see e.g.][]{Bigiel2011,Leroy2013,Sandstrom2013, Cormier2018}.  One advantage over a Cartesian grid is the equidistance to the neighboring pixels allowing for a more uniform sampling. Furthermore, since the beam shape is circular, given that the neighborhood for a hexagonal grid also grows with some circularity, beam effects are captured. Using a half beam-sized hexagonal grid, we have an oversampling factor\footnote{We define the oversampling factor by the ratio of the beam area over the pixel size: $N_\mathrm{s} = (1.13\times \theta_{\rm FWHM}^2)/A^{\rm hex}_{\rm pix}$. Since we sample the data hexagonally, the pixels have a hexagonal shape with the long diagonal being equal to half the beam FWHM.} of $N_\mathrm{s} = 4.6$. The spectra for all lines, sampled in the aforementioned way, are combined into a final data structure for further analysis.

We determine the velocity-integrated brightness temperature by integrating masked spectra of individual sight lines. For the innermost part of the galaxy up to a galactocentric radius of a CO scale length of $0.23r_{25}$ \citep{Leroy2008,Lisenfeld2011,Puschnig2020}\footnote{At a radial distance of $0.23r_{25}$, it has been found that the CO surface brightness has dropped, on average, by a factor of $1/e$ (see for more details \citealt{Puschnig2020}).}, we use the \chem{^{12}CO}{21} emission line as a prior for masking. For larger radii, where the CO emission becomes faint, we use the \mbox{21-cm} \hi\ emission line. By only integrating over the masked velocity region, we can improve the $\mathrm{S/N}$ significantly, which is essential to capture also the fainter line emission. The mask is produced by first flagging high $\mathrm{S/N}$ voxels in the \chem{^{12}CO}{21} or \hi\ data, respectively ($\mathrm{S/N} > 4$ for both). An additional lower $\mathrm{S/N}$ mask ($\mathrm{S/N} {>} 2$) is produced using the same lines and voxels from the high mask which are then expanded into the low mask. The velocity-integrated brightness temperature and its uncertainty, for both we are adopting units of K\,km\,s$^{-1}$ throughout this work, are given by:
\begin{align}
    W_{\rm line} &= \sum^{n_{\rm chan}} T_\mathrm{mb}(v) \cdot \Delta v_{\rm chan} \\
    \sigma_W &= \sqrt{n_{\rm chan}}\cdot  \sigma_{\rm rms} \cdot \Delta v_{\rm chan}~,
    \label{eq:unc}
\end{align}
where $n_{\rm chan}$ is the masked number of channels along a line of sight, $T_{\rm mb}$ is the surface brightness temperature of a given channel in~K,  $\sigma_{\rm rms}$ is the position-dependent $1\sigma$ root-mean-squared (rms) value of the noise in~K, and $\Delta v_{\rm chan}$ is the channel width in km\,s$^{-1}$. We calculate  $\sigma_{\rm rms}$ over the signal-free part of the spectrum using the \texttt{astropy} function \texttt{mad\_std}. It calculates the median absolute deviation and scales the result by a factor $1.4826$ to yield a standard deviation (the factor follows from the assumption that noise follows a Gaussian distribution).

\autoref{tab:lines} provides a summary of the lines observed by CLAWS, as well as information about the data such as the average rms or the $\mathrm{S/N}$ of the line detection over a certain aperture. \hyperref[fig:central_sightline]{Figure~\ref*{fig:central_sightline}} shows spectra of all these lines stacked over the central 1.5\,kpc region.
All the data of this project are made publicly available on the IRAM Large Programme website.\footnote{\url{https://www.iram-institute.org/EN/content-page-434-7-158-240-434-0.html}}



\section{Physical Parameter Estimation}
\label{lab:analysis}
We largely follow the methodology described in \citet{Cormier2018} investigating \chem{^{13}CO} line emission in EMPIRE galaxies and the EMPIRE survey paper by \citet{Donaire2019}. The line ratios are measured as a function of galactocentric radius, \chem{^{12}CO}{21} brightness temperature and total infrared surface brightness. For completeness, we also include the derived physical quantities of the molecular gas mass surface density ($\Sigma_{\rm mol}$) and the star formation rate surface density ($\Sigma_{\rm SFR}$), which we corrected for the inclination of M51 by applying the factor $\cos(i)$. We note, however, that the conversion from observed to physical quantity is subject to uncertainties \citep[see e.g.][]{Kennicutt2012, Bolatto2013, Usero2015}.

\subsection{Line Ratio}
\label{sec:ratio}
Line ratios are determined by taking the velocity-integrated brightness temperatures of two lines, in units of K\,km\,s$^{-1}$, and dividing them. This is done for every line of sight as well as for the stacked spectra (see \autoref{sec:stack}). The uncertainty of a line ratio, $R = W_{1}/W_{2}$, is given by the propagated uncertainty of the two lines:
\begin{equation}
    \sigma_R = \frac{W_{\rm 1}}{W_{\rm 2}}\sqrt{\left(\frac{\sigma_{W_{\rm 1}}}{W_{\rm 1}}\right)^2+\left(\frac{\sigma_{W_{\rm 2}}}{W_{\rm 2}}\right)^2}~.
\end{equation}
%

We generally express the average line ratio, $\langle R \rangle$, in terms of the \chem{^{12}CO}{21} brightness temperature weighted median, which is equivalent to the weighted 50th percentile.
Given the ordered set of line ratios~$R$ for $N$ lines of sight, with associated (but not necessarily ordered) \chem{^{12}CO}{21} surface brightness of $W_{^{\chem{^{12}CO}{21}}}$
\begin{equation}
    \left\{\left(R^i, W^i_{^{\chem{^{12}CO}{21}}}\right)\right\}_{i = 1,...,N} \text{ \ \ for \ \ } R^{i+1}_{^{\chem{^{12}CO}{21}}}\ge R^i_{^{\chem{^{12}CO}{21}}},
\end{equation}
and defining the relative weight as:
\begin{equation}
    w_i = \frac{W^{i}_{^{\chem{^{12}CO}{21}}}}{\sum_{i=1}^{N} W^{i}_{^{\chem{^{12}CO}{21}}}},
\end{equation}
%
%
the weighted $p$th percentile is given by the element  $R^k$:
\begin{equation}
    Q^{w}_{p}(R) = R^{k} \text{ \ \ satisfying } \sum_{i=1}^{k}w_i\le p \text{ and } \sum_{i=k+1}^{N}w_i\le 1-p~.
\end{equation}
Following this definition, we define the \chem{^{12}CO}{21} brightness temperature weighted median line ratio as:
\begin{equation}
\label{eq:w_median}
    \langle R \rangle \equiv Q^{w}_{0.5}(R)~. 
\end{equation}
The uncertainty of the weighted average line ratio is given by the 16th and 84th weighted percentile range throughout the study.

We construct the line ratios such that the generally brighter line is in the denominator while the, overall, fainter line is in the numerator. As a consequence, we will generally find often upper limits, while lower limits are very rare.

Furthermore, because we compare lines with different S/N, we try to estimate the region in the plot in which we cannot obtain any measurements. Such a censored region occurs, for example, if we investigate the line ratio of a fainter line to a brighter line. Given that we have the same observed sensitivity, lower line ratios will be ``censored'' since we reach the detection threshold for the faint line, while larger line ratios can still be observed, as they could originate from points with brighter emission of the faint line. To estimate the censored region, we first bin the line ratios by a certain quantity (such as, for example, the galactocentric radius). Since we constructed the line ratios such that the (generally) fainter line is in the numerator, we estimate the censored $1\sigma$ (or $3\sigma$) region by dividing the average rms (or $3{\times}$ this value) of the faint line per bin by the average brightness temperature of the brighter line. We notice that because the rms and the line brightness vary across the survey field, we do find also points within the censored region.
 
\subsection{Spectral Stacking}
\label{sec:stack}

In order to improve the $\mathrm{S/N}$, which is especially crucial for fainter emission lines, we apply a spectral stacking technique. A detailed description of the stacking technique is given in several previous studies \citep[e.g.][]{Cormier2018, Donaire2019, denbrok2021}. In short, the spectral axis of each cube is regridded such that the emission line of each sight line is centred at $v = 0$\,km\,s$^{-1}$. At $r_\mathrm{gal} < 0.23 r_{25}$, \chem{^{12}CO}{21} is used as the reference line, while at $r_\mathrm{gal} \ge 0.23 r_{25}$, \hi\ is used. We note that for sight lines with both \hi\ and CO detection, we find good agreement between their centroid velocities. With this approach, we can stack the lines of sight by a predefined quantity (for example galactocentric radius or \chem{^{12}CO} brightness temperature) and obtain an average spectrum for every bin. We stack all lines of sight, irrespective of their individual $\mathrm{S/N}$. We note that by stacking we disregard the intrinsic scatter of the data. Line ratios are then calculated from the velocity-integrated brightness temperatures of two (separately) stacked emission lines. Again, for an illustration of stacked spectra, we refer the reader to \autoref{fig:central_sightline}, which shows stacked spectra of the observed emission lines for the central region ($r_\mathrm{gal} < 1.5$\,kpc). With the exception of \chem{HC_3N}{109}, all emission lines are significantly detected at $\mathrm{S/N}>3$ within the centre.

\subsection{Total Infrared Surface Brightness}
\label{sec:sfr}

The total infrared (TIR) surface density can be used as a proxy of the local surface density of star formation. Following \citet{Galametz2013}, the TIR surface brightness ($\Sigma_{\rm TIR}$) takes the subsequent form:
\begin{equation}
   \Sigma_{\rm TIR} = \sum_i c_i\Sigma_i~,
\end{equation}
with $\Sigma_i$ being the surface brightness of the \textit{Herschel} bands and $c_i$ coefficients depending on the number of infrared bands available (see table~3 in \citealt{Galametz2013} for numerical values of the coefficients). 
We take the same approach as previous studies \citep[e.g.][]{Usero2015, Donaire2017, Cormier2018} by combining the \textit{Herschel} 70, 100 and 160\,$\mu$m bands to estimate the TIR surface brightness. The maps are first convolved to the common beam size of $30$\,arcsec using the kernels described in \citet{Aniano2011}.
In their study, \cite{Galametz2013} indicated that the combination of the \textit{Herschel} 70, 100 and 160\,$\mu$m bands has  a coefficient of determination\footnote{The coefficient of determination is equal to the square of Pearson's linear correlation coefficient.} of $R^2=0.97$, meaning that the calibration we use accounts for 97~per~cent of the total variation of the TIR surface brightness.

From the TIR surface brightness we can estimate the SFR surface density. We adopt the calibration given by \citet{Murphy2011}:
\begin{equation}
    \left(\frac{\Sigma_{\rm SFR}}{\rm M_\odot\,yr^{-1}\,kpc^{-2}}\right) = 1.48 \times 10^{-10}\left(\frac{\Sigma_{\rm TIR}}{\rm L_\odot\,kpc^{-2}}\right)~.
\end{equation}
Notice that compared to the FUV+24\,$\mu\mathrm{m}$ SFR prescription, a scatter of 40~per~cent is expected based on comparisons between the two prescriptions using resolved measurements in M33 \citep{Williams2018}.

\subsection{Molecular Gas Mass Surface Density}

The molecular gas mass surface density can be estimated from the \chem{^{12}CO}{10} line emission or from \chem{^{12}CO}{21} data using the well calibrated CO line ratio, $R_{21}^{\chem{^{12}CO}}$ (which we measure in this project). The conversion from CO emission to gas mass surface density relies on the CO-to-H$_2$ conversion coefficient, $\alpha_{\rm CO}$, as:
\begin{equation}
\begin{split}
    \left(\frac{\Sigma_{\rm mol}}{\rm M_\odot\,pc^{-2}}\right)
    & = \alpha_{\rm CO} W_{\chem{CO}{10}}\cos(i)\\
    & = \alpha_{\rm CO} \frac{W_{\chem{CO}{21}}}{R_{21}^{\chem{^{12}CO}}}\cos(i)~.
\end{split}
\end{equation}
For the $\alpha_{\rm CO}$ parameter usually the Milky Way value of is chosen $\alpha_{\rm CO} = 4.4$ $\mathrm{M_{\odot}\,pc^{-2}}$ $\mathrm{(K\,km\,s^{-1})^{-1}}$ (which includes a factor $1.36$ for helium) in the case of massive galaxies with solar metallicity \citep{Bolatto2013}.  Galaxy-to-galaxy as well as galaxy-internal variations of the conversion factor have been observed, but studies of nearby main sequence galaxies find largely values of similar order to those in the Milky Way \citep{Sandstrom2013,Cormier2018}. We note, however, that these $\alpha_{\rm CO}$ calibrations may not account for CO-dark gas, which could significantly impact the conversion factor in certain regions \citep[see e.g.][]{Gratier2017, Chevance2020,Madden2020}.



\section{Results}
\label{sec:result}

\subsection{CLAWS Line Emission}

\hyperref[fig:SB_maps]{Figure~\ref*{fig:SB_maps}} shows the velocity-integrated brightness temperature maps of the lines covered as part of this programme (see coloured panels), all convolved to a common working beam size of $15, 27 \text{ or } 34$\,arcsec (corresponding to $0.62, 1.1 \text{ or } 1.4$\,kpc, respectively). The \chem{^{12}CO} and \chem{^{13}CO} emission, and to some extent the \chem{C^{18}O}{10} emission are clearly extended and M51's spiral structure is visible by eye. The fainter \chem{C^{18}O}{21} and \chem{C^{17}O}{10} isotopologue lines are only detected in the centre of the galaxy. Besides the CO isotopologues, the other molecular lines are all confined to the centre of the galaxy. This is most likely just attributed to a lack of $\mathrm{S/N}$ at larger radii and not necessarily evidence of the molecule being truly more centrally concentrated.  We explore radial trends for the CO lines in the following sections. Regarding emission from the centre, we see that for the CO isotopologues, the emission peaks are slightly offset towards the western spiral arm of the galaxy. By contrast, for the dense gas tracers, the emission seems to peak directly in the centre. Radial stacking improves the significant detections of faint emission lines. Among the non-CO lines part of CLAWS, with the reached sensitivity of $1\sigma_\mathrm{rms}{\sim}4\,$mK at $4{\rm\,km\,s^{-1}}$, only the \chem{CN}{10} and \chem{CS}{21} emission show radial extension out to $r_\mathrm{gal} \sim 3$\,kpc (see \autoref{fig:SB_maps}).

\subsection{CO Emission Line Ratios}
\label{sec:line_ratios}

\begin{figure*}
    \centering
    \includegraphics[width =0.95 \textwidth]{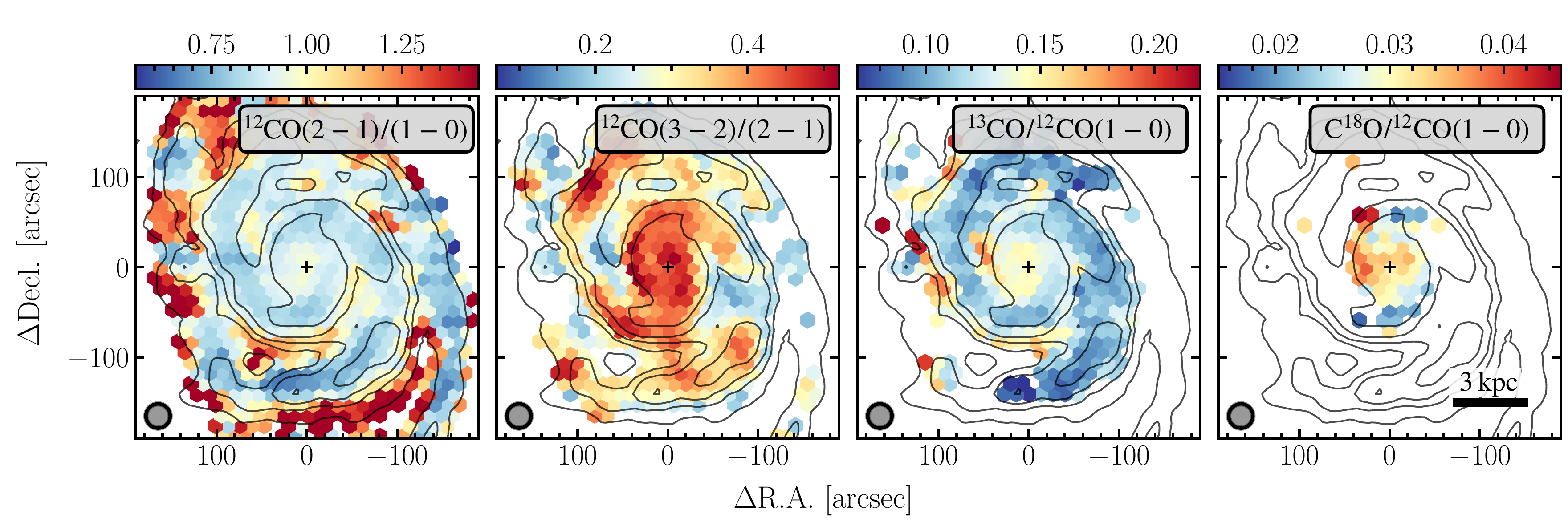}
    \caption{{\bf CO line ratio maps for four selected ratios.} All panels show the data convolved to $27$\,arcsec (indicated by circles in the lower left corners). Only lines of sight with $\mathrm{S/N} > 5$ in both lines are shown. Contours in each panel show the \chem{^{12}CO}{21} emission at 2, 4, 6, 10 and 20\,K\,km\,s$^{-1}$. A local enhancement of the line ratio in the centre of the galaxy (indicated by a plus sign) is clearly visible for $\chem{^{12}CO}{32}/\trans{21}$ \textit{(second to left panel)} and to some extent also for $\chem{^{12}CO}{21}/\trans{10}$ \textit{(left and second to right panels)}. The $\chem{C^{18}O}{21}/\chem{^{12}CO}{10}$ \textit{(right panel)} is only significantly detected in the centre of the galaxy.}
    \label{fig:ratio_maps}
\end{figure*}

\begin{figure*}
    \centering
    \includegraphics[width = 0.85\textwidth]{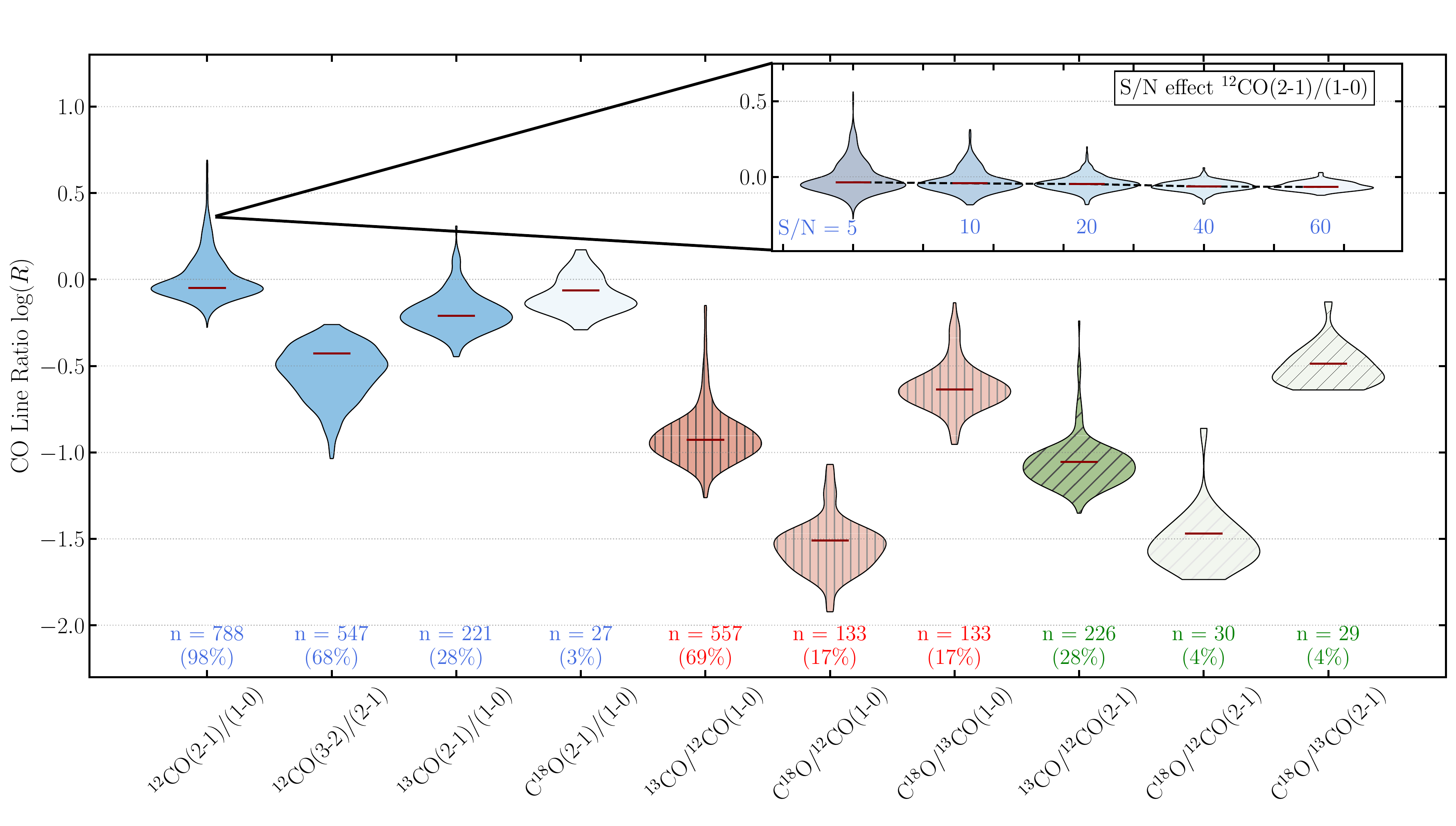}
    \caption{{\bf Distribution of CO line ratios at 27\,arcsec.}  The violin plots are colour coded: blue indicates ratios of fixed CO isotopologue but different transitions, while red (vertically hatched) and green (diagonally hatched) show ratios of different CO isotopologue but fixed transition. The number of significant ($\mathrm{S/N} > 5$) data points for each line ratio is indicated with~$n$ and the saturation of the colour of each violin. { Also the percentage in terms of total detection is indicated (with respect to $\rm S/N>5$ detected $^{12}$CO(1-0) data points).} 
    { In the inset panel, we show the effect of increasing the S/N cut from ${\rm  S/N}=5-60$ for the $^{12}$CO(2-1)/(1-0) line ratio. We do not find a significant difference in terms of the mean line ratio with increasing $\rm S/N$.}
    }
    \label{fig:violin_rat}
\end{figure*}

\begin{figure*}
    \centering
    \includegraphics[width = 0.95\textwidth]{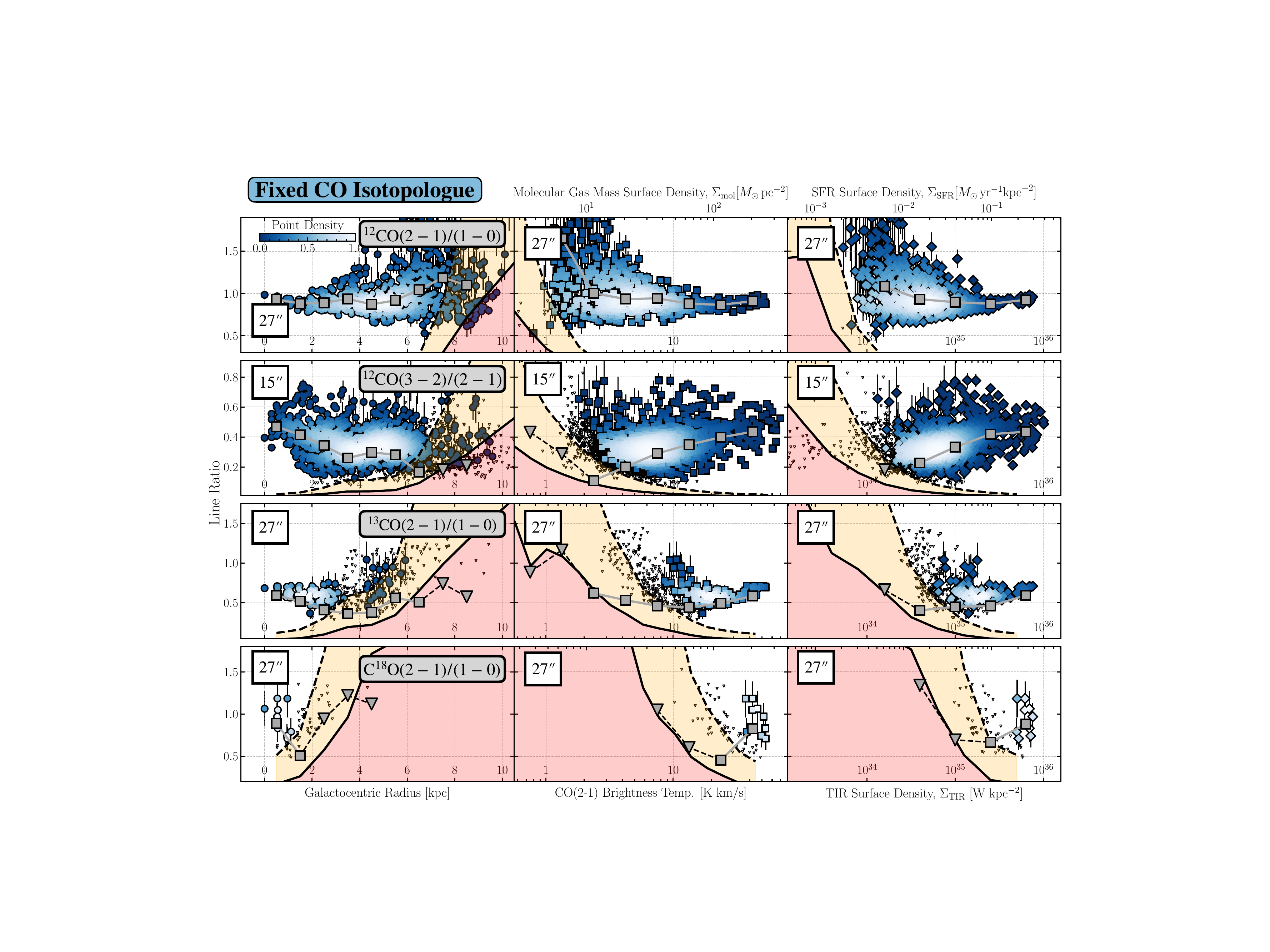}
    \caption{{\bf CO line ratios as function of galactocentric radius, \chem{CO}{21} brightness temperature and TIR surface brightness.}
    Each row shows the line ratio for individual lines of sight for fixed CO isotopologues but different transitions. 
    While each column shows the CO line ratio as function of galactocentric radius, \chem{CO}{21} brightness temperature and TIR surface brightness. 
    The coloured data points show the individual lines of sight with an  ${\rm S/N}> 5$ while the small black downward triangles indicate $3\sigma$ upper limits. { To illustrate the point density in the figure, we colour-code the points using a 2D kernel density estimation. Values are normalized to the peak density and range from most dense (1; white) to least dense (0;blue)  } The shaded area in each panel shows the $1\sigma$ (red) and $3\sigma$ (orange) censored regions for the individual lines of sight. They are an estimate of where we expect we cannot detect line ratios anymore (due to the lack of $\mathrm{S/N}$ of one of the lines).
    Marked in grey are the line ratios derived from the stacked spectra (downward triangles again mark upper limits). Error bars for significant points are indicated (generally, for the stacked data points, the error bars are not visible due to the plotted point size being larger).
    For each line ratio, we used the highest resolution possible (indicated in the upper left corner of the panel).
    To convert to $\Sigma_{\rm mol}$, which we provide for comparison on the top $x$-axis of each panel, we assume a constant $R_{21} = 0.89$. We note that the $x$- and $y$-axes are correlated if the \chem{^{12}CO}{21} emission is used in the line ratio. The stacked points allow us to probe the line ratios in fainter regions which sometimes confirms the trend suggested by the individual (significantly detected) sight line measurements (e.g. the $\chem{^{12}CO}{21}/\trans{10}$ in the top panel row) and sometimes reveals trends which could not be identified from the sight line measurements (e.g. trend in $\chem{^{12}CO}{32}/\trans{21}$ ratio in second row.)
    }
    \label{fig:ratio_fixed_iso}
\end{figure*}

\begin{figure*}
    \centering
    \includegraphics[width =0.95 \textwidth]{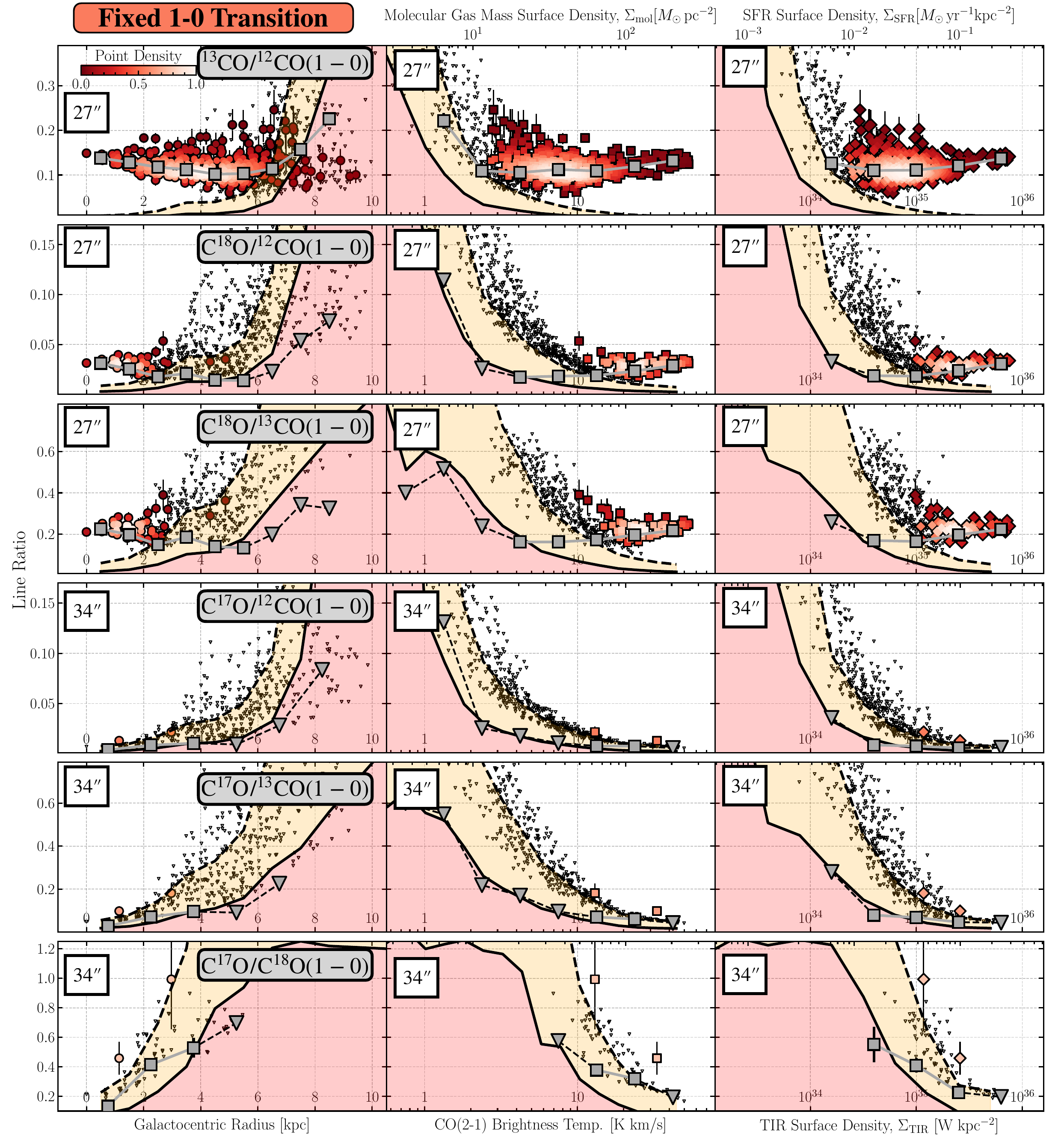}
    \caption{{\bf CO line ratios keeping a fixed \trans{10} transition but different CO isotopologues.} For description of the panels see \autoref{fig:ratio_fixed_iso}. 
    }
    \label{fig:ratio_fix_trans1}
\end{figure*}

\begin{figure*}
    \centering
    \includegraphics[width =0.95 \textwidth]{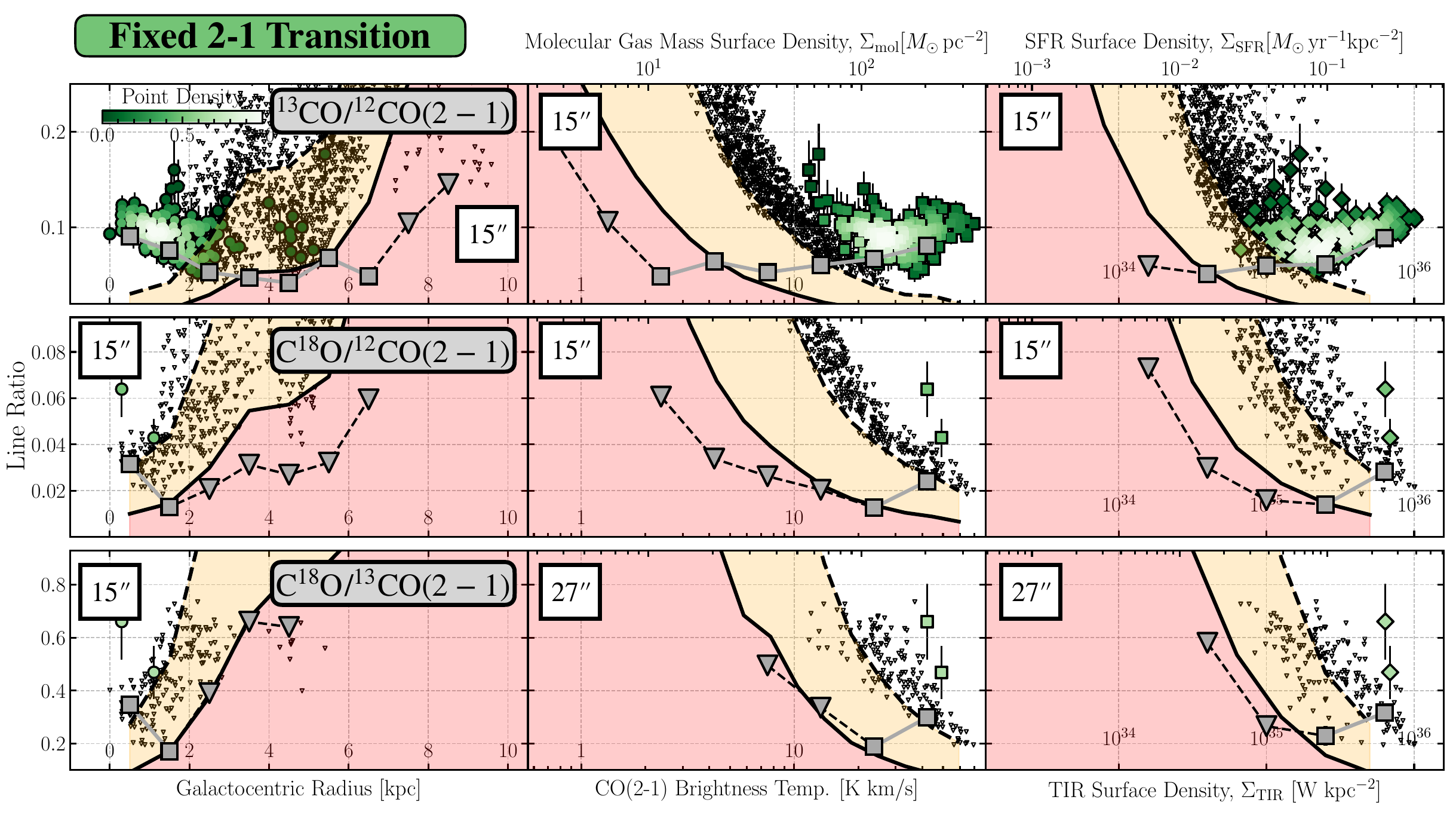}
    \caption{{\bf CO line ratio keeping a fixed \trans{21} transition but different CO isotopologues.} For description of the panels see \autoref{fig:ratio_fixed_iso}.}
    \label{fig:ratio_fix_trans2}
\end{figure*}

\begin{figure*}
    \centering
    \includegraphics[width = \textwidth]{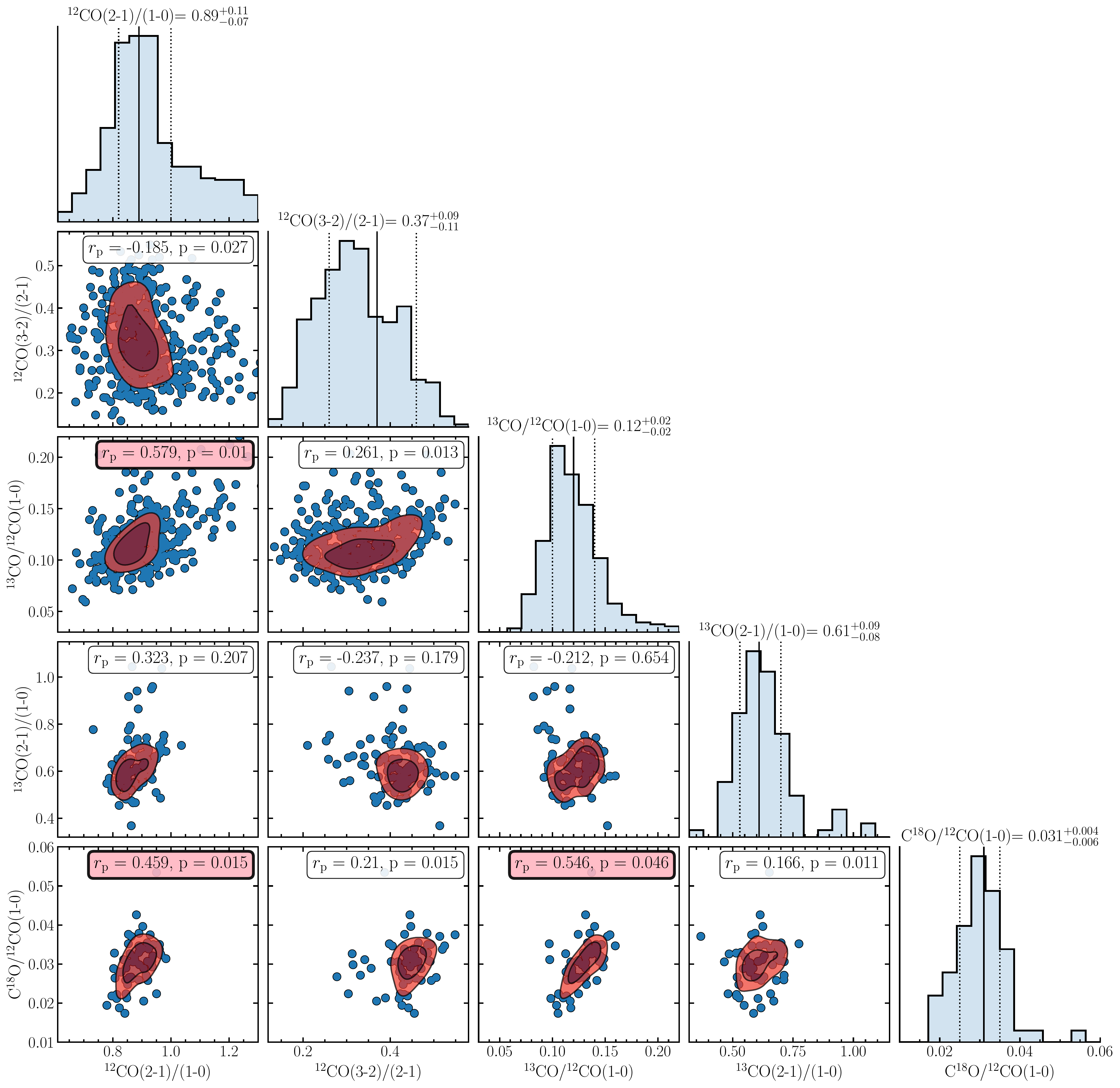}
    \caption{{\bf Comparing different CO line ratios against each other at 27\,arcsec.} Individual sight line, significantly detected CO line ratios vs.\ another line ratios are shown. Contours show the 50~per~cent and 75~per~cent inclusion region, based on a kernel density estimation (KDE). For each line ratio comparison, Pearson's~$r_\mathrm{P}$ correlation coefficients and its $p$-value are indicated in the panel. The histograms indicate the distribution of each line ratio using all significant data points. The \chem{CO}{21} weighted mean and weighted 16th and 84th percentiles. We find a moderate, significant linear correlation ($|r_\mathrm{P}|>0.4$ and $p<0.05$; panel labels are colour coded in case of a moderate linear correlation) only for ratio comparison in panels (b), (g) and~(i).}
    \label{fig:co_ratio_vs_ratio}
\end{figure*}

Line ratios can give insight into the physical conditions of the gas from which the emission originates. For example, under LTE conditions, contrasting optically thin and thick lines allows us to draw conclusions about the optical depth, while comparing two optically thin lines allows for investigating molecular abundances.

In \autoref{fig:ratio_maps}, we inspect spatial variations for a selection of four CO line ratios with a large number of significantly detected sight lines. The figure shows the line ratio maps for $\chem{^{12}CO}{21}/\trans{10}$, 
$\chem{^{12}CO}{32}/\trans{21}$, 
$\chem{^{13}CO}/\chem{^{12}CO}{10}$ and
$\chem{C^{18}O}/\chem{^{12}CO}{10}$ (only for sight lines detected above $5\sigma$ in both lines). On the one hand, the \chem{^{12}CO} lines are generally optically thick so their line ratios relate to gas density or temperature as well as optical depth. (For $\chem{^{12}CO}{21}/\trans{10}$, we find also a non-negligible fraction of values with line ratios $> 1$ at the edges of the map. Such line ratios would require optically thin \chem{^{12}CO} gas. However, at the edge of the map, $\mathrm{S/N}$ effects can drive up the ratio as well.) On the other hand, \chem{^{13}CO} and \chem{C^{18}O} are generally optically thinner, thus making it possible to investigate their optical depths. Just qualitatively assessing the spatial variations and trends, we see that there is a tendency of higher line ratios in the centre. Furthermore, $\chem{^{12}CO}{21}/\trans{10}$ shows clear arm--interarm differences. Along spiral arms, the ratio is lower (${\sim}0.85$) than in interarm regions (${\sim}0.93$; see \autoref{tab:env_sep}). We note that this finding is opposite to the results from \cite{Koda2012}, who found a larger line ratio in spiral arm regions (see \autoref{arm_interarm_result} and \ref{sec:Comp_Lit} for more details). For the $\chem{C^{18}O}/\chem{^{12}CO}{10}$ line ratio, we only detect significant sight lines in the centre. With the help of stacking, we can get significant detections also at larger galactocentric radii ($r_\mathrm{gal} < 6$\,kpc), showing a moderate negative radial trend.

In \autoref{fig:violin_rat}, we investigate the distributions of all CO isotopologue line ratios. These line ratios span a range of around 2\,dex. $\chem{^{12}CO}{21}/\trans{10}$ shows the highest line ratio with \chem{CO}{21} brightness temperature weighted mean of $0.89^{+0.11}_{-0.07}$. The lowest line ratio is given by $\chem{C^{18}O}/\chem{^{12}CO}{10}$ with weighted mean of $0.031^{+0.004}_{-0.006}$. The individual ratios show a 95~per~cent inclusion region of ${\sim}0.5$ to $1$\,dex. 
{ The inset panel shows the effect of increasing the S/N cut of the $^{12}$CO(2-1)/(1-0) line ratio. We do not find a significant difference in terms of the mean line ratio with increasing $\rm S/N$ cut.}

We also study the line ratios binned by galactocentric radius, \chem{CO}{21} brightness temperature and TIR surface brightness. We order the CO line ratios into three categories:
\begin{enumerate}
    \item {Fixed CO isotopologue -- Different Transition} (\autoref{fig:ratio_fixed_iso}).
    \item {Different CO isotopologue -- Fixed $J=1\!\rightarrow\!0$} (\autoref{fig:ratio_fix_trans1}).
    \item {Different CO isotopologue -- Fixed $J=2\!\rightarrow\!1$} (\autoref{fig:ratio_fix_trans2}).
\end{enumerate}
We use the highest working resolution possible for each line ratio. For ratios involving the \chem{C^{17}O}{10} line, we use the lowest resolution of $34$\,arcsec to maximise sensitivity. We define all line ratios such that the fainter line is in the numerator and the brighter one in the denominator. As a consequence, line ratios in faint regions may (frequently) appear as upper limits, while lower limits are very rare.

In \autoref{fig:ratio_fixed_iso}, \ref{fig:ratio_fix_trans1} and \ref{fig:ratio_fix_trans2}, we present our measurements in two ways: (i) we plot the line ratios of individual sight lines as function of galactocentric radius, \chem{CO}{21} brightness temperature and TIR surface brightness, and (ii) we stack the emission line spectra within bins of galactocentric radius, \chem{CO}{21} brightness temperature and TIR surface brightness, and plot the line ratio of these stacked spectra. We note that the sample of sight lines usually differs between case~(i) where only those sight lines are shown for which both lines have $\mathrm{S/N} > 5$ and case~(ii) where all sight lines are included in the stacked spectra while (again) only those line ratios are shown for which both stacked lines have $\mathrm{S/N} > 5$.  We stress that extrapolating these trends based on the stacked points to the individual lines of sight should be done with caution, as by stacking, we discard the intrinsic scatter of the data. This difference in sample implies that frequently the line ratio from stacked spectra does not coincide with the `middle' of the line ratio distribution for individual sight lines (which provide a biased view as long as non-detections are neglected). The censored regions are described in \autoref{sec:ratio} and show regions in the ratio space which we cannot sample due to the limited sensitivity. In these censored regions line ratios from stacked spectra should be considered as these provide a robust and unbiased measurement (since they include non-detected sight lines).


\autoref{tab:spearman} lists Kendall's~$\tau$ rank correlation coefficient as well as its significance, the $p$-value, of various line ratios, based on the stacked data points which have $\mathrm{S/N} > 5$. 
{ We employ the Kendall's~$\tau$ rank correlation coefficient to measure a monotonic increasing, non-linear relationship in our data. It is more robust to error and discrepancies in the data \citep{Croux2010} than the widely used Spearman's rank correlation coefficient. We also do not use the Spearman's rank correlation coefficient, since it is more sensitive to our choice of binning of the stacked datapoints.}
For the calculation of the coefficients for the CO line ratios as a function of galactocentric radius, we only include stacked points at $r_\mathrm{gal} \le 5$\,kpc.
Since \chem{^{12}CO}{10} has the highest sensitivity and this transition appears in the denominator of our line ratios, this leads to line ratio for individual sight lines to turn upwards at larger radii or at fainter \chem{CO}{21} brightness temperatures. To avoid this, we only include stacks with $W_{\chem{CO}{21}} > 2\,\mathrm{K\,km\,s^{-1}}$ for computation of Kendall's~$\tau$ for the CO line ratios as a function of \chem{CO}{21} brightness temperature.
We do not compute the correlation coefficients for CO line ratios with only two or less significantly stacked points.

\begingroup
\setlength{\tabcolsep}{10pt} 
\renewcommand{\arraystretch}{1.5}
\begin{table*}
    \caption{{\bf CO line ratios and galactic environment.} The \chem{CO}{21} brightness temperature weighted mean line ratios and weighted 16th and 84th percentiles are given for different galactic environments.}
    \label{tab:env_sep}
    \centering
    \begin{tabular}{c c c c c }
         \hline
         Line Ratio & Global & Central $45\arcsec$ & Arm & Interarm  \\ \hline 
         $\chem{^{12}CO}{21}/\trans{10}$& $0.89^{+0.11}_{-0.07}$&$0.91^{+0.03}_{-0.05}$&$0.85^{+0.09}_{-0.05}$&$0.93^{+0.11}_{-0.07}$\\
         $\chem{^{12}CO}{32}/\trans{21}$& $0.37^{+0.09}_{-0.11}$&$0.42^{+0.04}_{-0.05}$&$0.31^{+0.06}_{-0.07}$&$0.26^{+0.07}_{-0.05}$\\
         $\chem{^{13}CO}/\chem{^{12}CO}{10}$& $0.12^{+0.02}_{-0.02}$&$0.14^{+0.01}_{-0.02}$&$0.11^{+0.01}_{-0.02}$&$0.12^{+0.03}_{-0.02}$\\ \hline
    \end{tabular}
\end{table*}
\endgroup

\begingroup
\renewcommand{\arraystretch}{1.5} 
\begin{table*}
    \centering
    \caption{{\bf Mean values and Kendall's~$\bm{\tau}$ rank correlation coefficient ($\bm{p}$-value given in parenthesis).} Measured for the line ratios of stacked spectra as function of galactocentric radius, \chem{CO}{21} brightness temperature and TIR surface brightness (see \autoref{fig:ratio_fixed_iso}, \ref{fig:ratio_fix_trans1} and~\ref{fig:ratio_fix_trans2}). Only stacked points with $\mathrm{S/N} > 5$ are considered. A~dash indicates that only two or less significant measurements exist for the specific line ratio.}
    \label{tab:spearman}
    \begin{tabular}{p{0.6cm} c c c | c c c} \hline
    &&&& \multicolumn{3}{c}{Kendall's~$\tau$ rank correlation coefficient}\\
    &Line Ratio&$\langle R \rangle$&$\langle R \rangle^{\rm equal}$& Radius & $W_{\chem{CO}{21}}$ & $\Sigma_{\rm TIR}$  \\
    &&(1)&(2)&(3) & (4) & (5)\\\hline \hline
    &$\chem{^{12}CO}{21}/\trans{10}$&$0.89^{+0.11}_{-0.07}$&$0.9^{+0.3}_{-0.1}$&$-0.20\ (0.82)$&$-0.6\ (0.14)$&$-0.6\ (0.23)$\\
    &$\chem{^{12}CO}{32}/\trans{21}$&$0.37^{+0.09}_{-0.11}$&$0.32^{+0.10}_{-0.09}$&$-0.80\ (0.083)$&$1.0\ (0.003)$&$1.0\ (0.083)$\\
    &$\chem{^{13}CO}{21}/\trans{10}$&$0.61^{+0.09}_{-0.08}$&$0.61^{+0.10}_{-0.08}$&$-0.80\ (0.083)$&$-0.2\ (0.7)$&$1.0\ (0.083)$\\
    &$\chem{C^{18}O}{21}/\trans{10}$&$0.87^{+0.24}_{-0.15}$&$1.0^{+0.2}_{-0.2}$&$-$&$-$&$-$\\ \hline
    
    \multirow{6}{*}{
        \shortstack[l]{fixed-\textit{J} \\ $1\!\rightarrow\!0$}
    }&$\chem{^{13}CO}/\chem{^{12}CO}$&$0.12^{+0.02}_{-0.02}$&$0.12^{+0.02}_{-0.02}$&$-1.0\ (0.017)$&$0.6\ (0.13)$&$0.40\ (0.48)$\\
    &$\chem{C^{18}O}/\chem{^{12}CO}$&$0.031^{+0.004}_{-0.006}$&$0.031^{+0.004}_{-0.008}$&$-0.80\ (0.083)$&$1.0\ (0.017)$&$0.67\ (0.33)$\\
    &$\chem{C^{18}O}/\chem{^{13}CO}$&$0.23^{+0.02}_{-0.03}$&$0.23^{+0.02}_{-0.03}$&$-0.80 \ (0.083)$&$0.80\ (0.083)$&$0.67\ (033)$\\
    &$\chem{C^{17}O}/\chem{^{12}CO}$&--&--&$1.0\ (0.3)$&$-$&$-1.0\ (0.33)$\\
    &$\chem{C^{17}O}/\chem{^{13}CO}$&--&--&$1.0\ (0.3$)&$-$&$-1.0\ (0.33)$\\ 
    &$\chem{C^{17}O}/\chem{C^{18}O}$&--&--&$1.0\ (0.3)$&$-$&$-1.00\ (0.33)$\\ \hline
    \multirow{3}{*}{
         \shortstack[l]{fixed-\textit{J} \\ $2\!\rightarrow\!1$}
    }&$\chem{^{13}CO}/\chem{^{12}CO}$&$0.09^{+0.01}_{-0.02}$&$0.09^{+0.02}_{-0.02}$&$-1.0\ (0.017)$&$0.73\ (0.056)$&$1.0\ (0.083)$\\
    &$\chem{C^{18}O}/\chem{^{12}CO}$&$0.034^{+0.006}_{-0.008}$&$0.034^{+0.008}_{-0.005}$&$-$&$-$&$-$\\
    &$\chem{C^{18}O}/\chem{^{13}CO}$&$0.36^{+0.06}_{-0.06}$&$0.33^{+0.08}_{-0.04}$&$-$&$-$&$-$\\ \hline
    \end{tabular}
    
    {\raggedright {\bf Notes:} (1) $\langle R \rangle$ indicates the average line ratio weighted by \chem{^{12}CO}{21} brightness temperature (see \autoref{eq:w_median}). The uncertainty for each line ratio is given by the weighted 16th and 84th percentiles. (2) The volume weighted median line ratio and 16th and 84th percentiles (since all pixel have the same size, this corresponds to weighing all points equally). (3) For the correlation coefficient computation, we only include stacked points at $r_\mathrm{gal} < 5$\,kpc. (4) Only includes stacked points with $W_{\chem{CO}{21}} > 2\,\mathrm{K\,km\,s^{-1}}$. (5) The correlation coefficient computation does not include any additional constraints.}
\end{table*}
\endgroup

\subsubsection{Fixed CO Isotopologue -- Different Transitions}

\hyperref[fig:ratio_fixed_iso]{Figure~\ref*{fig:ratio_fixed_iso}} shows fixed CO isotopologue line ratios for the three lowest rotational transitions. Such line ratios can give insight into the excitation state of the interstellar medium. Looking at the line ratios as a function of galactocentric radius, we see that especially the stacked measurements of $\chem{^{12}CO}{32}/\trans{21}$ and $\chem{^{13}CO}{21}/\trans{10}$ show negative trends with both having a Kendall's rank correlation coefficient of $\tau = -0.80$ $(p=0.083)$. Within the central region ($r_\mathrm{gal} < 1$\,kpc), the line ratios are enhanced by 37~per~cent for $\chem{^{12}CO}{32}/\trans{21}$ and 45~per~cent for $\chem{^{13}CO}{21}/\trans{10}$ as compared to the average stacked line ratio for $r_\mathrm{gal} < 5$\,kpc. An increase of the line ratio within the centre of the galaxy is also qualitatively visible in \autoref{fig:ratio_maps} (see second to left panel for $\chem{^{12}CO}{32}/\trans{21}$). Conversely, the $\chem{^{12}CO}{21}/\trans{10}$ line ratio does not show a significant trend with galactocentric radius, it shows only a slight enhancement of 5~per~cent in the centre over the average within 5\,kpc. For $\chem{C^{18}O}{21}/\trans{10}$, we do not have enough significantly detected stacked spectra to investigate a trend with radius.

For the line ratios as a function of the \chem{^{12}CO}{21} brightness temperature, only  $\chem{^{12}CO}{32}/\trans{21}$ shows a significant positive trend with Kendall's $\tau = 1.00$ ($p = 0.003$). However, we note that especially for lower CO brightness temperatures, the effect of correlated axes might enhance such a trend. Both the $\chem{^{12}CO}{21}/\trans{10}$ and $\chem{^{13}CO}{21}/\trans{10}$ line ratios do not show any clear trend with CO brightness temperature. However, for $\chem{^{13}CO}{21}/\trans{10}$, we find a positive trend in the range $W_{\chem{CO}{21}} > 2\,{\rm K\,km\,s^{-1}}$. 

Finally, considering the line ratios as function of the TIR surface brightness, we again find positive trends for $\chem{^{12}CO}{32}/\trans{21}$ with $\tau = 1.0$ $(p=0.083)$. While, for $\chem{^{12}CO}{21}/\trans{10}$, we find a flat trend with respect to $\Sigma_{\rm TIR}$. Due to the faintness of the \chem{C^{18}O}{21} line, we do not have enough significant points for the $\chem{C^{18}O}{21}/\trans{10}$ line ratio to determine if any trend exists.

Of particular conspicuousness is the fact that the stacked $\chem{^{12}CO}{32}/\trans{21}$ line ratio shows opposite trends in every column (i.e. with galactocentric radius, CO brightness temperature and TIR surface brightness) compared to the $\chem{^{12}CO}{21}/\trans{10}$ line ratio. This is opposite to the notion that both ratios show a similar behaviour, but could be coupled to the fact that \chem{^{12}CO}{32} is more constrained to the denser regions of the molecular interstellar medium.

\subsubsection{Different CO Isotopologue -- Fixed-\textit{J} Transition}

Looking at different CO isotopologue line ratios with fixed rotational transitions can give insight into various physical quantities such as the abundance of the molecule or its optical depth, depending on whether the emission line is optically thin or thick \citep{Davis2014, Donaire2017, Cormier2018}.
\hyperref[fig:ratio_fix_trans1]{Figures~\ref*{fig:ratio_fix_trans1}} and~\ref{fig:ratio_fix_trans2} show CO line ratios for different CO isotopologues but at fixed-\textit{J} transition. The $\chem{^{13}CO}/\chem{^{12}CO}$ line ratio shows for both the \trans{10} and \trans{21} transition the same significant correlation with galactocentric radius (negative), \chem{^{12}CO} brightness temperature (positive) and TIR surface brightness (positive). The \trans{10} line ratio shows also indication of an enhancement in the centre of the galaxy of order 20~per~cent (see also centre right panel in \autoref{fig:ratio_maps}). The $\chem{C^{18}O}/\chem{^{12}CO}{10}$ line ratio shows a negative trend with galactocentric radius, but positive trends with \chem{^{12}CO} brightness temperature and TIR surface brightness. For the \chem{C^{17}O} line, we only have two sight lines with significant detection at $34$\,arcsec resolution. We are very limited by the censored region and from the individual sightlines alone, we cannot infer any trends with radius, CO brightness and/or the TIR surface density. For \chem{C^{17}O}/\chem{^{12}CO}(1-0) we find an average value of ${\sim}0.02$, for \chem{C^{17}O}/\chem{^{13}CO}(1-0) ${\sim}0.15$, and for \chem{C^{17}O}/\chem{C^{12}O}(1-0) we find ${\sim}0.7$.  Using stacking, we identify a positive trend with radius out to 4\,kpc for the \chem{C^{17}O}/\chem{^{12}CO}(1-0), \chem{C^{17}O}/\chem{^{13}CO}(1-0) and \chem{C^{17}O}/\chem{C^{18}O}(1-0) ratios. Since the individual sightlines are affected by the censoring, we also provide the average line ratios using the stacked spectra: For stacked \chem{C^{17}O}/\chem{^{12}CO}(1-0) we find $0.009\pm0.002$, for the stacked \chem{C^{17}O}/\chem{^{13}CO}(1-0) we have $0.07\pm0.03$ and for the stacked \chem{C^{17}O}/\chem{C^{18}O}(1-0) we measure $0.4\pm0.1$.

\subsubsection{Comparing Different CO Line Ratios}

As previously highlighted, different CO line ratios trace different gas characteristics and physical properties, depending on the optical depth of the two lines. By comparing different line ratios which trace different conditions (e.g. the optical depth or the chemical abundance), we can gain insight into the relation between various physical conditions. \hyperref[fig:co_ratio_vs_ratio]{Figure~\ref*{fig:co_ratio_vs_ratio}} shows a corner plot comparing different CO isotopologue line ratio combinations against each other. Plotted are the individual sight lines which show significantly detected line ratios in both axes, as well as the 50~per~cent and 75~per~cent inclusion contours using a kernel density estimation (KDE). We note that we do not account for censoring effects which can affect the wider distribution of the data points.  To test for a linear correlation between the different line ratios, we compute the Pearson's $r_{\rm p}$ coefficient.

We do not find a clear linear trend for most combinations of line ratios at our working resolution of $1$ to $2$\,kpc. A~moderate trend ($|r_\mathrm{P}|>0.4$) of high significance ($p<0.05$) can only be seen for $\chem{^{12}CO}{21}/\trans{10}$ versus $\chem{^{13}CO}/\chem{^{12}CO}{10}$, 
$\chem{^{12}CO}{21}/\trans{10}$ versus $\chem{C^{18}O}/\chem{^{12}CO}{10}$ and 
$\chem{^{13}CO}/\chem{^{12}CO}{10}$ versus $\chem{C^{18}O}/\chem{^{12}CO}{10}$ (ratio comparisons in question are marked in \autoref{fig:co_ratio_vs_ratio}). However, these are the combinations that both have the \chem{^{12}CO}{10} line in the denominator. Consequently, the trends may hint more at a correlation between the two numerators. Such trends are expected, given that all CO isotopologues have a similar spatial distribution within the galaxy, so we expect a correlation in the strength of their emission. For all other combinations, no significant linear trend can be determined.

\begin{figure}
    \centering
    \includegraphics[width =0.95 \columnwidth]{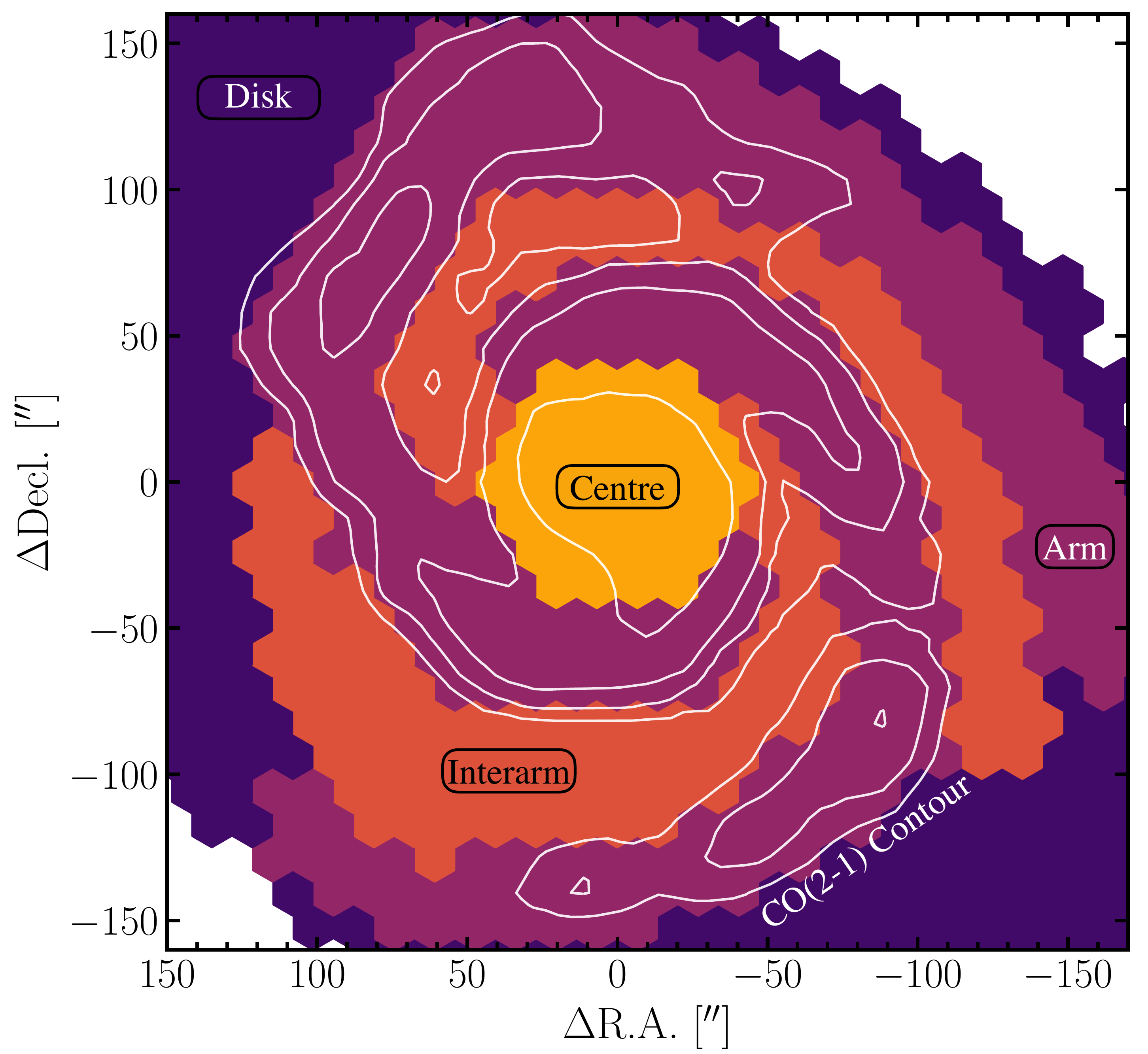}
    \caption{{\bf Environmental mask of M51.} Based on visual inspection of optical \textit{HST} data and \chem{^{12}CO}{21} emission, we generated the following environmental mask. {The pixels are colour-coded by their associated region. This includes the centre (bright orange), interarm (dark orange), spiral arm (magenta) and disk (purple).} The centre includes data points within a $45$\,arcsec aperture. White contours indicate \chem{^{12}CO}{21} emission at 7, 10, 15 and 30\,K\,km\,s$^{-1}$. }
    \label{fig:env_mask}
\end{figure}

\begin{figure*}
    \centering
    \includegraphics[width = \textwidth]{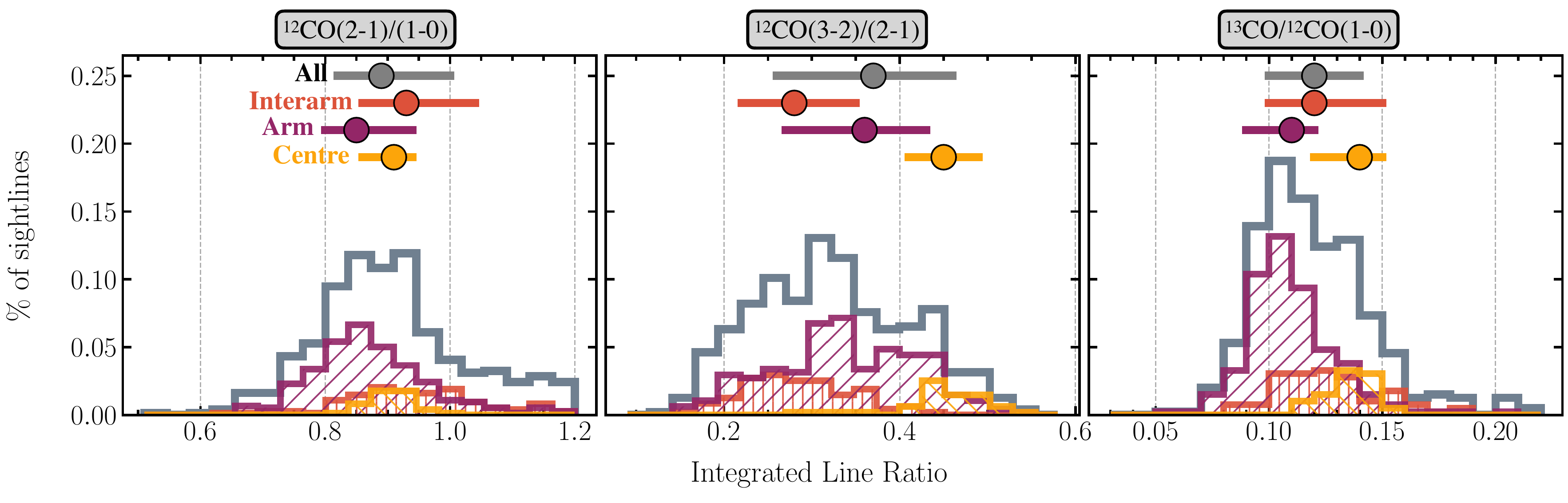}
    \caption{{\bf Distributions for three selected CO line ratio in different galactic environments.} The locations of the various galactic regions are shown in \autoref{fig:env_mask}. For each region, the \chem{CO}{21} brightness temperature weighted median (see \autoref{eq:w_median}) {is indicated by the circle} and the 16th and 84th percentile range { by the bars}. The numerical values are given in \autoref{tab:env_sep}. For the $\chem{^{12}CO}{21}/\trans{10}$ line ratio, we find evidence of larger values in the interarm than in the spiral arm. For $\chem{^{12}CO}{32}/\trans{21}$ we find the opposite trend. $\chem{^{12}CO}{32}/\trans{21}$ and $\chem{^{13}CO}/\chem{^{12}CO}{10}$ show a clear enhancement of the line ratio in the galaxy centre.}
    \label{fig:environment}
\end{figure*}

\subsection{Line Ratios and the Galactic Environment}

For several CO line ratios, an enhancement towards the centre of the galaxy is observed (see \autoref{sec:line_ratios}). To analyse whether galaxy morphological features, such as the galactic centre, spiral arms or interarm regions, have an impact on the molecular gas properties, we classify each sight line as centre, spiral arm, interarm or general disc (see \autoref{fig:env_mask}). This classification is based on visual inspection of optical \textit{HST} data as well as the extent of the \chem{^{12}CO}{21} emission. In our study, the `centre' refers to the central ($r_\mathrm{gal} \le 45$\,arcsec $\approx 2$\,kpc) region of the galaxy.
 
\hyperref[fig:environment]{Figure~\ref*{fig:environment}} shows histograms of the CO line ratio distributions for $\chem{^{12}CO}{21}/\trans{10}$, $\chem{^{12}CO}{32}/\trans{21}$ and $\chem{^{13}CO}{21}/\trans{10}$, separated by the different environments. These three line ratios show significant pixel detections in the centre, spiral arm and interarm regions. The numerical values of the mean and 16th and 84th percentiles for each environment are given in \autoref{tab:env_sep}. We find again that all three line ratios show higher values in the central region, but this enhancement is significant only in the case of $\chem{^{12}CO}{32}/\trans{21}$. Furthermore, we find evidence for a difference in the line ratios between arm and interarm regions for $\chem{^{12}CO}{21}/\trans{10}$ and $\chem{^{12}CO}{32}/\trans{21}$. We note that the line ratios in the interarm regions are higher for $\chem{^{12}CO}{21}/\trans{10}$ and lower for $\chem{^{12}CO}{32}/\trans{21}$ than the spiral arm regions.
For the $\chem{^{13}CO}/\chem{^{12}CO}{10}$ line ratio, we also find a larger average value in the interarm ($0.12^{+0.03}_{-0.02}$) than in the spiral arm region ($0.11^{+0.01}_{-0.02}$), though the two medians lie within the margin of error of each other.

\subsection{Spiral Arm and Interarm Variations}
\label{arm_interarm_result}

\begin{figure*}
    \centering
    \includegraphics[width = \textwidth]{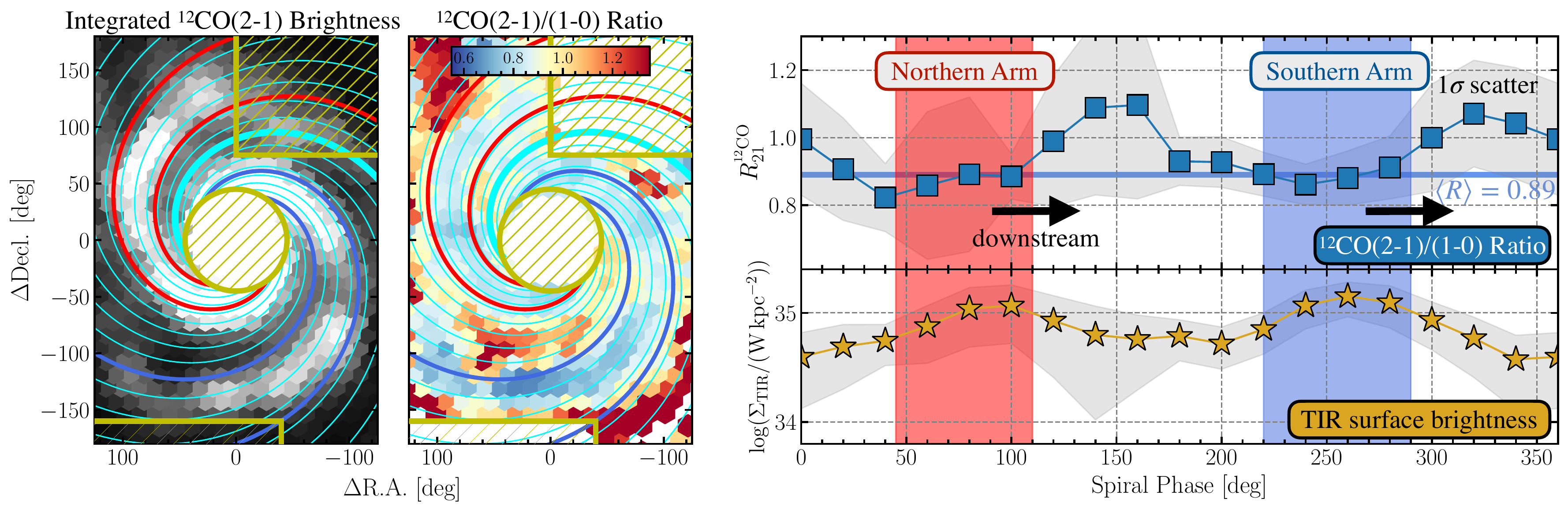}
    \caption{{\bf Spiral arm--interarm CO line ratio variation.} This figure follows closely the layout of fig.\,4 in \protect\cite{Koda2012}. (\textit{Left}) \chem{^{12}CO}{21} brightness temperature map with the logarithmic spiral pattern. The logarithmic spiral has a pitch angle of $20^\circ$. The bold cyan line indicates the starting spiral phase ($\psi = 0^\circ$). The spiral phases increase counterclockwise with 20$^\circ$ increments. The red and blue coloured spirals indicate the bounds of the molecular gas spiral arm (at $\psi = 45^\circ - 110^\circ$ and $220^\circ - 290^\circ$). We ignore the top right and bottom left regions in our analysis, where the molecular arm deviates from the logarithmic spiral pattern. Furthermore, we exclude the central (radius of $45$\,arcsec) region (indicated by the yellow hashed region). (\textit{Centre}) The $\chem{^{12}CO}{21}/\trans{10}$ line ratio map. We only show sight lines at $\mathrm{S/N} > 5$ and not within the exclusion regions. (\textit{Right}) The average line ratio for each spiral phase bin on top (which has a width of $40^\circ$ and increases by $20^\circ$ every step). The blue horizontal line shows the galaxy-wide, luminosity weighted median line ratio of $\langle R_{21}^{\chem{^{12}CO}} \rangle$. The bottom panel shows the TIR surface density (which is proportional to the SFR surface density) to highlight the spiral arm. The gray shaded area indicates the standard deviation. The location of the Northern (red) and Southern (blue) spiral arm are indicated. { The downstream direction of the spiral arm is indicated by the black arrow.}}
    \label{fig:arm_interarm}
\end{figure*}

\begin{figure}
    \centering
    \includegraphics[width = 0.9\columnwidth]{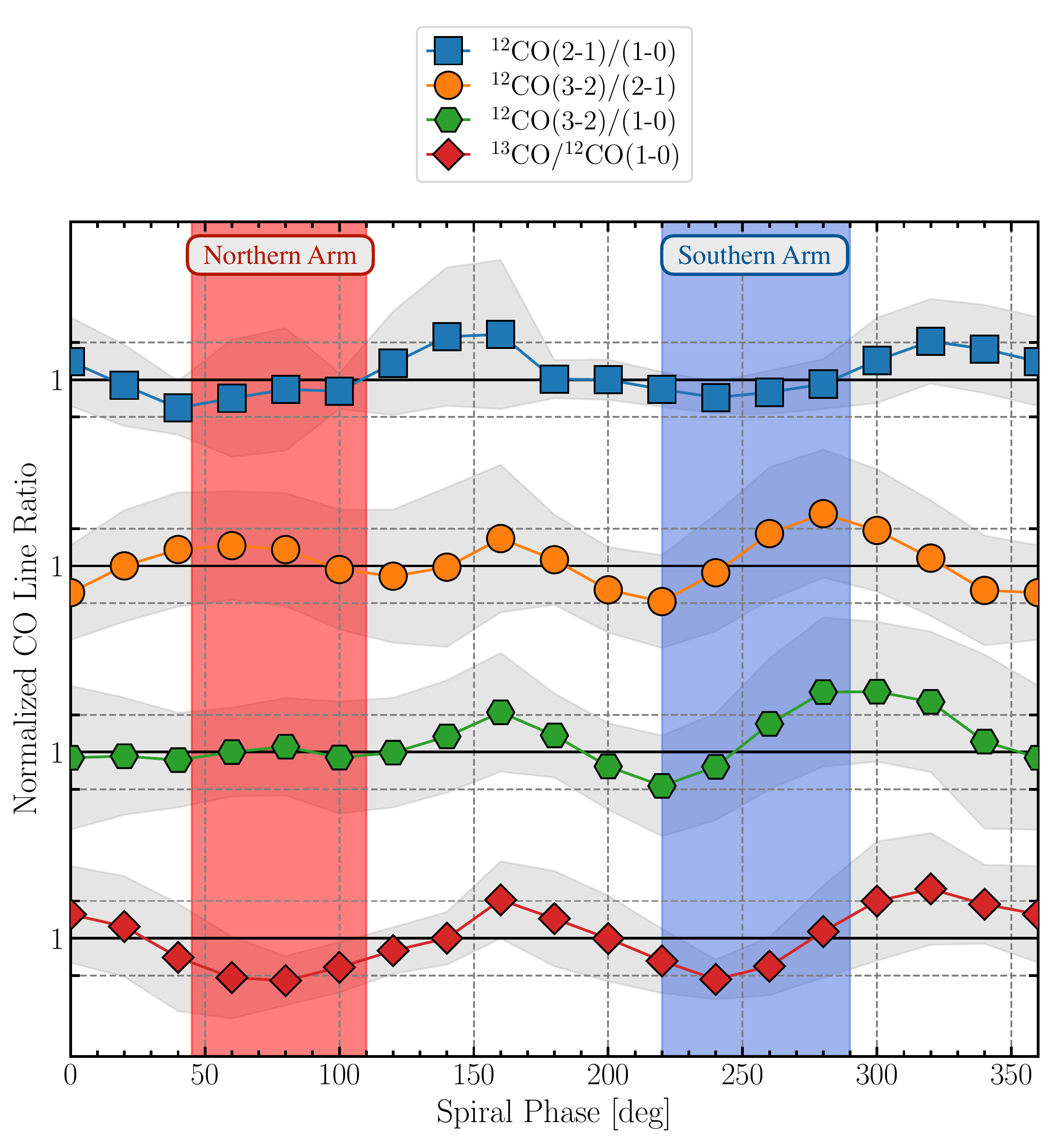}
    \caption{{\bf Spiral arm--interarm line ratio variations for three different CO line ratios.} We only include sight lines with $\mathrm{S/N} > 5$ for each of the emission lines of the respective line ratio. We use a working resolution of $27$\,arcsec. We normalise the line ratio by the median value. The horizontal grey dotted lines indicate $\pm 15$~per~cent deviations from the median value.}
    \label{fig:arm-inter_all}
\end{figure}

In the following section, we further investigate the arm/interarm contrast, as it stands against previous studies, which found a larger $\chem{^{12}CO}{21}/\trans{10}$ line ratio in the spiral arm region than the interarm region \citep{Koda2012}.
To further study the systematic variations of the $\chem{^{12}CO}{21}/\trans{10}$, $\chem{^{12}CO}{32}/\trans{21}$ and $\chem{^{13}CO}/\chem{^{12}CO}{10}$ line ratios across the spiral arm and interarm regions, we use a similar approach as described in \citet{Koda2012}. The method is illustrated for the $\chem{^{12}CO}{21}/\trans{10}$ line ratio in \autoref{fig:arm_interarm}, while a comparison of all four line ratios is shown in \autoref{fig:arm-inter_all}. The data are binned using logarithmic spirals, which are described by:
\begin{equation}
    r = \mathrm{e}^{k\cdot \psi}~,
\end{equation}
where $k \equiv \tan(\theta)$ with the pitch angle~$\theta$ and the spiral phase~$\psi$. For M51, we use a pitch angle\footnote{To be consistent with the study of \cite{Koda2012}, we use the value of $20\degr$ for the pitch angle. This value is similar to the value of $21\fdg1$ given by \cite{Shetty2007} and the value of $18\fdg5$ found in \cite{Pineda2020}.} of $\theta = 20\degr$. Each segment spans over $40\degr$ and we increment in steps of $\Delta\psi = 20\degr$ counterclockwise. For the analysis, we exclude the central ($r_\mathrm{gal} \le 45$\,arcsec) region and only bin points with $\mathrm{S/N} > 5$. The logarithmic spiral segments can be seen in the left and central panels of \autoref{fig:arm_interarm}. The red and blue lines indicate the molecular spiral arms ($\psi = 40\degr - 100\degr$ and $220\degr - 290\degr$). We note that the shaded regions in the upper right and bottom left corners are also excluded, as there the molecular arm starts to deviate significantly from a simple logarithmic spiral. 

The right panel of \autoref{fig:arm_interarm} shows the result of the line ratios binned by spiral phase angle. The two spiral arms are based on molecular gas emission. The figure shows the binned arithmetic average of each logarithmic spiral arm segment. Using the same data, we can confirm the larger $\chem{^{12}CO}{21}/\trans{10}$ line ratios in the interarm regions compared to the arm regions found by \cite{denbrok2021}. The average line ratio in the central segment of the molecular arm region is $0.85$, while in the interarm region the average is ${\sim}0.93$. We note that the enhancement of the line ratio in the interarm is more concentrated towards the downstream, convex part of the spiral region. A similar variation is also seen in the $\chem{^{13}CO}/\chem{^{12}CO}{10}$ ratio (see \autoref{fig:arm-inter_all}). 
Note that \autoref{tab:env_sep} also lists the line ratios separated into arm and interarm. But the value in the table are $\chem{^{12}CO}{21}$ weighted and combine data points spanning a wider range of spiral phases. Consequently, we concentrate the analysis of the azimuthal variation on the analysis plotting the line ratios as function of the spiral phase.

\begin{figure}
    \centering
    \includegraphics[width = \columnwidth]{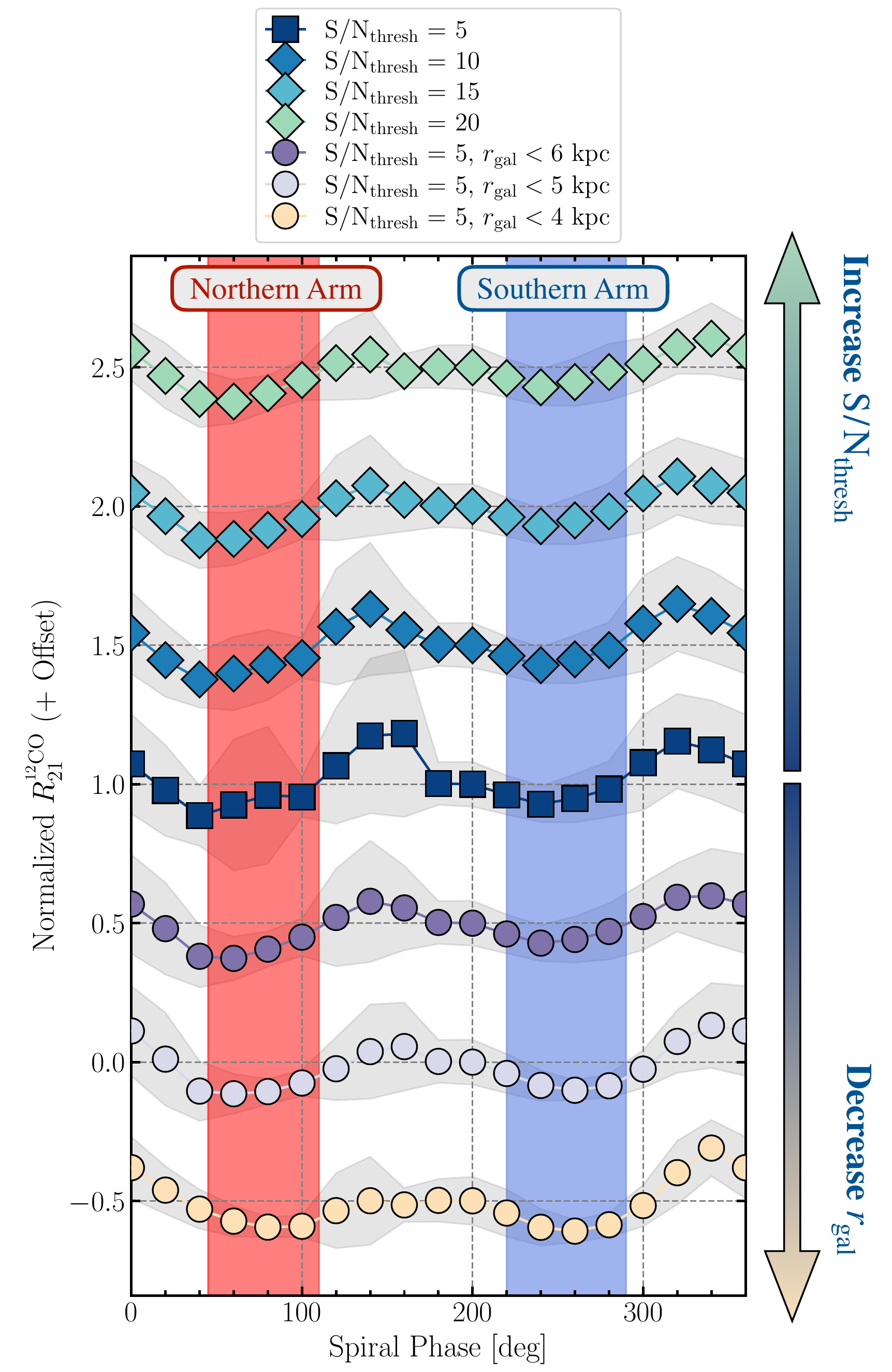}
    \caption{{\bf Spiral arm--interarm CO line ratio variation consistency.} For the arm--interarm variation in \autoref{fig:arm_interarm}, we include all sight lines with $\mathrm{S/N} > 5$. Here, we increase the $\mathrm{S/N}$ threshold from~5 to~20 in steps of~5 to investigate whether the higher line ratios are still preserved in the interarm region. Furthermore, we add a radial constraint, only including sight lines within a given galactocentric radius of 4, 5 and 6\,kpc. We normalise the line ratios by the global \chem{CO}{21} brightness weighted median. We offset different line ratios in steps of~$0.5$ (in positive direction for increasing $\mathrm{S/N}$ and in negative steps for decreasing radial constraint). We see that the trend of higher line ratios in the interarm region is persistent for all computations shown in this figure.}
    \label{fig:cons_check}
\end{figure}

To make sure the trend we find is not due to higher line ratios at larger radii, we vary the $\mathrm{S/N}$ threshold and also add a constraint to only include sight lines within a given galactocentric radius. The results are shown in \autoref{fig:cons_check}, where besides the $\mathrm{S/N} > 5$ threshold (which is used in \autoref{fig:arm_interarm}), we also include thresholds of $\mathrm{S/N} > 10, 15, 20$. For the radial thresholds, we include $r_\mathrm{gal} < 6, 5, 4$\,kpc. We see that the finding of larger $\chem{^{12}CO}{21}/\trans{10}$ line ratios in the interarm region is independent of the threshold implemented.

We note that in particular the 3mm lines in the interarm region are subject to significant error beam contribution, which to some degree can drive the larger  $\chem{^{12}CO}{21}/\trans{10}$ line ratio in the interarm region. We discuss the effect in more detail in \hyperref[sec:errcontr]{Appendix~\ref{sec:errcontr}}.


\section{Discussion}
\label{sec:discussion}

The rich data set of different CO isotopologues 
allows us to address a multitude of science questions. 
First, in \autoref{sec:Comp_Lit}, we will look at the various trends we find in the CO isotopologue line ratios. We will compare them to the literature and then, in \autoref{sec:imps_ratio}, investigate what the observed trends in the line ratios may imply. In \autoref{sec:Comp_Lit}, we will also look into a particular environmental variation of the line ratio -- the arm--interarm differences we find most notably in the $\chem{^{12}CO}{21}/\trans{10}$ line ratio -- and constrain the cause for this variation. The galactic environment appears to have an impact on the line ratio, so we will also investigate how environment affects the CO SLED in \autoref{sec:disc_COSLED}.

\subsection{Comparing \texorpdfstring{$R_{21}$}{L} with Previous Studies}
\label{sec:Comp_Lit}


A first question to address is whether the galaxy M51 conforms to the trends seen in other galaxies regarding the CO line ratios and what we can learn from such trends. When looking at the \chem{^{12}CO} line ratios, a few peculiarities are evident for M51. The well-studied line ratio of $\chem{^{12}CO}{21}/\trans{10}$ (denoted hereafter as $R_{21}^{\chem{^{12}CO}}$) is rather high with a luminosity-weighted average value of $R_{21}^{\chem{^{12}CO}} = 0.89^{+0.11}_{-0.07}$. How does this compare to findings from other studies using different data sets?
In their \chem{^{12}CO} line ratio study, comparing literature values of a larger set of spiral disc galaxies, \citet{denbrok2021} found a common line ratio of $R_{21}^{\chem{^{12}CO}} = 0.59\pm0.1$, making the line ratio we find for M51 clearly stand out. Similarly, \cite{Leroy2021_ratio}, studying the low-\textit{J} \chem{^{12}CO} line ratios using single-dish CO mapping surveys and the PHANGS-ALMA survey find a mean value and respective 16th and 84th percentile of $R_{21}^{\chem{^{12}CO}} = 0.65^{+0.18}_{-0.14}$. Other studies that investigated this line ratio in M51 found lower global values. \citet{Koda2012} found an average value of $R_{21}^{\chem{^{12}CO}} \approx 0.7 $, while they identified enhanced values in the spiral arm regions which are closer to our mean value. \citet{Vlahakis2013} also investigated the spatial variation of the \chem{^{12}CO} line ratios in M51. They found an overall higher median $R_{21}^{\chem{^{12}CO}}$ of~$0.8$, which is close to our luminosity-weighted value. 

In~\autoref{fig:arm_interarm}, we study the spatial variation of the $\chem{^{12}CO}{21}/\trans{10}$ line ratio across the arm and interarm regions of M51. The main result is the higher line ratios in the interarm regions (${\sim}0.93$) as opposed to the arm regions~(${\sim}0.85$), as also seen in \autoref{tab:env_sep}. This trend stands against previous findings e.g.\ from \citet{Sakamoto1997} who found the opposite trend in the Milky Way. \citet{Koda2020} reported higher line ratios in arm as opposed to interarm regions for the spiral galaxy M83. In~fact, \cite{Koda2012} and \cite{Vlahakis2013} (using the same data) studied the arm--interarm variation in M51 and found opposing trends with respect to this study.

We have investigated in detail the origin of this discrepancy of the arm--interarm variation in \citet{denbrok2021}. The study by \cite{Koda2012} used different \chem{CO}{10} and \chem{CO}{21} data than used here. Their \chem{CO}{21} data were taken from the HERACLES survey \citep{Leroy2011}, and the \chem{CO}{10} data came from NRO observations \citep{Koda2011}. Performing the same analysis as shown in \autoref{fig:arm_interarm} for all the different combinations of the \chem{CO}{10} and \chem{CO}{21} data sets, 
\cite{denbrok2021} found that the cause of the discrepancy comes from substituting the \chem{CO}{10} data and not from substituting \chem{CO}{21} data. This demonstrates that the opposite arm--interarm trend is not an artifact of the new \chem{CO}{21} data taken with the IRAM \mbox{30-m} telescope as part of this programme and that care should be taken combining data sets from different telescopes and receivers when carrying out measurements of molecular line ratios.

To verify if using \chem{^{12}CO}{21} with a different sensitivity and flux calibration uncertainty changes the outcome, we substitute the CLAWS \chem{CO}{21} data with the HERACLES \chem{CO}{21} data \citep{Leroy2009} as an experiment. We also find a lower line ratio ($R_{21}^{\chem{^{12}CO}} \approx 0.7$; see \hyperref[sec:CLAWS-HERA]{Appendix~\ref{sec:CLAWS-HERA}}). Such a dependence of the absolute value on the specific data set has been discussed in \cite{denbrok2021} by comparing HERACLES, ALMA, IRAM \mbox{30-m} and NRO \chem{CO}{10} and \chem{CO}{21} data. A variation of order 20~per~cent has been found between observations of the same line from different telescopes. This variation is attributed to uncertainties in the absolute flux calibration of the individual telescopes, which can vary between 5~per~cent to 20~per~cent. However, the flux calibration generally affects the observational data globally, so while the absolute values may differ, we would not expect to find different galaxy-wide \emph{trends} when comparing CO ratios using different single-dish telescope data (e.g.\ \citealt{denbrok2021} find the same arm/interarm $R_{21}^{\chem{^{12}CO}}$ -- i.e. higher values in the interarm region relative to the arm region -- trend in M51 when substituting CLAWS with HERACLES data).

The discrepancy could also be explained due to instrumental reasons. As previously mentioned, uncertainties in the absolute flux calibration of order up to 20~per~cent can lead to differences in the absolute value found when comparing observations from different telescopes and/or observing runs. Such variations are, however, expected to lead to global differences, so we do not expect local variation. Furthermore, contributions from the error beam might significantly affect the detected emission, in particular when observing interarm positions, if emission from brighter spiral arm regions enters via the error beam. In \hyperref[sec:errcontr]{Appendix~\ref{sec:errcontr}}, we investigate these effects for this survey and show that we still find the same arm--interarm trend even when taking account of these instrumental effects.

\begin{figure*}
    \centering
    \includegraphics[width = \textwidth]{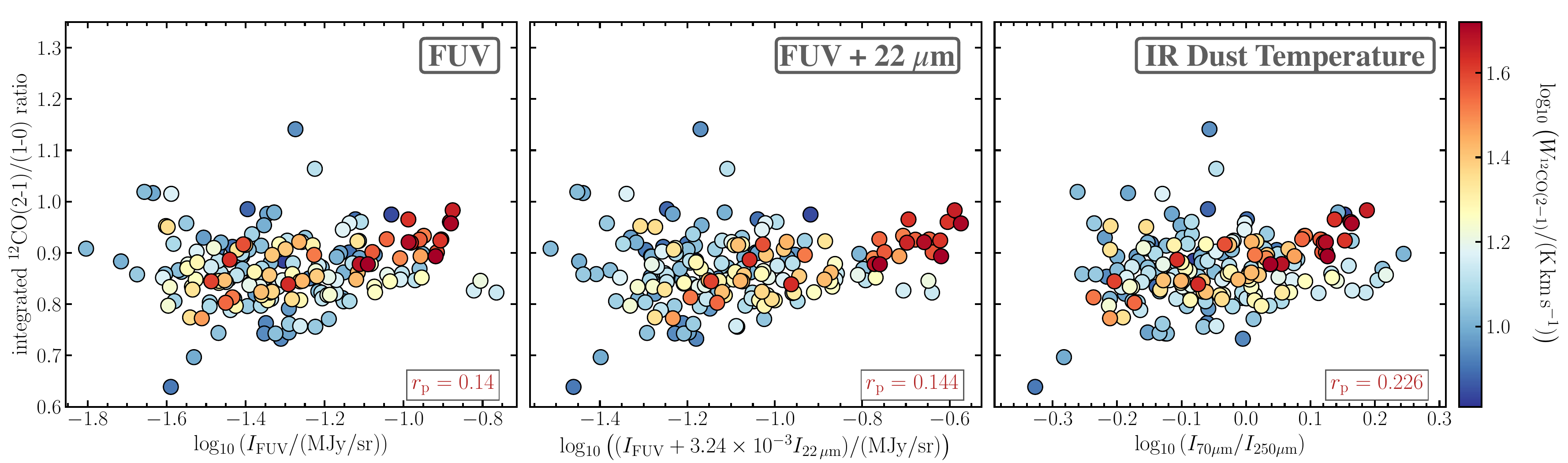}
    \caption{{\bf $\chem{^{12}CO}{21}/\trans{10}$ line ratio compared to FUV intensity (\textit{left}), FUV+22\,$\mu$m (\textit{centre}) and infrared colour (\textit{right}).} We only include sight lines with $\mathrm{S/N} > 5$ for the line ratio to minimise scatter from faint \chem{^{12}CO} lines. We use a working resolution of $27$\,arcsec. FUV emission follows to some extend young (${\le}100$\,Myr) stellar populations. Because FUV can be heavily obscured, we correct using 22\,$\mu$m data from \textit{WISE} band-4 observations. To combine FUV and 22\,$\mu$m, we multiply the 22\,$\mu$m by a factor $3{-}24\times10^{-3}$. The resulting sum is proportional to the SFR surface density \citep{Leroy2019}. The IR $70\,\mu\mathrm{m} / 250\,\mu\mathrm{m}$ colour is a proxy for the dust temperature. We provide the Pearson's linear correlation coefficient, $r_{\rm p}$, in each panel. We do not find any evident correlation of the line ratio with either quantity.}
    \label{fig:corr_dust}
\end{figure*}

The $\chem{CO}{21}/\trans{10}$ line ratio is a tracer of both the gas density and excitation temperature. There is strong evidence that interarm regions do not host higher density molecular gas \citep{Sun2020}. Consequently, a higher line ratio in interarm regions would indicate the presence of molecular gas with a higher excitation temperature. If the UV and optical attenuation is lower in a certain region, one can assume that the gas is less shielded and the molecular gas can reach higher excitation temperatures, producing a higher line ratio. By contrast, regions with higher density gas are better shielded and thus showing a lower excitation temperature. We note, however, that \cite{Koda2020} did not exclusively found larger line ratios in spiral arm regions in M83. To be more precise, they found larger $R_{21}^{\chem{^{12}CO}}$ ratios towards the downstream, convex part of the spiral arm, spanning into the interarm region (see fig.\,1 in \citealt{Koda2020}). We also find enhanced line ratios more towards the downstream, convex part of the spiral arm. Further, they postulated that there is a direct or indirect link between dust heating via the interstellar radiation field and molecular cloud conditions, which may explain the trend. The large CO line ratio could then potentially be explained due to the evolution of massive stars after leaving the spiral arm downstream and consequently contributing more heavily to the dust heating because they are no longer obscured by their birth cloud. Based on the timescales of this evolution, we would expect the location to vary within spiral arm and interarm regions. Looking at \autoref{fig:corr_dust}, we do not find any apparent global correlation between FUV, the sum\footnote{We use \textit{WISE} band-4 22\,$\mu$m observations. For the sum, we normalise the IR intensity by a factor $3.24\times10^{-3}$. The resulting sum is proportional to the SFR surface density \citep{Leroy2019}.} FUV+22\,$\mu$m or dust colour and $R_{21}^{\chem{^{12}CO}}$. Consequently, the explanation provided by \cite{Koda2020} for the M83 is at least not straightforward to apply to explain the arm--interarm line ratio variations observed in M51.
Alternatively, the presence of diffuse emission can impact the line ratio. The result of a lower  $R_{21}^{\chem{^{12}CO}}$ in the spiral arm region could suggest the presence of a diffuse CO gas component, which boosts the \chem{^{12}CO}{10} emission \citep{Cormier2018}. Such a diffuse component would be in accordance with the diffuse component found by \cite{Pety2013}.

\subsection{{Interpreting Variations of Other Line Ratios}}

For the $\chem{^{12}CO}{32}/\trans{21}$ line ratio, \citet{Vlahakis2013} found a global value of $R_{32}^{\chem{^{12}CO}} = 0.5\pm0.14$, which is higher than the value we find ($R_{32}^{\chem{^{12}CO}} = 0.37^{+0.09}_{-0.11}$), but within the margin of error. Similar to our finding, they identified a trend of larger values in the arm ($0.5$) than in the interarm ($0.4$), but their absolute values again are higher than the values we find.

Due to the high sensitivity of the observations we also significantly detect \chem{C^{17}O}{10} emission towards the centre of the galaxy, if we use the $34\arcsec$ spatial resolution data (see \autoref{fig:ratio_fix_trans1}). While we find a positive trend with the galactocentric radius for the \chem{C^{17}O}/\chem{^{12}CO}{10},  \chem{C^{17}O}/\chem{^{13}CO}{10}, and \chem{C^{17}O}/\chem{C^{18}O}{10} ratios, we note, however, the caveat, that this could also be an artefact due to the fact, that are constrained by data points within the censored region. For \chem{C^{17}O}/\chem{C^{18}O}{10}, when stacking by radius, we find an average value of $0.4\pm0.1$. This is slightly larger than but of similar order of the value of $0.24\pm0.01$ found for the solar neighborhood \citep{Wouterloot2005}.

We also compare the azimuthal trend for the $\chem{^{12}CO}{32}/\trans{21}$, $\chem{^{12}CO}{32}/\trans{10}$ and $\chem{^{13}CO}/\chem{^{12}CO}{10}$ line ratios. We select these line ratios because we already investigated them in \autoref{fig:environment}. For all three line ratios in \autoref{fig:arm-inter_all}, we only include sight lines which have $\mathrm{S/N} > 5$ for both lines for each spiral phase. We notice that particularly the $\chem{^{13}CO}/\chem{^{12}CO}{10}$ line ratio clearly shows larger values in interarm regions compared to spiral arm regions. Again, such a trend could be attributed to either an increased \chem{^{13}CO} abundance or variation in the optical depth of the gas. Based on our discussion in \autoref{sec:imps_ratio}, in which we argue that \chem{^{13}CO} abundance variations cannot explain the variation in the different CO isotopologue line ratios, we therefore conclude that most likely changes in the optical depth explain the observed difference in the arm--interarm values for the $\chem{^{13}CO}/\chem{^{12}CO}$ line ratio.

The $\chem{^{12}CO}{32}/\trans{21}$ line ratio also shows clear azimuthal variations. 
{ We see that the line ratio also peaks in the interarm region at spiral phase of ${\sim}150^\circ$ (see \autoref{fig:arm-inter_all}). We note that from the lines analysed in the figure, it is the only one that does not have $^{12}$CO(1-0) in the denominator. The fact that the line ratio also peaks in the interarm region further underlines the presumption that the discrepancy to previous results comes from the use of different $^{12}$CO(1-0) datasest (PAWS and NRO; see discussion in \autoref{sec:Comp_Lit}).} But for  $\chem{^{12}CO}{32}/\trans{21}$, the line ratio value also peaks in the spiral arm regions. \cite{Vlahakis2013} attributed the larger values in spiral arm compared to interarm regions to the presence of warmer and/or denser molecular gas. This is consistent with the detection of \chem{HCN}{10} in spiral arms \citep{Querejeta2019}, but it does not explain the opposite trend for the $\chem{^{12}CO}{21}/\trans{10}$ line ratio. We have also see opposite stacked $\chem{^{12}CO}{32}/\trans{21}$ ratio trends with galactocentric radius, \chem{^{12}CO}{21} brightness temperature and TIR surface brightness when comparing to stacked $\chem{^{12}CO}{21}/\trans{10}$ ratio trends (see \autoref{fig:ratio_fixed_iso}). This could be explained by the fact that, because of its higher excitation, \chem{^{12}CO}{32} emission is more constrained to denser regions within the molecular gas and the detected emission does not have a large diffuse component.

{ In \autoref{fig:co_ratio_vs_ratio}, we compared different CO line ratios with each other. We do not find any clear ratio-to-ratio trends. We recognise that in particular noise effects and uncertainties (e.g. flux calibrational uncertainty) can wash out any minor existing trends. Furthermore, due to the censoring effect of line ratios with a low value, potential correlations could non-trivially be suppressed in this analysis. }

\subsection{Comparing CO Line Ratios to Simulations}

To better understand what we can learn from specific CO isotopologue lines, we will compare our findings to simulations. \citet{Penaloza2017} studied the utility of the $\chem{^{12}CO}{21}/\trans{10}$ line ratio for uncovering the physical and chemical properties of molecular clouds. In their study, they carried out a high resolution smoothed particle hydrodynamics (SPH) simulation of an isolated molecular cloud using the \texttt{GADGET-2}) SPH code supplemented with a model for the time-dependent H$_{2}$ and CO chemistry. The results of this simulation were resampled onto a hierarchical Cartesian mesh and then post-processed with the \texttt{RADMC-3D} radiative transfer code \citep{2012ascl.soft02015D} to generate synthetic maps of \chem{^{12}CO}{21} and \trans{10} line emission. We note that the spatial resolution used in the study is on parsec scale, while our observations are on kiloparsec scales. Consequently, any comparison has to be taken with caution, since differences in the spatial scale can also impact the relations. \citet{Penaloza2017} found a bimodality in $R_{21}^{\chem{^{12}CO}}$, with one peak at ${\sim}0.4$ and another one at ${\sim}0.7$, which are below the value we see (${\sim}0.89$). They attributed the high peak to emission from cold ($T\leq40\,{\rm K}$) and denser ($n\geq10^3\,{\rm cm^{-3}}$) molecular gas, and the lower peak to faint emission from warmer ($T\geq40\,{\rm K}$) and diffuse ($n\leq10^3\,{\rm cm^{-3}}$) molecular gas. We do not see such a bimodality in $R_{21}^{\chem{^{12}CO}}$ (see e.g.\ \autoref{fig:violin_rat}). However, we cannot resolve individual giant molecular clouds with our resolution ($27$\,arcsec or ${\sim}1{-}2$\,kpc). As the lower ratio peak comes from diffuse emission, we would need high sensitivity observations for a secure detection of this component and high spatial resolution to separate it from molecular clouds. In a follow-up study, \cite{Penaloza2018} investigated the impact of the galactic environment on `cloud-averaged' CO line ratios $R_{21}^{\chem{^{12}CO}}$ and $R_{32}^{\chem{^{12}CO}}$. They performed simulations similar to \cite{Penaloza2017}, but involving a much broader range of model clouds, and analysed the post-processed CO emission. For $R_{32}^{\chem{^{12}CO}}$, they reported larger values (${\sim}0.6$) than what we find (${\sim}0.4$). However, their inferred scatter is $\pm0.2$ for both $R_{21}^{\chem{^{12}CO}}$ and $R_{32}^{\chem{^{12}CO}}$ when averaging the line ratio over whole clouds. Based on their simulations, they suggested that the scatter is mainly driven by variations in the interstellar radiation field (ISRF) and the cosmic ray ionisation rate (CRIR). Our data show a similar range of scatter, despite the large difference in spatial scale, so variations of the ISRF and CRIR could be a potential explanation for the observed change in $R_{21}^{\chem{^{12}CO}}$ and $R_{32}^{\chem{^{12}CO}}$. 

The ratios of $R_{21}^{\chem{^{12}CO}}$ and $R_{32}^{\chem{^{12}CO}}$ across galactic environments were also investigated in the recent study by \cite{Bisbas2021}. They rely on three-dimensional thermo\-chemical simulations and synthetic observations of magnetised, turbulent, self-gravitating molecular clouds. They found a remarkably flat trend for both $R_{21}^{\chem{^{12}CO}}$ and $R_{32}^{\chem{^{12}CO}}$ with respect to several galactic parameters, such as CRIR, FUV emission and metallicity. They suggested that the flat trend is mainly due to the fact that all the environmental factors they investigated affect the \chem{^{12}CO} transitions equally. Matching to this prediction, we do not find any significant systematic variation for $R_{21}^{\chem{^{12}CO}}$ (see \autoref{fig:ratio_fixed_iso}). However, we see a clear negative trend with galactocentric radius and SFR surface density for $R_{32}^{\chem{^{12}CO}}$. Consequently, this hints at another driving factor besides changes in ISRF and CRIR. But we note that \cite{Bisbas2021} only simulated molecular clouds and not the diffuse medium, which could explain the observed discrepancy or it can again be attributed to the fact, that in their study, they use a much higher spatial scale (${\sim}$pc scales). 

So far, we have only discussed \chem{^{12}CO} line ratios. In our further analysis, we will now include the other CO isotopologue lines as well to study the impact of the galactic environment.


\subsection{Implications from CO Isotopologue Line Ratio Trends}
\label{sec:imps_ratio}

As discussed in the previous section, we find clear evidence for variations within the galaxy for several combinations of different CO isotopologue line ratios (see \autoref{fig:ratio_fixed_iso}, \ref{fig:ratio_fix_trans1} and~\ref{fig:ratio_fix_trans2}). We remind as a caveat that our observations have a spatial resolution of ${\sim}1-2$\, kpc. Consequently, we study beam-averaged emission, so sub-beam variations can play a role (see \ref{sec:patterplot}). In the subsequent discussion, we will focus on the following line ratios:
\begin{description}
\item $\chem{^{12}CO}{21}/\trans{10}$
\item $\chem{^{13}CO}{21}/\trans{10}$
\item $\chem{^{13}CO}/\chem{^{12}CO}{10}$ 
\item $\chem{C^{18}O}/\chem{^{12}CO}{10}$
\item $\chem{C^{18}O}/\chem{^{13}CO}{10}$
\end{description}
There are several potential explanations for the observed variations and trends. In a study of the $\chem{^{13}CO}/\chem{^{12}CO}$ line ratio variations in a sample of early-type galaxies, \cite{Davis2014} offers three major explanations for systematic changes in the line ratio between different isotopologues. Summarised, variations can be explained by (i) different excitation processes for the individual CO isotopologues, (ii) fractional abundance variations of \chem{^{13}CO} and \chem{C^{18}O} relative to \chem{^{12}CO}, and (iii) changes in the optical depth of the gas for one of the CO isotopologues. We note that, while \chem{^{12}CO} is generally optically thick, \chem{^{13}CO} and \chem{C^{18}O} are mostly optically thin \citep{Cormier2018}, so changes in abundance will impact the emission of these lines. Using data from our survey, we can, to first order, try to isolate the main driver for the observed line ratio variations.

CO is mainly excited to higher rotational states by collisions with H$_2$ or He, or through photon trapping \citep{Narayanan2014}. We do assume that the excitation is subthermal for most parts of our observations. Because all CO isotopologues seem to trace the same spatial region at the spatial resolution of this study (${\sim}20\,\mathrm{arcsec}/1$\,kpc), and the $\chem{^{13}CO}/\chem{^{12}CO}{10}$ and \trans{21} line ratios both exhibit negative trends with galactocentric radius, and positive trends with CO brightness temperature and TIR surface brightness, it is unlikely that different excitation processes, that we will investigate in more detail below, cause the observed variations.

Potential mechanisms that can cause variations in the relative fractional abundance of \chem{^{13}CO} and \chem{C^{18}O} are selective photodissociation, chemical fractionation or selective nucleosynthesis. In a recent letter, \citet{Donaire2017} argued that CO isotopologue trends with galactocentric radius and SFR surface density, found across a sample of nearby spiral galaxies, are consistent with fractional abundance variations expected due to fractionation. Chemical fractionation is a process that can enrich \chem{^{13}CO}, but it is highly temperature dependent \citep{Watson1976, Keene1998} and favourable in cold conditions. This process occurs in cold regions such as predominantly in the outskirts of galaxies. If this were the main cause of the observed variations, we would find lower \chem{^{13}CO} abundances in warmer regions. Due to more heating from young stars, it is expected that the SFR surface density correlates to a certain degree with gas temperature. Consequently, we expect a decrease of \chem{^{13}CO} abundance with increasing $\Sigma_{\rm SFR}$. As \chem{C^{18}O} is not affected by fractionation, the decreasing radial trend that we find for the $\chem{^{13}CO}/\chem{C^{18}O}$ line ratio, is in agreement with this explanation. However, we also see an increasing trend with increasing $\Sigma_{\rm SFR}$ for the $\chem{^{13}CO}/\chem{^{12}CO}$ line ratio, which contradicts the explanation that abundance variations are due to chemical fractionation alone, as we would expect the opposite trend: A larger line ratio at larger radii and smaller SFR surface densities would be due to an increase in the abundance of \chem{^{13}CO} in these environments.

Also selective photodissociation cannot explain the observed trend in $\chem{^{13}CO}/\chem{^{12}CO}$. The process occurs in strong interstellar radiation fields. While on the one hand, \chem{^{12}CO} more shielded, \chem{^{13}CO} on the other hand is less well protected over large areas, so strong UV radiation will destroy the molecule \citep{Dishoeck1988}. As high star formation is linked to the presence of OB~stars, a negative trend with star formation rate would have been expected. 
Abundance variations due to selective nucleosynthesis could explain the observed decreasing trend in the $\chem{^{13}CO}/\chem{C^{18}O}$ line ratio with SFR surface density. While \chem{^{12}C} and \chem{^{18}O} are mainly produced in massive stars, \chem{^{13}C} is primarily produced in low-mass stars \citep{Sage1991}, as it is converted further to \chem{^{14}N} in high-mass stars (${>}8\,\mathrm{M}_\odot$), which would lead to only a very small replenishing rate of \chem{^{13}C} in the ISM \citep{Prantzos1996}. But in low- and inter\-mediate-mass stars \chem{^{13}C} can surface, due to convection, during the red giant phase and consequently enhance the abundance \citep{Wilson1992}. In their study, \citet{Brown2019} attributed the extremely low $\chem{^{13}CO}/\chem{C^{18}O}$ line ratio in ultra\-luminous infrared galaxies to an excess in massive star formation. Furthermore, an increase of $^{12}$C/$^{13}$C and $^{16}$O/$^{18}$O with galactic radius is observed \citep{Langer1990, Milam2005}. Such a trend arises if we assume an inside-out formation scenario for galaxies \citep{Tang2019}. If M51 exhibits similar carbon and oxygen isotope trends, also the observed $\chem{^{13}CO}/\chem{^{12}CO}$ and $\chem{C^{18}O}/\chem{^{12}CO}$ variation can be explained.

Finally, variations in the optical depth or changes in physical conditions (e.g., gas temperature or density) of the CO isotopologues will also cause variations in the observed line ratios. For example, in local thermodynamic equilibrium (LTE), the \chem{^{12}CO} optical depth depends on the gas column density, $N_\mathrm{c}$, the gas velocity dispersion, $\sigma$, and the gas kinetic temperature, $T_\mathrm{k}$, via $\tau \propto N_\mathrm{c}/(\sigma T_\mathrm{k})$ \citep{Paglione2001}. Increased turbulence in the centre, for example, could decrease the optical depth of \chem{^{12}CO}{10}. The trends we observe in the CO isotopologue line ratios are all consistent with changes in the optical depth.  

Given the trends we find in the $\chem{^{13}CO}/\chem{C^{18}O}$ and $\chem{^{13}CO}/\chem{^{12}CO}$ line ratios, we reach a similar conclusion to \cite{Cormier2018}: Abundance variations due to nucleosynthesis (assuming similar isotope trends as in the Milky Way) and/or changes in the physical conditions (temperature, density, opacity) of the gas can explain the global observed CO isotopologue line ratio variations. We recognise that most likely we are seeing a combination of effect that depends further on the SF history and chemical enrichment.

\begin{figure*}
    \centering
    \includegraphics[width = \textwidth]{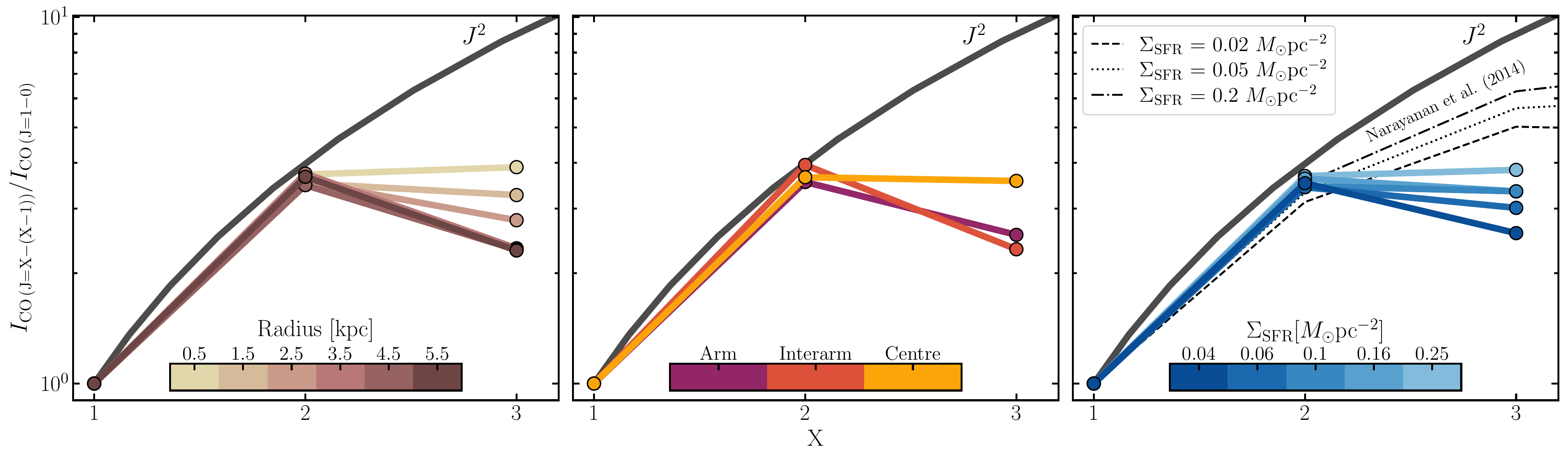}
    \caption{{\bf CO spectral line energy distribution (CO SLED) variation in M51.} (\textit{Left}) CO SLED binned by 1\,kpc radial bins ranging up to 6\,kpc. With the exception of the most central 1\,kpc bin, we see that the CO SLED peak is already reached for $J=2\!\rightarrow\!1$. (\textit{Middle}) Variation with galactic environment. The spiral arm and interarm regions show a similar shape of the CO SLED. (\textit{Right}) Variation as a function of SFR surface density. We include predictions for $\Sigma_{\rm SFR} = 0.2, 0.05 \text{ and } 0.02 \, {\rm M_\odot\,{\rm pc^{-2}}}$ based on the model code provided in \citet{Narayanan2014}. We use their unresolved model, which takes into account different beam filling factors for the three CO transitions. 
    For this analysis, we use the data at $27$\,arcsec working resolution.
    }
    \label{fig:CO_sled}
\end{figure*}

\subsection{CO Spectral Line Energy Distribution}
\label{sec:disc_COSLED}

In the previous sections, we have discussed variations we find in the CO line ratios. Here, we want to further analyse in particular \chem{^{12}CO} excitation, which is relevant, for example, to accurately estimate molecular gas masses at high redshift where CO low-\textit{J} transitions are difficult to obtain. Converting CO luminosities to H$_2$ gas masses generally relies on observations of the \chem{^{12}CO}{10} transition, or via down-conversion of observed higher-\textit{J} observations. In particular high redshift studies rely on such down-conversions \citep{Solomon2005, Carilli2013}. A~proper down-conversion requires a good understanding of the CO spectral line energy distribution (CO SLED).
Using the three rotational transitions of \chem{^{12}CO} presented in this study, $J =1\!\rightarrow\!0$, $2\!\rightarrow\!1$ and $3\!\rightarrow\!2$, we can investigate the CO SLED. As described by \cite{Narayanan2014}, the CO excitation and, thus, the precise shape of the SLED, is expected to depend on gas temperatures, densities and optical depth within the ISM. Consequently, we do not expect that the CO SLED is constant across the galaxy. In their study, \cite{Narayanan2014} provided a parametrisation of the CO SLED as a function of SFR surface density. 
In \autoref{fig:CO_sled}, we investigate variations of the CO SLED with galactocentric radius, galactic morphology and SFR surface density. We see that, with the exception of sight lines within the central 1\,kpc radius (which are also impacted by the central AGN), the CO SLED has reached its turning point already at $J=2\!\rightarrow\!1$. The shapes of the CO SLEDs for arm and interarm regions are very similar. But we see that variations of the CO SLED as a function of $\Sigma_{\rm SFR}$ do not follow the model presented in \cite{Narayanan2014}, as in particular the \chem{^{12}CO}{32} line seems to be fainter than the predictions of the models (see right panel in \autoref{fig:CO_sled}). In their study, they compared their model mainly to high-$z$ sub\-milli\-metre galaxies (SMGs). As SMGs exhibit more extreme star formation, the ISM conditions are very likely different in terms of gas temperature and density from the conditions present in M51, which could explain the discrepancy of the CO SLED in M51 and predicted by the models. 

We note that there are also other explanations for a depression of the CO SLED relative to the models. If the beam filling factor differs significantly between the three \chem{^{12}CO} transitions, with the beam filling factor for \chem{^{12}CO}{32} being the smallest, beam dilution could drive down the CO SLED. The model presented in \cite{Narayanan2014} takes into account different beam filling factors for the different CO transitions. In the case of the three \chem{^{12}CO} transitions we analyse, the effect is of order 10~per~cent decrease if we apply the model which account for beam filling factors.  Only for higher-\textit{J} transitions, the effect will increase as their emission is coming from more compact regions due to the larger excitation. For confirmation that the assumed differences in beam filling factors of the model are appropriate in the case of M51, we would need higher resolution observations for all the three low-\textit{J} transitions.
\cite{Narayanan2014} described other effects that may lead to discrepancy between model and observation. In particular, dust extinction, which affects the very high-\textit{J} transitions more, can explain the depression of the CO SLED relative to the prediction from the model. However, as our CO lines are observed in the millimetre regime, dust effects will be minimal in a normal star-forming disc galaxy like M51. Consequently, we do not believe that the divergence of the observations from the model can be attributed to the effects of dust.

\begin{figure}
    \centering
    \includegraphics[width =0.95 \columnwidth]{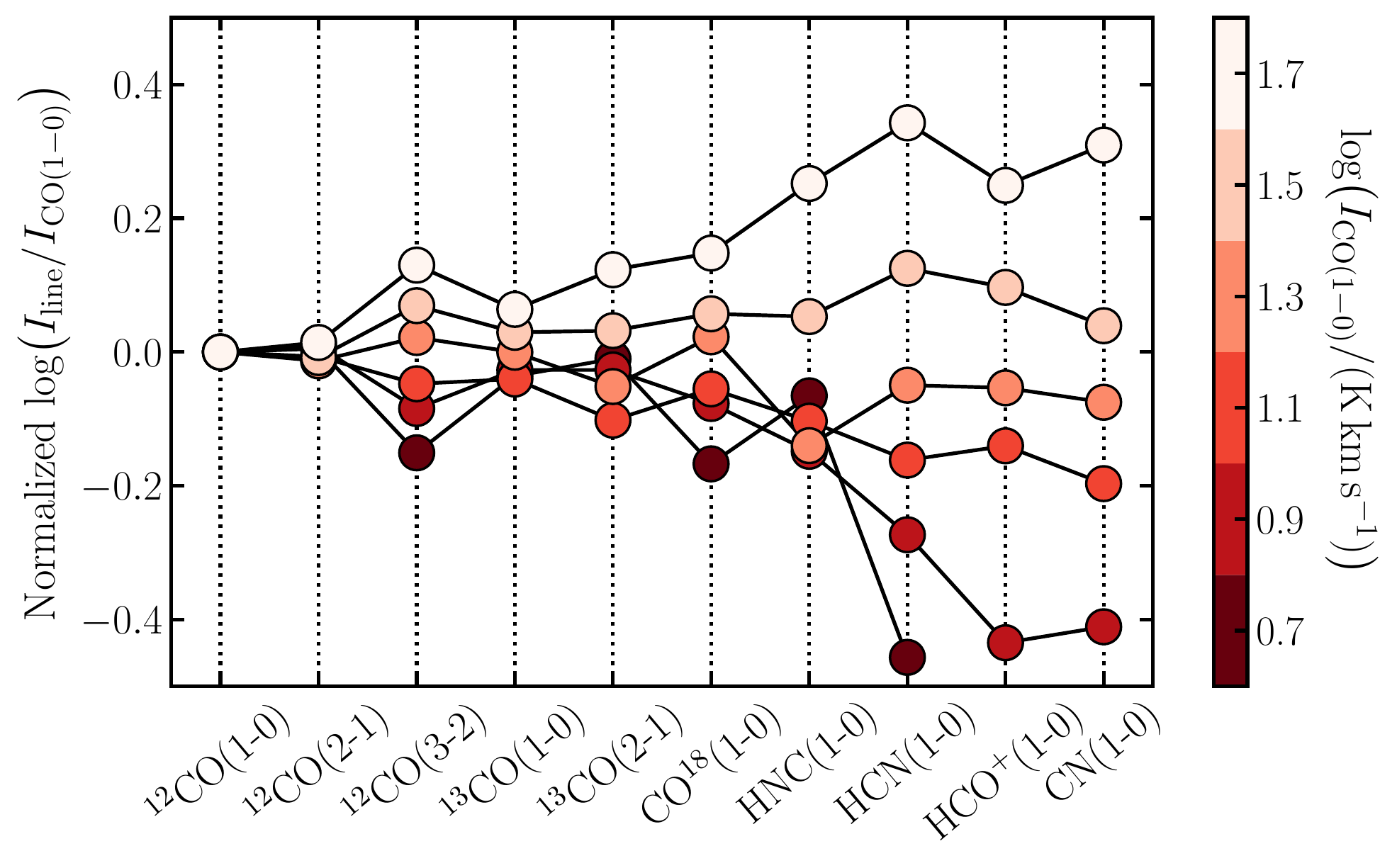}
    \caption{{\bf Patterns of molecular line ratios for M51.} Each molecular line is stacked by \chem{CO}{10} and the line ratio is normalised by the mean stacked line ratio with \chem{CO}{10}. We colour coded by the \chem{CO}{10} line brightness of the stacked bin. The line ratios is roughly ordered from left to right by increasing critical density of the line entering in the numerator. We find an increase of the variation of the line ratios towards the right end of the plot.  Such a trend is in agreement with the results from the models presented in \citet{Leroy2017_density}.}
    \label{fig:pattern_plot}
\end{figure}

\subsection{Molecular Lines and Systematic Density Variation}
\label{sec:patterplot}
In our discussion so far, we have only focused on the CO isotopologue lines. However, also molecular line ratios including denser gas tracers (such as \chem{HCN}, \chem{HNC} or \chem{HCO^+}) can be used to study the underlying molecular gas density. 
The picture is complex: Our working resolution of ${>}15\,\mathrm{arcsec}/600$\,pc is substantially larger than the dense star-forming cores, which have sizes of $0.1{-}1$\,pc \citep[e.g.][]{Lada2003, Andre2014}. Consequently, within our beam, a large range of gas volume densities are included. Therefore, the sub-beam density distribution affects the overall emission averaged over the full beam size, because certain lines emit more efficiently for a particular density distribution. Generally speaking, emission lines originate not just at or above the critical density, but the wide density distribution within our coarse observations has to be taken into account.

\cite{Leroy2017_density} showed by applying non-LTE radiative transfer models on a range of underlying density distributions that the line ratios of high-to-low critical density lines (i.e. emission lines with higher critical density\footnote{For our qualitative discussion in this section, we do not go into detail regarding the various definitions of critical densities. See \citet{Leroy2017_density} or \citet{Shirley2015} for a more in-depth discussion of the various definitions and their advantage and disadvantages in describing the conditions of efficient line emission.}) are more sensitive to changes in gas density. For such lines, a substantial fraction of the emission can originate from regions that have a density below the nominal critical density. They based their predictions on basic radiative transfer models and a parametrised density probability distribution to characterise the effect of sub-beam density variations on the observed beam-averaged emission. The crucial point is that a line can still be emitted at densities much below the critical density, just with a reduced emissivity. Consequently, a slight increase of the gas density can significantly increase the emissivity of the emission line. This is not the case for lines with low critical density. The gas density generally exceeds the critical density already for such lines, so a variation in the gas density will not significantly impact the emissivity of the emission line.

The lines observed as part of our sample as well as the ancillary lines can be used to test whether lines with higher critical density show a larger scatter in the line ratio with respect to \chem{CO}{10}. We stack the data by \chem{CO}{10} brightness temperature and compute the line ratio with respect to \chem{CO}{10}. In \autoref{fig:pattern_plot}, we normalise the line ratios by the median value for each line ratio. Lines with a higher critical density are shown towards the right end of the plot. We colour-code the points by binned \chem{CO}{10} brightness temperature, which we use as a shorthand for gas density. There is a distinct increase of the variation (a~`flaring') of the line ratio pattern towards higher density tracers, in agreement with the predictions from \cite{Leroy2017_density}, as these lines are more sensitive to variations in the sub-beam gas density distribution. This is already apparent from looking at the \chem{^{12}CO} transitions. The critical density of \chem{^{12}CO}{32} is about an order of magnitude larger than for the \trans{21} and \trans{10} transitions (in the optically thin case; \citealt{Carilli2013}) and we clearly find a larger variation in the ratio with \chem{^{12}CO}{32}/\trans{10} than in the \chem{^{12}CO}{21}/\trans{10} line ratio.

With this analysis, we thus find a larger dynamical range of line ratios of lines with a large difference in critical density. This supports the idea that the flarring pattern seen in \autoref{fig:pattern_plot} is in agreement to higher sensitivity of these line ratios with respect to the mean gas density. Line ratio patterns of such a diverse suite of lines are thus a powerful tool to constrain the molecular gas physical conditions.


        
\section{Conclusion}

In this paper, we present observations on several CO isotopologues in the galaxy M51 obtained with the IRAM \mbox{30-m} telescope. Besides $J=1\!\rightarrow\!0$ and $J=2\!\rightarrow\!1$ transitions of \chem{^{13}CO} and \chem{C^{18}O}, we also detect \chem{C^{17}O}{10} emission as well as supplementary lines, such as \chem{CN}{10}, \chem{CS}{21}, \chem{N_2H^+}{10} and \chem{CH_3OH}{21}. 
\begin{enumerate}
\item We study the CO isotopologue line ratios  as a function of galactocentric radius, \chem{^{12}CO}{21} intensities (which translates to molecular mass surface density) and total infrared surface brightness (which is correlated with the star formation rate surface density). Several line ratios, such as $\chem{^{12}CO}{21}/\trans{10}$, $\chem{^{12}CO}{32}/\trans{21}$ or $\chem{^{13}CO}{21}/\trans{10}$, show a significant increase of order $5$ to $40$~per~cent towards the centre of the galaxy, compared to their disc-averaged line ratios.

\item Galactic morphology, such as spiral arm and interarm regions, seems to affect several line ratios. Besides increased line ratios in the centre, $\chem{^{12}CO}{21}/\trans{10}$ and $\chem{^{13}CO}/\chem{^{12}CO}\trans{10}$ show indications of larger values in interarm regions than in spiral arm regions. Previous studies attributed an increase of the $\chem{^{12}CO}{21}/\trans{10}$ line ratio to the more efficient dust heating by bright, young (${<}100$\,Myr) stars at the convex, downstream end of the spiral arm. We do, however, not see any trend of the line ratio with neither UV nor IR radiation.

\item We investigate the potential cause for the observed variations in the $\chem{^{13}CO}/\chem{^{12}CO}$, $\chem{C^{18}O}/\chem{^{12}CO}$ and $\chem{C^{18}O}/\chem{^{13}CO}$ line ratios. A change in optical depth most likely explains the trend seen with galactocentric radius and total infrared surface density, which argues against abundance variations of the CO isotopologues as the cause.

\item The shape of the CO spectral energy distribution, varies with galactic environment. We find a relation between the shape of the CO SLED and the star formation rate surface density, but the turning point of the CO SLED shape is at lower-\textit{J} than predicted by recent models, which are mostly calibrated on high-$z$ sub\-milli\-metre galaxies.

\end{enumerate}

As a potential  future study, the high-quality observations at low spatial resolutions can be combined with higher spatially resolved observations to study the diffuse CO component also for the CO isotopologues and furthermore investigate line ratio variation at different spatial scales.
In addition, performing non-LTE analyses using modelling tools, it can be possible to gain further constraints on the CO--to--H$_2$ conversion factor or -- in combination with dense gas data from EMPIRE (HCN, HCO$^+$ and HNC \mbox{(1--0)}) -- the probability density function. This will improve constraints on the average volume density, which constituted on of the major science goals of the EMPIRE survey.

\begin{acknowledgements}
      JdB, FB, JP, ATB and IB acknowledge funding from the European Research Council (ERC) under the European Union’s Horizon 2020 research and innovation programme (grant agreement No.726384/Empire). TS, ES, HAP and TGW acknowledge funding from the European Research Council (ERC) under the European Union’s Horizon 2020 research and innovation programme (grant agreement No. 694343). HAP further acknowledges the Ministry of Science and Technology (MOST) of Taiwan under grant 110-2112-M-032-020-MY3. The work of AKL is partially supported by the National Science Foundation under Grants No. 1615105, 1615109 and 1653300. 
      AU acknowledges support from the Spanish grants PGC2018-094671-B-I00, funded by MCIN/AEI/10.13039/501100011033 and by ``ERDF A way of making Europe'', and PID2019-108765GB-I00, funded by MCIN/AEI/10.13039/501100011033. 
      ER acknowledges the support of the Natural Sciences and Engineering Research Council of Canada (NSERC), funding reference number RGPIN-2017-03987. CE acknowledges funding from the Deutsche Forschungsgemeinschaft (DFG) Sachbeihilfe, grant number BI1546/3-1. SCOG and RSK acknowledge support from the DFG via SFB 881 “The Milky Way System” (sub-projects B1, B2 and B8) and from the Heidelberg cluster of excellence EXC 2181-390900948 “STRUCTURES: A unifying approach to emergent phenomena in the physical world, mathematics, and complex data”, funded by the German Excellence Strategy. RSK furthermore thanks for financial support from the European Research Council via the  Synergy Grant ``ECOGAL'' (grant 855130). JMDK gratefully acknowledges funding from the Deutsche Forschungsgemeinschaft (DFG, German Research Foundation) through an Emmy Noether Research Group (grant number KR4801/1-1) and the DFG Sachbeihilfe (grant number KR4801/2-1), as well as from the European Research Council (ERC) under the European Union's Horizon 2020 research and innovation programme via the ERC Starting Grant MUSTANG (grant agreement number 714907).
      
     {For our research, we made use of Astropy and affiliated packages. Astropy is a community-developed core Python package for Astronomy \citep{Astropy2018}. Furthermore, we employed the Python package NumPy \citep{harris2020array}, SciPy \citep{2020SciPy-NMeth}, and APLpy, an open-source plotting package for Python \citep{Robitaille2012}. }
\end{acknowledgements}

%
%

\bibliographystyle{aa}
\bibliography{references.bib}

\begin{appendix}
\section{IRAM \texorpdfstring{\mbox{30-\lowercase{m}}}{30-m} Error Beam Contribution}
\label{sec:errcontr}
In this appendix, we analyse the impact of the IRAM \mbox{30-m} error beam on the detection of extended emission in our data set. The response of the telescope to a point source is not a single perfect Gaussian, but has an additional contribution from the so-called error beams. \citet{Greve2009} characterised these error beams as a series of 2D Gaussians broader than the main beam but with a lesser contribution to the telescope power. The telescope beam patter was characterised again by  \citet{Kramer2013} after the last major upgrade. The most recent characterization implies that a point source of $1$\,Jy will only provide about $0.8$ and $0.6$\,Jy in the telescope main beam at 3 and 1\,mm, respectively, the flux remainder being scattered in the error beams. The image of a point source by the telescope will appear fainter at its actual position and the point source will contribute a faint extended brightness halo around it. This can be potentially critical when observing fainter positions inside a galaxy (e.g. interarm positions), as emission from brighter central parts of the galaxy will boost the detected line brightness. For similar observations with the  IRAM \mbox{30-m} telescope, \citet{Pety2013} first modelled the contributions of bright M51 sources on the interarm signal (see their appendix~C), and \citet{Leroy2015} proposed a first iterative deconvolution solution. We note that other deconvolution schemes had been proposed in the past \citep[e.g.,][]{Westerhout1973,Bensch1997, Lundgren2004}. We will describe here the method we use to extract the contribution of the error beam to the emission and investigate its extent. This method will be precised in P.~Tarrio et al.\ (in prep.).
Notice that we perform the succeeding error beam estimation after attempt to correct for the main beam efficiency, but that this correction had assumed a signal free error beam, which may not be correct. So here we will account for what happens when there is emission from the galaxy in the (assumed empty) error beam.

\subsection{Model of the Error Beam}

The exact pattern and shape of the error beam is hard to measure. It evolves as a function of the telescope's elevation because of gravitational deformation of the primary dish. It depends also on the evolution of the thermal environment, in particular at sunrise and sunset. We will rely on the beam pattern characterization by \citet{Kramer2013}. This characterization comes from On-The-Fly measurements of the Moon edge at the IRAM \mbox{30-m} optimal elevation of ${\sim}50\deg$. \autoref{tab:err_param} lists the details of the error beam parameters used here.

In essence, we are interested in the underlying, ideal, error beam corrected main brightness temperature $\hat T_{\rm mb}$.  The main brightness temperature is not to be confused with the intrinsic brightness temperature in the sky, $T$. They are related via:
\begin{equation}
    \hat T_{\rm mb} = G_{0}\otimes T,
\end{equation}
where we indicate the main beam, which has a shape of a 2D circular Gaussian, by $G_0$. Similarly, following \cite{Kramer2013}, we assume that the error beam consists also of a set of wide 2D Gaussian beams which, indicate with $G_i$, where $i=1,2,3$ (see \autoref{tab:err_param}).

With the telescope, we only have access to the measured brightness in $T_{\rm A}^\star$ unit, which we initially converted to $T_{\rm mb}$ (the brightness temperature we use in the main text) under the simplifying assumption of an empty error beam:
\begin{equation}
    \frac{F_{\rm eff}}{B_{\rm eff}}\times T_{\rm A}^\star =  T_{\rm mb},
\end{equation}
with the forward efficiency $F_{\rm eff}$ and the beam efficiency $B_{\rm eff}$.

\begin{figure}
    \centering
    \includegraphics[width = \columnwidth]{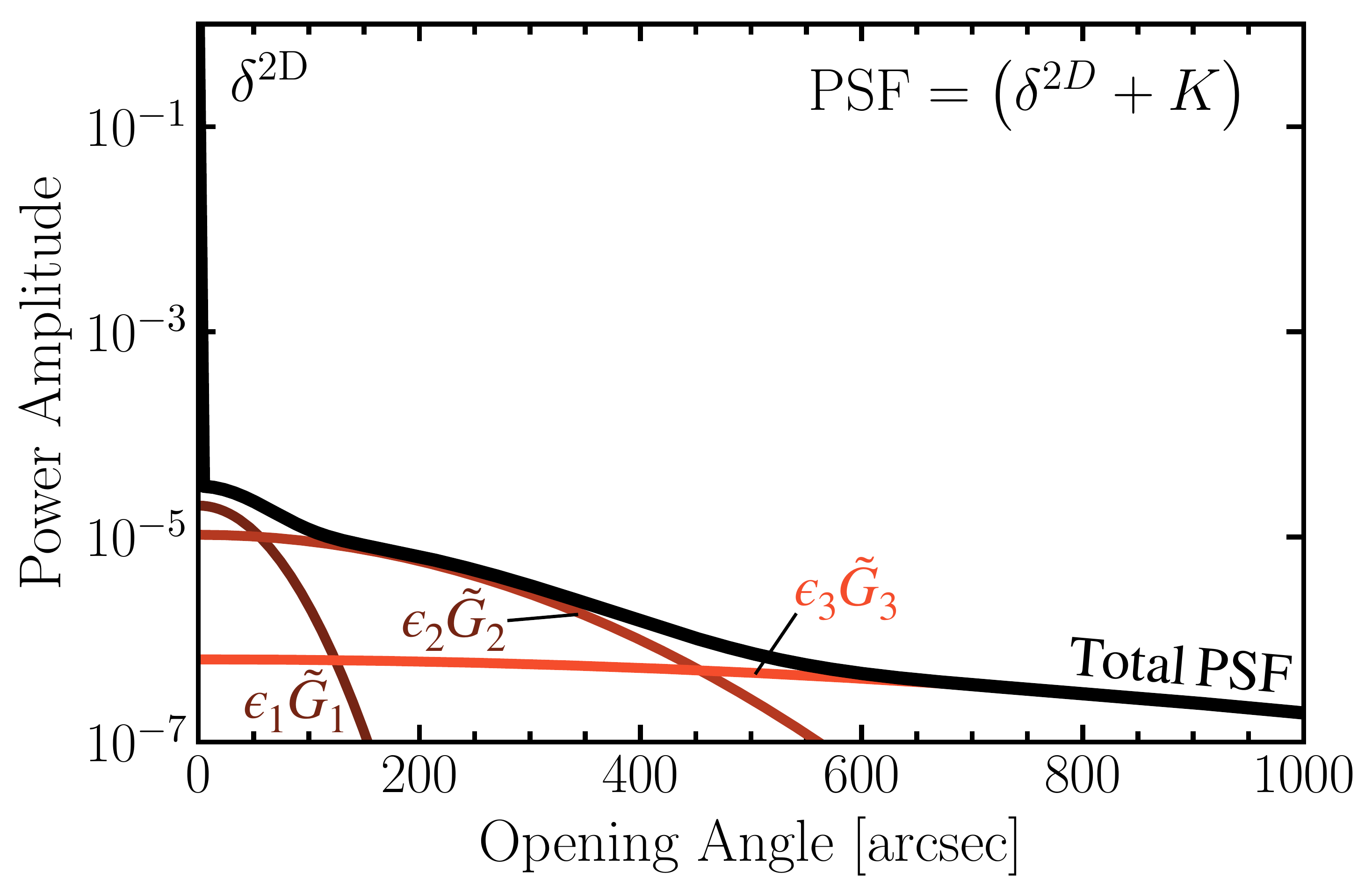}
    \caption{IRAM \mbox{30-m} full kernel for the error beam contribution analysis, { after deconvolution of the main beam,  $G_{0}$ for 3\,mm}. The kernel includes the three components of the error beam { ($\tilde G_i$ indicates the Gaussian 2D profile after deconvolution with $G_0$). The ratio of the relative beam power ($P_i$) with respect to the relative main beam power ($P_0$) is indicate by $\epsilon_i = P_i/P_0$. }}
    \label{fig:psf}
\end{figure}

The error beam corrected brightness temperature $\hat T_{\rm mb}$ is related to $T_{\rm mb}$ via the convolution kernel $K$ as follows (see \autoref{fig:psf}):
\begin{equation}
\label{eq:err_beam}
    T_{\rm mb} = (\delta^{2\rm D} + K) \otimes \hat T_{\rm mb}~,
\end{equation}
where $\delta^{2\rm D}$ is the Dirac 2D distribution and the kernel~$K$ is the sum of the error beams components, after deconvolution by the main beam:
\begin{equation}
    K = \sum_{i=1}^{3}\epsilon_i \tilde G_i~,
\end{equation}
where $\epsilon_i = P_i/P_0$ with $P_i$ being the relative beam power and $\tilde G_i$ the deconvolution with the main beam ($G_i = \tilde G_i\otimes G_0$, $\tilde \theta_i^2 = \theta_i^2-\theta_0^2$). 
Because the error beam consists of very wide 
2D Gaussians, we can deconvolve the narrower main beam and get a well behaved function.
Finally, to estimate the error beam contribution we have to perform a further deconvolution on \autoref{eq:err_beam}, to determine $\hat T_{\rm mb}$.

\begin{table*}
    \centering
    \caption{Error Beam parameters based on a cubic interpolation from table~1 in \citet{Kramer2013}.}
    \label{tab:err_param}
    
    \begin{tabular}{r c c c c }
        \hline \hline
         &  Main Beam & 1. Error Beam & 2. Error Beam & 3. Error Beam\\ \hline
         \textbf{115.3 GHz}&\\ 
         Beam Width $\theta$& 21\arcsec & 113\arcsec & 434\arcsec & 1518\arcsec \\
         Integrated relative power $P$ [\%]&84 & 1 & 9 & 6 \\\hline
         \textbf{230.5 GHz}&\\ 
         Beam Width $\theta$&10\arcsec & 56\arcsec & 217\arcsec & 759\arcsec \\
         Integrated relative power $P$ [\%]&69 & 5 & 13 & 13 \\\hline
    \end{tabular}
\end{table*}

\begin{figure}
    \centering
    \includegraphics[width = \columnwidth]{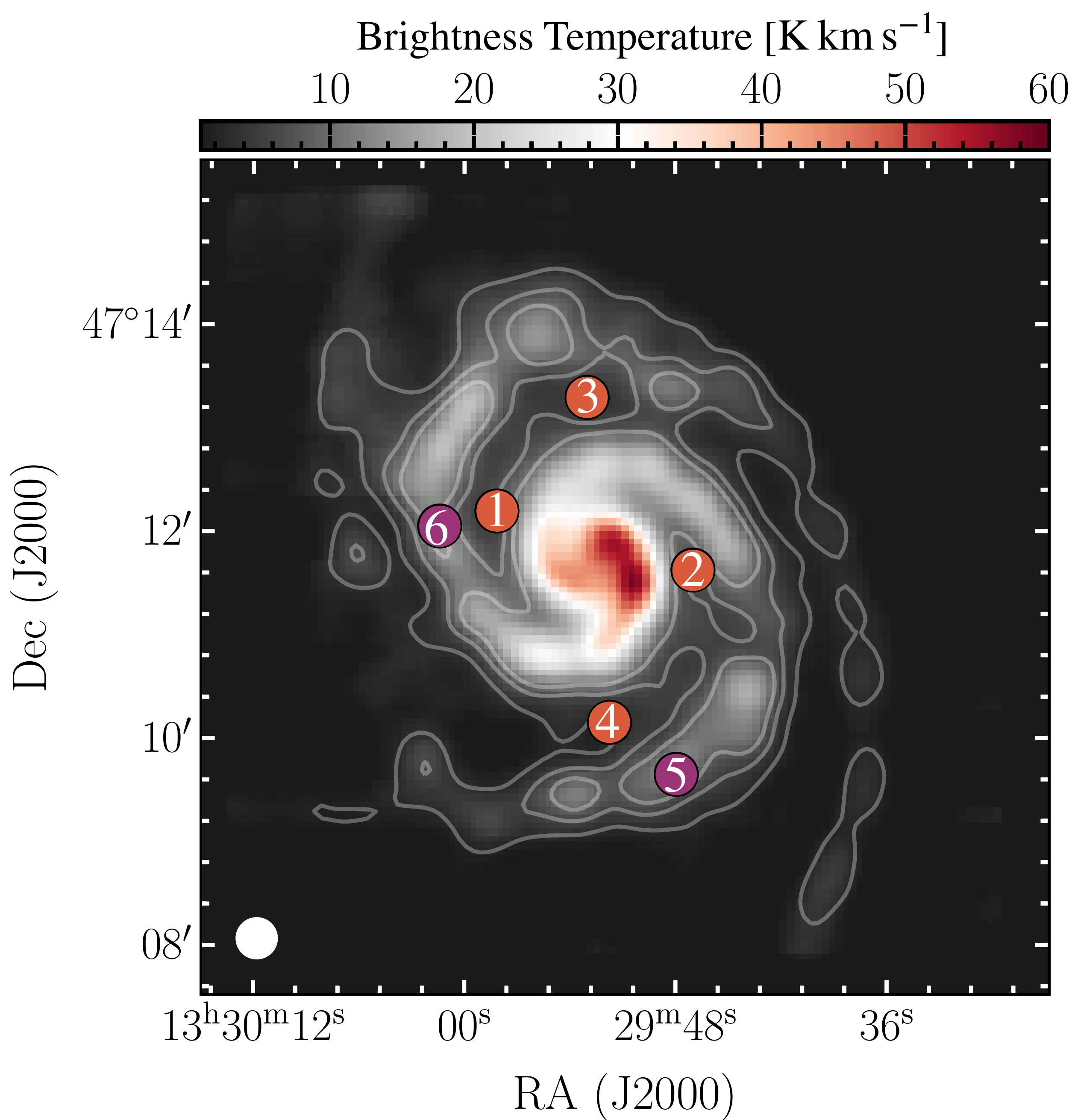}
    \caption{{\bf \chem{CO}{21} line brightness temperature map.} Indicated are the six pointings for which we performed the error beam correction analysis in \autoref{fig:err_contr}. Pointings 1--4 are located in interarm regions (marked in orange). Pointings~5 and~6 are located in spiral arm regions (marked in purple). Contours drawn at $\mathrm{S/N} = 10, 20, 30$. For the IRAM \mbox{30-m} DDT project E02-20, we observed these six pointings (see \hyperref[sec:DDT]{Appendix~\ref{sec:DDT}}).}
    \label{fig:pointing}
\end{figure}

\begin{figure*}
    \centering
    \includegraphics[width = \textwidth]{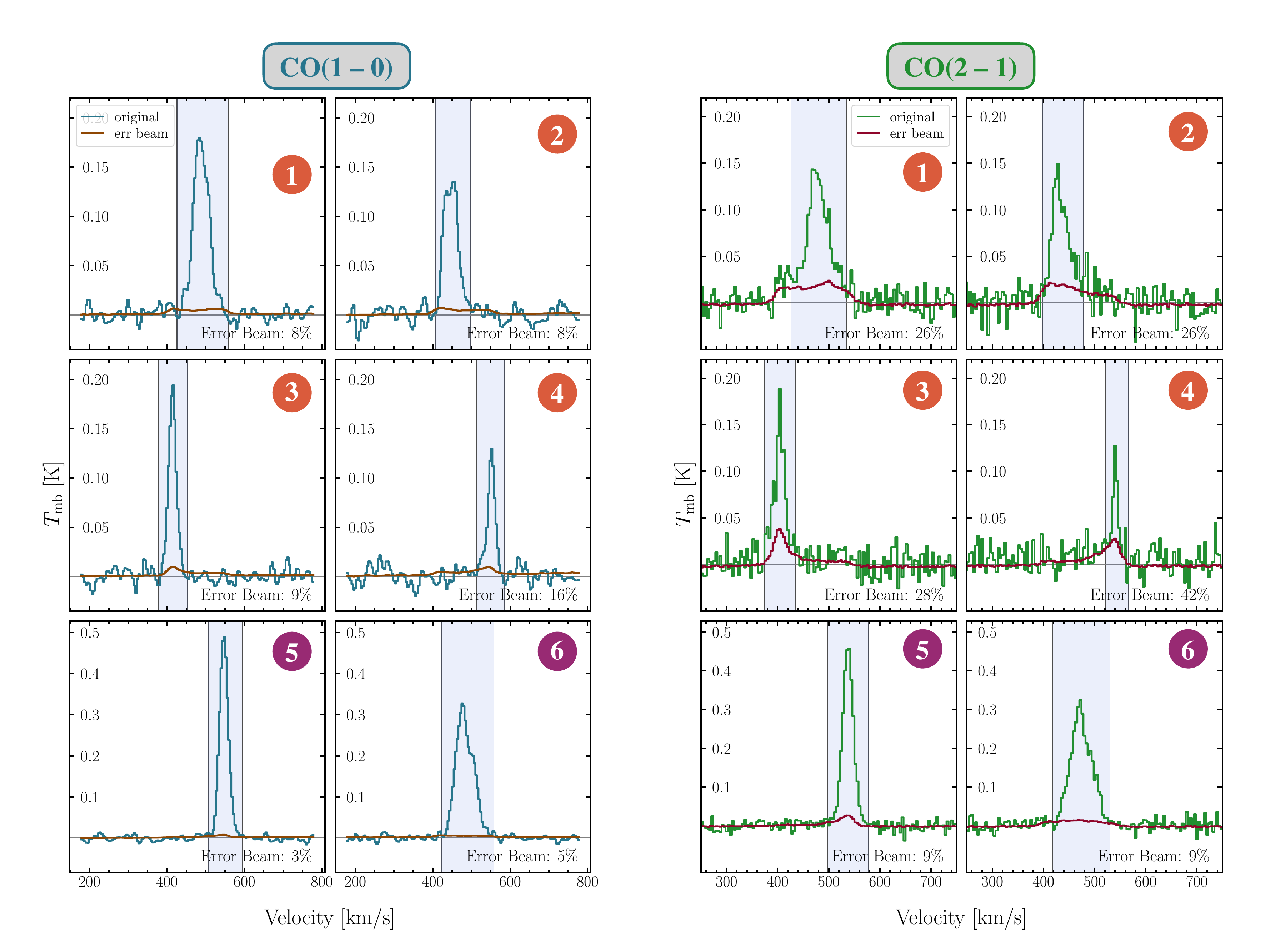}
    \caption{{\bf Error beam contribution analysis for individual pointings.} We performed the deconvolution using the exact approach via Fourier deconvolution. After deconvolution, we performed a baseline correction. In the panels, we show the observed line brightness ($T_{\rm mb}$) as well as the contribution from the error beam $T_{\rm mb}-\hat T_{\rm mb}$. For each panel, we indicate the contribution from the error beam to the emission with $\mathrm{S/N} > 3$ (see masked region). The position of the pointings is indicated in \autoref{fig:pointing}.}
    \label{fig:err_contr}
\end{figure*}

\subsection{Deconvolution}
\label{sec:deconv}
There are two ways in which we can approximate the error beam free source brightness temperature $\hat T_\mathrm{mb}$. 
\begin{enumerate}
    \item Iterative solution in the image plane: this approach, first described in \cite{Leroy2015}, elaborates on the statement by \citet{Pety2013} that the bright intensity part of a galaxy can be approximated by the measured brightness in $T_{\rm mb}$ unit. It is possible to determine the error beam contribution by convolving the measured brightness in $T_{\rm mb}$ unit with the error beam part of the PSF of the telescope. This gives another estimation of the source brightness that can be then used iteratively to improve the solution. In particular, we define the $N^{\rm th}$ approximate solution via the following recursion:
    \begin{equation}
    \hat T_{{\rm mb}|N} = T_{\rm mb} - K \otimes \hat T_{{\rm mb}|N-1},
    \end{equation}
    with $ \hat T_{{\rm mb}|0} \equiv  T_{\rm mb}$.
    This iterative process is stopped when the difference between two estimations becomes smaller than a give criterion.
    \item Fourier plane solution: Performing a 2D Fourier transform, it follows from \autoref{eq:err_beam}
    \begin{equation}
    \hat T_\mathrm{mb} = \mathrm{FT}^{-1}\left(\frac{\mathrm{FT}( T_\mathrm{mb})}{1+\mathrm{FT}(K)}\right)~.
    \end{equation}
    We implemented the approach in \texttt{Python} using the unsupervised Wiener-Hunt deconvolution (based on the Wiener-Hunt approach and estimating the hyperparameters automatically).
\end{enumerate}


\begin{figure*}
    \centering
    \includegraphics[width = 0.85\textwidth]{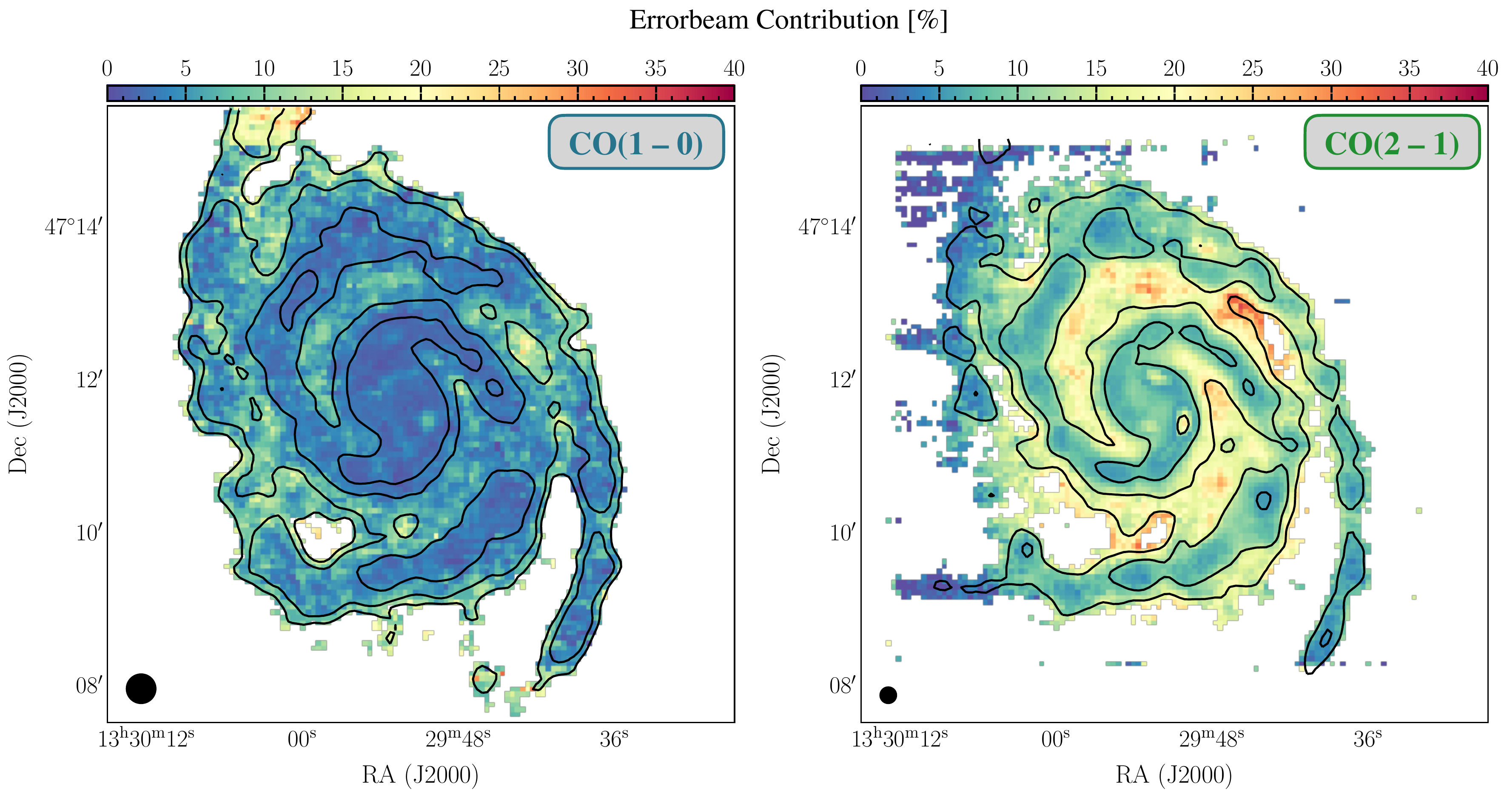}
    \caption{{\bf 2D distribution of the error beam contribution.} (\textit{Left}) \chem{^{12}CO}{10} map. Contours drawn at $\mathrm{S/N} = 10, 20, 30$ of \chem{CO}{10} emission. (\textit{Right}) \chem{^{12}CO}{21} map. We see that the error beam contribution is more pronounced for \chem{CO}{21}, in particular in the interarm regions of the  galaxy. }.
    \label{fig:err_contr_map}
\end{figure*}

\subsection{Result and Implication}

We perform the error beam contribution analysis for both the PAWS \chem{^{12}CO}{10} and CLAWS \chem{^{12}CO}{21} observations. To see the effect on the spectrum, we investigate six pointings with a $23$\,arcsec aperture (see \autoref{fig:pointing}). Four pointings (1--4) are located in the fainter interarm region and two pointings (5~and~6) are situated in the brighter spiral arm region. The result of the deconvolution can be seen in \autoref{fig:err_contr} for the six pointings. The blue and green spectrum shows $ T_\mathrm{mb}$ (the brightness temperature we derive from $T_{\rm A}^\star$ assuming no error beam contributions) for \chem{^{12}CO}{10} and \chem{^{12}CO}{21} respectively. In red, we indicate the contribution to the spectrum from the error beam (i.e. $T_\mathrm{mb}-\hat T_\mathrm{mb}$). We indicated the percentage contribution to the ${T}_{\rm mb}$ integrated intensity for each pointing. This contribution is calculated only for the spectral range where $\mathrm{S/N} > 3$. We performed both methods described in the \hyperref[sec:deconv]{Appendix~\ref{sec:deconv}}. Both methods yield a similar percentage contribution ( $<1\%$ point difference). We continue using the exact approach via Fourier deconvolution, since it is easier to implement.
We find that for the \chem{^{12}CO}{10} line, the impact is minor, with the contribution ranging from ${\sim}15$~per~cent in the interarm and to only 4~per~cent in the spiral arm region. For the \chem{^{12}CO}{21} emission, because the main beam efficiency is smaller (B$_{\rm eff}^{230\rm\,GHz} \approx 60$~per~cent), the contribution is more significant.  In the interarm, the contribution is up to 40~per~cent. This is mainly due to emission from the brighter regions in the galaxy, such as the central region, entering the observation via the different error beam components. \hyperref[fig:err_contr_map]{Figure~\ref*{fig:err_contr_map}} shows the full 2D map. For every pixel, we computed the error beam contribution along its spectral axis. We again see a larger effect for the \chem{^{12}CO}{21} emission, in particular for the interarm regions (up to $30{-}35$~per~cent).

Because the 1 mm lines are affected more than the 3 mm lines by the error beam contribution, we expect the corrected $R_{21}^{\chem{^{12}CO}}$ to be lower. In \autoref{tab:res_e0220_ratio}, we indicate the line ratio before and after correcting for the error beam. Except for pointing 4, we still find larger $R_{21}^{\chem{^{12}CO}}$ values in the interarm than in the spiral arm. Consequently, even though in certain instances the error beam contribution is far from negligible, it alone cannot explain the arm/interarm trend.

We reiterate that the preceding error beam analysis is subject to many uncertainties: The exact shape of the error beam is difficult to measure and subject to temporal and positional (e.g., the elevation of the telescope) variation. Furthermore, the approach we described will generally in fact overestimate the effect of the error beam in the case of single dish maps of a galaxy. 
Since the error beam will be comparable to the size of the galaxy, the individual spectra will include a component consisting of a strongly convolved spectrum of the full galaxy. The baseline fitting procedure we performed will then subtract such low and broad emission in resolved observations of galaxies. 
So the estimated value for the error beam contribution for the different positions should be interpreted with caution. However, we believe that our measurement constitute a reasonable upper limit for the order of magnitude of the error beam contribution.
We refrain from suggesting a particular constant percentage uncertainty value for general observations since the error beam contribution is not constant across the galaxy and depends strongly on the galaxy morphology. But generally, one should be aware that an additional uncertainty of order 20\%-40\% can be possible.

\section{Flux Calibration Uncertainties}
\label{sec:CLAWS-HERA}

The flux measurements from various telescopes are subject to various degrees of calibrational uncertainties. \cite{denbrok2021} discuss in detail the impact such calibrational uncertainties can have by comparing ALMA, IRAM \mbox{30-m} and NRO data. \cite{Donaire2019} find a flux calibration uncertainty rms of order 7~per~cent for the EMIR observations from line calibrator monitoring. Finally, based on jack-knifing several HERA data sets, \cite{Leroy2009} estimate that their HERACLES \chem{CO}{21} observations are subject to up to a typical 20~per~cent uncertainty in rms. As the data for CLAWS were observed using the EMIR instrument, which has more stable calibration than HERA, we assume our data to have an uncertainty under 10~per~cent, as reported by IRAM.\footnote{\url{https://publicwiki.iram.es/EmirforAstronomers\#Telescope_efficiencies}} 


In this section, we describe the results from the IRAM \mbox{30-m} DDT proposal, in which we observed the six pointings shown in \autoref{fig:pointing} to address the flux stability in the arm and interarm to understand its impact on the arm--interarm CO line ratio.

\subsection{DDT Proposal E02-20}
\label{sec:DDT}

As we have seen in the previous section, comparing data sets from different telescopes/instruments taken at different times needs extra care as uncertainties in the flux calibration can affect absolute values of line emission and ratios. \cite{denbrok2021} determined that the arm--interarm CO line ratio discrepancy contrast (and the line ratio itself) is sensitive to combining data sets from different telescopes and instruments. For example, \cite{Koda2012} find a different line ratio in the interarm region of M51 using NRO \chem{^{12}CO}{10} compared to \cite{denbrok2021} and this study. Here we address the question of whether the stability of the flux calibration could explain this discrepancy.

To address this question we obtained 6\,h DDT IRAM \mbox{30-m} time to observe six carefully selected pointings (see \autoref{fig:pointing}), four in the interarm and two in the spiral arm region. As we observe \chem{CO}{10} and \COIItoI\ simultaneously, any time dependence is removed when investigating the line ratio. 

Observations were carried out on 2021 February~27 and on the night of 2021 March~8. We cannot simply take the line ratio, as the \COIItoI\ beam is smaller than the \chem{^{12}CO}{10} beam.  
To estimate how to scale the high resolution \COIItoI\ spectrum when convolving it to the resolution of \chem{^{12}CO}{10}, we first extract a spectrum using a $11.5$\,arcsec (i.e. beam-sized) aperture from the CLAWS \COIItoI\ map. We then convolve the CLAWS \COIItoI\ to the lower angular resolution of $23$\,arcsec (i.e. PAWS resolution). We now extract a \COIItoI\ from the same position in the convolved CLAWS map, but use a $23$\,arcsec aperture. By comparing the two extracted \COIItoI\ spectra, we can determine a scaling factor, which we can apply to the \COIItoI\ spectrum obtained from the DDT program.

\begin{figure}
    \centering
    \includegraphics[width = 0.95\columnwidth]{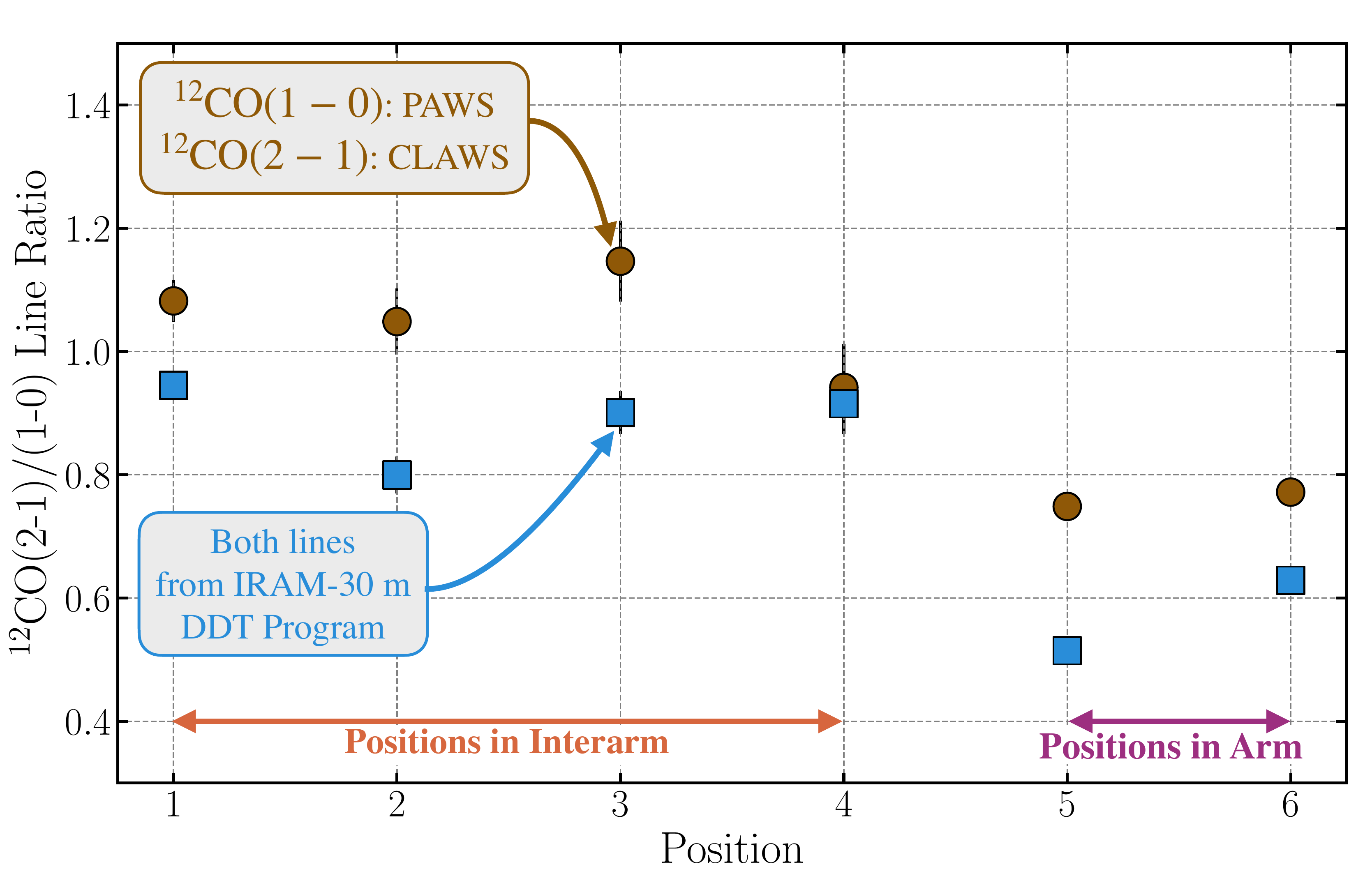}
    \caption{{\bf Arm--interarm CO line ratio analysis.} Circles indicate the $\chem{^{12}CO}{21}/\trans{10}$ line ratio for lines extracted over the apertures shown in \autoref{fig:pointing}. The squares show the CO line ratio from the IRAM \mbox{30-m} DDT project E02-20. Positions 1--4 are within the interarm of the galaxy and position~5 and~6 are in the spiral arm region. We see that both show larger line ratios in the interarm regions, leading us to conclude, that the trend we find is not due to issues with the flux calibration uncertainties. The numerical values of the individual points are listed in \autoref{tab:res_e0220_ratio}. }
    \label{fig:result_ddt}
\end{figure}

The comparison of the line ratios is shown in \autoref{fig:result_ddt} and the numerical values are listed in \autoref{tab:res_e0220_ratio}. Circles represent the line ratio using the PAWS and CLAWS data. We notice that the line ratio is elevated in the positions of the spiral arm (1--4). Blue rectangles indicate the line ratio using the acquired DDT observations. While there is a global offset between the ratios measured in both experiments, the trend of larger line ratios in the interarm region still remains. The offset of order $20$ to $30$~per~cent between the two data sets is mainly due to an overall calibration difference. We find that the \chem{CO}{10} line intensities are systematically higher by $20$ to $30$~per~cent for the DDT EMIR observations compared to the \chem{CO}{10} data from PAWS, which reduces the line ratio overall (notice, that the discrepancy between NRO and the PAWS \chem{CO}{10} is still grater. Especially in the interarm region, the NRO data are larger by a factor ${\sim}2$). This is related to a change of the calibration strategy of the EMIR receivers which happened in February 2017 when the calibration software swapped from \texttt{MIRA} to \texttt{MRTCAL}. To first order, the calibration factor applied to the spectrometer data is proportional to the measured system temperature computed on the calibration scan. While \texttt{MIRA} was computing this system temperature on spectral chunks of 4\,GHz, \texttt{MRTCAL} computes it every 20\,MHz. \citet{Marka2017} shows that this leads to an overestimation of the system temperature for lines that lies at the edges of the atmospheric windows as this is the case for \chem{^{12}CO}{10} whose rest frequency lies inside the wings of the di-oxygen telluric line. In this case, calibrating the PAWS data with \texttt{MRTCAL} would lead to higher (and more accurate) system temperature, and thus higher line brightnesses.

Relative flux calibration can be significant. But our analysis leads us to conclude that relative flux calibration is not the main cause of the arm--interarm trend, since we find also larger line ratio values in the interarm region using the newer IRAM \mbox{30-m} DDT observations.

 \begin{table}
     \centering
     \caption{Comparison of the $^{12}$CO\,\mbox{(2--1)/(1--0)} line ratio within the selected positions 1--6 using either the CLAWS \chem{^{12}CO}{21} and PAWS \chem{^{12}CO}{10} dataset or the newly acquired IRAM \mbox{30-m} data of project E02-20. For the CLAWS/PAWS $R_{21}^{\chem{^{12}CO}}$, we provide the value without (raw) and with error beam correction (based on error beam contribution indicated in \autoref{fig:err_contr}).}
     \label{tab:res_e0220_ratio}
     \begin{tabular}{c c c  c c}
        \hline
          & Position & \multicolumn{3}{c}{$R_{21}^{\chem{^{12}CO}}$}  \\
          & & \multicolumn{2}{c}{CLAWS/PAWS data} & E02-20 data \\
          &&raw&e.b. corrected&\\\hline \hline
          \multirow{4}{*}{Interarm} & 1 & $1.08\pm0.03$&$0.94\pm0.03$&$0.87\pm0.02$ \\
          &2&$1.05\pm0.05$&$0.84\pm0.05$&$0.80\pm0.03$ \\
          &3& $1.15\pm0.07$&$0.91\pm0.06$&$0.90\pm0.04$ \\
          &4& $0.94\pm0.07$&$0.65\pm0.05$&$0.92\pm0.05$ \\\hline
           \multirow{2}{*}{Arm} &5& $0.75\pm0.01$&$0.70\pm0.01$&$0.52\pm0.01$ \\
          &6 & $0.77\pm0.02$&$0.74\pm0.01$&$0.63\pm0.01$ \\ \hline
     \end{tabular}
 \end{table}
         
    

\section{Products for Public Data Release}

Along with this survey paper, we provide several data products for the various spectral lines. The data products are stored on the IRAM server.\footnote{\url{https://www.iram-institute.org/EN/content-page-434-7-158-240-434-0.html}} The data has been processed following the methodology adopted for the IRAM Large Programmes EMPIRE \citep{Donaire2019} and HERACLES \citep{Leroy2009}. The IRAM repository for Large Programmes provides the following, non-error beam corrected data products: 
\begin{itemize}
    \item For all lines: 
    \begin{itemize}
        \item 3D data cubes
        \item RMS/uncertainty maps
        \item integrated brightness temperature (moment-0) maps
    \end{itemize}
    \item For \chem{^{12}CO}{21} and \chem{^{13}CO}{10} only:
    \begin{itemize}
        \item intensity-weighted velocity (moment-1) maps
        \item equivalent width maps
        \item peak temperature maps
    \end{itemize}
\end{itemize} 

We refer the reader to the \texttt{Readme} file at the IRAM data repository for more detailed information. When using this dataset or parts of it, please cite this paper.

\end{appendix}

\end{document}